\documentclass[aps,twocolumn,superscriptaddress]{revtex4-2}
\usepackage{graphicx} 
\usepackage{amsmath,amssymb}  
\usepackage{bm}  
\usepackage[pdftex,bookmarks,colorlinks,breaklinks]{hyperref}  

\usepackage{subfigure}
\usepackage{float}
\makeatletter
\let\newfloat\newfloat@ltx
\makeatother

\usepackage{algorithm}
\usepackage[noend]{algpseudocode}

\DeclareUnicodeCharacter{2009}{\,} 

\newcommand{\rev}[1]{#1}

\begin{document}

\title{High-performance cellular automaton decoders for quantum repetition and toric code}

\author{D. Winter}
\email{d.winter@fz-juelich.de}
\affiliation{Peter Gr{\"u}nberg Institute, Theoretical Nanoelectronics, Forschungszentrum J{\"u}lich, D-52425 J{\"u}lich, Germany}
\affiliation{Institute for Quantum Information, RWTH Aachen University, D-52056 Aachen, Germany}

\author{T. L. M. Guedes}
\affiliation{Peter Gr{\"u}nberg Institute, Theoretical Nanoelectronics, Forschungszentrum J{\"u}lich, D-52425 J{\"u}lich, Germany}
\affiliation{Institute for Quantum Information, RWTH Aachen University, D-52056 Aachen, Germany}
\affiliation{International Institute of Physics, Federal University of Rio
Grande do Norte, Natal 59078-970, Brazil}

\author{M. M{\"u}ller}
\affiliation{Peter Gr{\"u}nberg Institute, Theoretical Nanoelectronics, Forschungszentrum J{\"u}lich, D-52425 J{\"u}lich, Germany}
\affiliation{Institute for Quantum Information, RWTH Aachen University, D-52056 Aachen, Germany}

\begin{abstract}
Execution of quantum algorithms on large-scale quantum computers will require extremely low logical error rates and scalable decoding architectures. Local decoders are promising candidates for this task because they avoid the communication and data-processing bottlenecks inherent in global decoding strategies. Here, we present \texttt{SCALA} (Signaling CA with Local Attraction), a non-hierarchical cellular automaton decoder for quantum repetition and toric codes, and compare it with the hierarchical CA decoder proposed by Harrington \cite{harrington_analysis_2004}. \texttt{SCALA} combines constant local memory and simple nearest-neighbor signaling with a system-size-dependent global reset schedule. We find \rev{a code-capacity threshold estimate of $p_c\simeq7.40(5)\%$} and sub-threshold scaling approaching $p_L\propto p^{d/4}$ for the toric code. 
\rev{For the finite system sizes accessible numerically, and despite numerical evidence for the absence of a finite critical threshold, \texttt{SCALA} also provides strong logical-error suppression under phenomenological noise while retaining its low local computational overhead. \texttt{SCALA} furthermore remains robust when its internal signals are themselves noisy.}
Together with its simple local dynamics, this makes \texttt{SCALA} particularly interesting for low-overhead real-time decoding and motivates its exploration as a quantum cellular automaton for measurement-free quantum error correction.
\end{abstract}

\maketitle

\section{Introduction}\label{intro}

A fundamental challenge in quantum computing is the high sensitivity of qubits to environmental noise. To extend the lifetime of quantum information under noisy conditions, quantum error correction (QEC) techniques have been developed \cite{nielsen_quantum_2010, terhal_quantum_2015, roffe_quantum_2019, campbell_roads_2017, demacedoguedes_short_2026}. In QEC, quantum information is redundantly encoded into a larger system of physical qubits --- a quantum error correcting code (QECC) --- characterized by its code distance $d$, which represents the minimum number of physical operations required to change the logical state. This redundancy ensures that the encoded logical information is subjected to a much lower logical error rate ($p_L$) compared to the physical noise rate ($p$), provided $p$ is below a certain critical threshold ($p_c$). For many code families, increasing the code distance $d$ can lead to an exponential suppression of $p_L$ below threshold, typically following a scaling of $\mathcal{O}(p^{\lambda(d)})$. The parameter $\lambda(d)$ represents the effective code distance, characterizing the error-correction capability of the system. This value depends on both the specific code and the decoder, and is at most linear in the code distance $d$.

For many applications \cite{shor_algorithms_1994, grover_fast_1996}, logical error rates below $10^{-10}$ are required \cite{kivlichan_improved_2020, campbell_early_2021, acharya_quantum_2025}, which is significantly smaller than the error rates of today's most accurate physical entangling gates ($p \approx 10^{-3}$) \cite{decross_computational_2025, mckay_benchmarking_2023}. To bridge this gap, which will persist even taking into account future quantum hardware improvements, one needs large QECCs with many physical qubits and correspondingly scalable decoding algorithms. Realizing scalable decoding, however, presents a major challenge. Conventional decoders like Minimum-Weight Perfect Matching (MWPM) \cite{dennis_topological_2002, edmonds_paths_1965, fowler_practical_2012} exhibit a computational complexity that scales polynomially with system size. While formally efficient, this still poses demanding requirements for real-time decoding, in particular for fast quantum processors such as those based on superconducting qubits \cite{acharya_quantum_2025, choucair_rigetti_2024, skoric_parallel_2023, battistel_realtime_2023, ziad_local_2025, senior_scalable_2025, lee_scalable_2025, caune_demonstrating_2024}. Tensor-network decoders \cite{piveteau_tensor_2024} suffer from exponential computational cost with increasing code size, and neural network decoders \cite{torlai_neural_2017, varbanov_neural_2025} require demanding training and inference resources, while often lacking the flexibility to be applied to codes of varying sizes. Local decoders \cite{holmes_nisq_2020, ravi_better_2023, ueno_qecool_2022, chan_actis_2023}, by contrast, process classical error syndrome information from a fixed-size neighborhood, thereby decoupling their computational complexity from the system size. Nevertheless, local decoding strategies face the challenge of achieving competitive performance in terms of critical error thresholds and sub-threshold scaling due to their locality constraint as compared to global decoding strategies.

In this work, we focus on a particular class of local decoders, cellular automata (CA) \cite{vonneumann_theory_1966, wolfram_cellular_1984, wolfram_computation_1984, wolfram_statistical_1983}. A CA is a discrete computational model consisting of a grid of cells, where each cell's state evolves based on its local neighborhood. CA decoders can be classified into hierarchical \cite{harrington_analysis_2004, breuckmann_local_2017, balasubramanian_local_2024} and non-hierarchical \cite{vasmer_cellular_2021, lake_fast_2025, herold_cellular_2017, herold_cellularautomaton_2015}. While hierarchical CAs use a dynamical coarse-graining procedure to correct errors of increasing size at increasing hierarchy levels, non-hierarchical automata correct errors of all sizes through a collective, decentralized evolution.

Beyond scalability, a critical prerequisite for large-scale quantum computation is real-time operation. Decoders must correct errors faster than they can accumulate to prevent the generation of uncorrectable errors between certain gate operations \cite{terhal_quantum_2015}. Recent experiments have demonstrated first real-time decoders \cite{battistel_realtime_2023} on small-scale systems, typically using parallelized MWPM, predecoding, or simple lookup tables \cite{acharya_quantum_2025, riverlane_introducing_2025, zhao_realization_2022, paetznick_demonstration_2024, alavisamani_promatch_2024}. While these approaches are highly effective for current devices, scaling them to larger codes presents significant challenges regarding computational overhead and integration with low-latency hardware. These constraints motivate the exploration of alternative paradigms, such as local decoders, which are designed to maintain high throughput as the number of physical qubits increases.

To evaluate the potential of different CA strategies, we compare the architectural trade-offs between hierarchical and non-hierarchical designs. \rev{We also compare \texttt{SCALA} with other recent local decoders in terms of their error suppression, local resource requirements, and robustness to noise.} For the hierarchical case, we examine a version of the decoder proposed by Harrington \cite{harrington_analysis_2004}, a seminal reference for renormalization-group (RG)-type \cite{duclos-cianci_renormalization_2010, rozendaal_analysis_2024} CA decoding, incorporating the specific refinements and rule set detailed in Ref.~\cite{breuckmann_local_2017}. \rev{Harrington's decoder provides a useful reference for studying how a hierarchical architecture affects decoding performance, scalability, and robustness. A one-dimensional reduction presented in App.~\ref{app:har1d} makes its concatenated-decoding structure explicit and shows how lower-level decoding failures can propagate through the hierarchy. It further illustrates how noise in the decoder's internal signaling dynamics can generate extended error chains and severely degrade its performance.}

To address these limitations, we develop and benchmark \texttt{SCALA}, a non-hierarchical CA decoder that treats all error sizes within a single, decentralized computational layer. \rev{We show that this architecture combines strong logical-error suppression with constant local memory and simple local dynamics.} Crucially, \texttt{SCALA}'s local, attractive dynamics exhibit significantly stronger resilience against signaling noise, which, combined with a locally modular architecture where local resources are independent of system size, \rev{offers a low-overhead path toward high-throughput online decoding}. \rev{The required system-size-dependent reset is a global synchronization operation rather than an increase in local decoder complexity, providing a practically inexpensive relaxation of strict locality on architectures with native global control.} Finally, while the local, noise-robust nature of this dynamics suggests a potential path toward measurement-free decoding realized as a quantum cellular automaton (QCA) \cite{farrelly_review_2020, arrighi_overview_2019, guedes_quantum_2024, wintermantel_unitary_2020, watrous_onedimensional_1995}, such an implementation is in itself highly non-trivial and beyond the scope of this work.

\subsection{Summary of results}\label{summary_results}

In this work, we explore the architectural trade-offs between CA decoding strategies and show that non-hierarchical cellular automaton decoders can offer distinct advantages over hierarchical ones in terms of scaling and noise resilience. \rev{Harrington's decoder \cite{harrington_analysis_2004} was a pioneering proposal for local decoding of topological quantum codes, introducing a hierarchical architecture for processing syndrome information and coordinating corrections through local classical operations. Despite its foundational role, its quantitative performance and robustness across different noise models have remained comparatively underexplored. We therefore use Harrington's decoder as a representative hierarchical CA and compare it with our non-hierarchical \texttt{SCALA} decoder in three main regards: performance, scalability, and robustness.}

\begin{enumerate}
    \item \textit{Performance.} As a representative of renormalization-group decoders, the Harrington CA has an expected asymptotic effective-distance scaling $\lambda(d)=d^{\log_3 2}$ and a code-capacity threshold of \rev{$p_c=4.523(2)\%$}. This sublinear scaling is common to hierarchical structures, where error propagation between levels limits logical error suppression. Our non-hierarchical CA decoder, on the other hand, has a code-capacity threshold \rev{estimated as $p_c\simeq7.40(5)\%$} and an effective distance that scales asymptotically linearly with $d$. In the one-dimensional setting, we provide numerical evidence that our decoder is optimal under code-capacity noise. \rev{Under phenomenological noise, however, the finite-size crossover points move toward lower physical error rates as the code distance increases. In \texttt{SCALA1D}, we trace this behavior to screening of the signal-mediated attraction as continuously generated defects interfere with each other's signals.}
    \item \textit{Scalability.} When increasing the code size, the computational resources needed per CA cell in Harrington's decoder also grow, whereas in our non-hierarchical design, the memory requirement per cell stays constant, yielding a modular local design. \rev{The remaining system-size dependence of \texttt{SCALA} enters only through a global reset schedule rather than through growing local computational resources. Such global control operations can be comparatively inexpensive on platforms that natively support parallel control.}
    \item \textit{Robustness.} Moreover, we show that our non-hierarchical design remains functional even when we allow for noise inside the CA dynamics, whereas Harrington's decoder breaks down in this scenario. \rev{For \texttt{SCALA}, signal noise primarily introduces an offset in logical-error performance, while the distance-dependent scaling remains approximately unchanged within the accessible system sizes.} \rev{This robustness reduces the requirement for perfectly reliable decoder hardware and motivates a future exploration whether the local \texttt{SCALA} dynamics can be realized as a QCA for measurement-free QEC.}
\end{enumerate}

The remainder of this paper is structured as follows. In Sec.~\ref{sec:qecc}, we briefly review the quantum repetition and toric codes, followed by a brief definition of cellular automata which provide the basis for Harrington's decoder in Sec.~\ref{sec:har2d}. \rev{We use locality to denote constant-size local state and nearest-neighbor communication, unless stated otherwise. A one-dimensional reduction of Harrington's decoder is presented in App.~\ref{app:har1d}, where we analyze its hierarchical structure, logical failure mechanisms, and robustness to noise in the CA dynamics.} Readers primarily interested in our new decoder may proceed directly to Sec.~\ref{sec:scala1d}, \rev{where we introduce \texttt{SCALA1D} and analyze its performance on the repetition code.} Subsequently, in Sec.~\ref{sec:scala2d}, we extend these dynamics to two dimensions and obtain the \texttt{SCALA2D} decoder for the toric code. We provide numerical simulations and analyze \rev{its performance, scalability, and robustness compared to Harrington's decoder.} Finally, we conclude and provide an outlook on future work in Sec.~\ref{sec:conc}.

\subsection{Repetition and toric codes}\label{sec:qecc}

In the following, we briefly review the two QECCs which the one and two-dimensional versions of Harrington's decoder and \texttt{SCALA} attempt to protect, the quantum repetition code and the toric code. Both of these codes have been extensively studied \cite{fowler_practical_2012, raussendorf_topological_2007, dennis_topological_2002} and experimentally realized across multiple platforms, including superconducting circuits \cite{acharya_quantum_2025, krinner_realizing_2022} and, more recently, reconfigurable neutral atom arrays \cite{bluvstein_faulttolerant_2026, ebadi_quantum_2021, graham_multiqubit_2022, evered_highfidelity_2023, bluvstein_logical_2024} in the form of planar surface codes. We start with the repetition code with periodic boundary conditions. The $n$-qubit, one-dimensional quantum repetition code is constructed by placing qubits on the edges of a cycle. One logical qubit is encoded into $n$ physical qubits by mapping the basis states $|0\rangle\rightarrow |0_L\rangle \equiv |0\rangle^{\otimes n}$ and $|1\rangle\rightarrow |1_L\rangle \equiv |1\rangle^{\otimes n}$, such that any state $|\psi_L\rangle=\alpha |0_L\rangle + \beta |1_L\rangle$ becomes a valid code word. The quantum repetition code can either protect against bit-flip (Pauli-$X$) or phase-flip (Pauli-$Z$) errors but not both. We focus on the bit-flip quantum repetition code with logical operator $X_L=X^{\otimes n}$, flipping the logical states $|0_L\rangle \leftrightarrow |1_L\rangle$, i.e., $X_L|0_L\rangle=|1_L\rangle$. On the other hand, any Pauli-$Z$ operator (or an odd number thereof) acts as a logical-$Z$ operator on $|\psi_L\rangle$. The qubit support of the smallest logical operator determines the distance $d$ of a code. When we only consider bit flips, the bit-flip repetition code has a distance of $d=n$. The stabilizer generators of the repetition code, $S_i=Z_iZ_{i+1}$, correspond to the $Z$-parity between pairs of nearest neighbor qubits for $i\in \mathbb{Z}_{n-1}$. Measuring the generators has no effect on the encoded information but yields valuable information about the errors which occurred. The collection of stabilizer generator measurements is called the error \textit{syndrome}. Each non-trivial measurement outcome within the syndrome is referred to as \textit{defect}.

The decoding problem is to correctly deduce the logical error class (i.e., the equivalence class of physical errors modulo stabilizers) given the measured syndrome. An optimal decoder, also called \textit{maximum-likelihood (ML) decoder} \cite{nielsen_quantum_2010}, is a solution to the decoding problem which finds the most probable equivalence class of physical errors (and hence the corresponding logical error class) consistent with the syndrome. For the repetition code, a perfect matching of defects, minimizing the sum of distances between defect pairs, is an ML decoder. A \textit{correction} then involves applying Pauli-$X$ operators to each physical qubit enclosed by a matched defect pair. A correction can have two outcomes. Either all errors that occurred are canceled out or the correction operators and the errors form the logical operator $X_L$, corresponding to a \textit{logical error}. The minimum-weight perfect matching (MWPM) decoder \cite{edmonds_paths_1965, fowler_practical_2012} is equivalent to performing a majority vote on the $X$-errors of the code. If more than half of $n$ qubits have an error, where $n$ is restricted to odd integers for simplicity, the decoder output will lead to a logical error. The \textit{logical error rate} $p_L$ as a function of odd $n$ and the bit-flip error probability $p$ in the code capacity setting, i.e., bit-flip noise only on the initial state, can thereby be obtained from the binomial distribution,

\begin{equation}
    p_L = \sum_{k=\lceil n/2 \rceil}^{n}{n\choose{k}}p^k(1-p)^{n-k}
    \label{eq:pL_ML}.
\end{equation}

\begin{figure}[hbtp]
    \centering
    \includegraphics[scale=0.7]{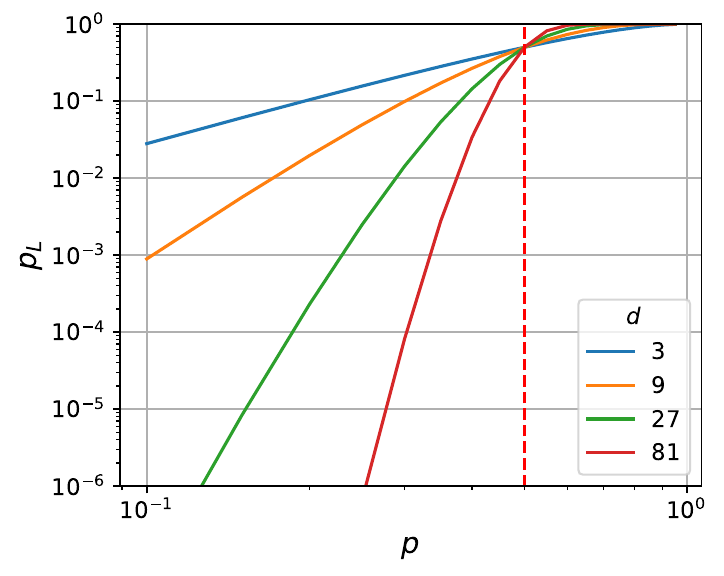}
    \caption{Analytical logical error rate $p_L$ as a function of bit-flip error rate $p$ for the ML decoder (Eq.~\ref{eq:pL_ML}) of the repetition code. Different colors correspond to different code distances $d$ (equivalent to number of qubits $n$ making up the repetition code), the dashed red line marks the QEC threshold $p_c=0.5$.}
    \label{fig:rep_mwpm}
\end{figure}

Fig.~\ref{fig:rep_mwpm} shows the logical error rate $p_L$ for various system sizes. In the sub-threshold regime ($p<0.5$), increasing the system size provides better protection, as errors are suppressed more effectively with additional physical qubits. This value $p_c\equiv p=0.5$ is called the QEC \textit{threshold}. In this regime, $p_L$ scales with $p^{\lambda(d)}$, with $\lambda(d)=(d+1)/2$ and $d=n$, corresponding to the minimum number of physical errors for which the matching decoder gives a logical error. 

\begin{figure}[hbtp]
    \centering
    \includegraphics[scale=0.15]{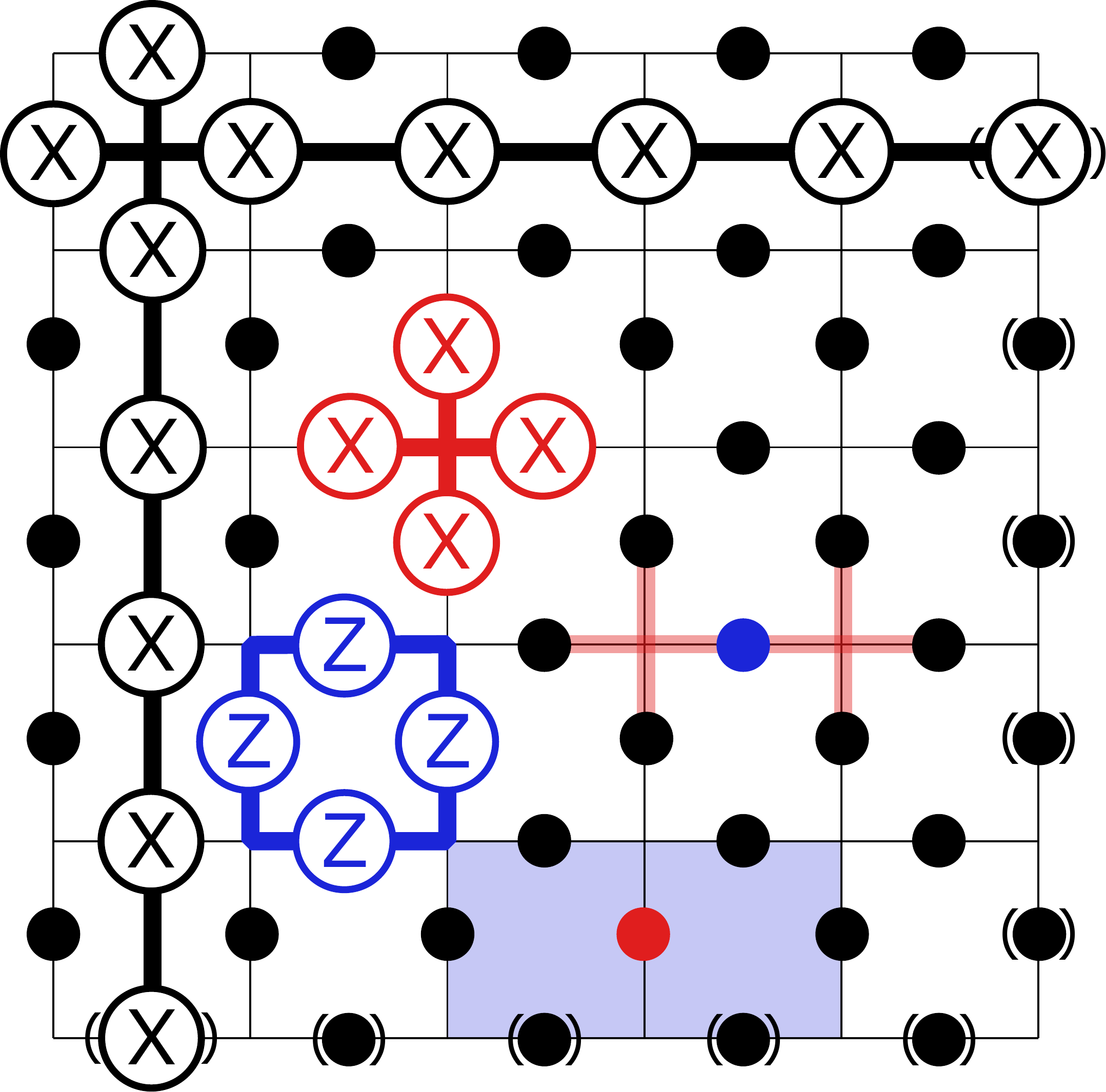}
    \caption{Illustration of a distance $d=5$ toric code with qubits (black circles) on edges of a regular $d\times d$ lattice under periodic boundaries. Qubits in parenthesis wrap around the torus. The minimum-weight logical operators $X_L^1$ and $X_L^2$ are vertical and horizontal strings of single-qubit Pauli-$X$ operators. Logical $Z$-operators are omitted. The stabilizers of the toric code are either of pure $X$- (red stars) or $Z$-type (blue plaquettes). A $X$-error (red dot) anti-commutes with two $Z$-stabilizers, causing two defects (blue shades). A $Z$-error (blue dot) causes the star syndrome (red shades). Any $X$ error configuration which forms a (product of) stabilizer generator(s) commutes with the $Z$-stabilizers, thus yields a trivial syndrome.}
    \label{fig:toric_code}
\end{figure}

Another code which, however, is capable to protect against both $Z$- and $X$-errors is the toric code \cite{kitaev_faulttolerant_2003}, shown in Fig.~\ref{fig:toric_code}. The toric code is defined on a $n\times n$ regular square lattice with periodic boundary conditions, i.e., a torus. Physical qubits are placed on the $2n^2$ edges of the lattice, encoding two logical qubits. The stabilizer generators come in two flavors, of pure $X$- or $Z$-type, making the toric code a \textit{Calderbank-Steane-Shor (CSS)} code \cite{calderbank_good_1996, steane_multipleparticle_1997, steane_error_1996}. This property allows one to decode the $X$- and $Z$-syndrome independently. In present work, we focus solely on bit-flip errors which are detected by the $Z$-type generators, which are four-body Pauli-$Z$ operators (blue) acting on the qubits surrounding a face of the lattice. The $X$-type stabilizers are four-body Pauli-$X$ operators (red) acting on the qubits adjacent to a vertex of the lattice. Due to their shape, $Z$-stabilizers are referred to as \textit{plaquette} and $X$-stabilizers as \textit{star} operators \cite{dennis_topological_2002}. Owing to the symmetry between $X$ and $Z$ sectors, we describe the logical operators and errors in terms of the $X$-type; an analogous description applies to the $Z$-type on the dual lattice.

Logical-$X$ operators of minimal weight are formed by horizontal and vertical non-trivial loops of Pauli-$X$ operators (black) wrapping around the torus, corresponding to the logical generators of two logical qubits, which have even overlap (and therefore commute) with all $Z$ stabilizers. Since the minimal number of qubits needed to form a non-trivial loop is $n$, the distance of the toric code is $d=n$. Logical operators in the toric code are not unique. Any product of a logical operator with a stabilizer is also a logical operator of the same class. Similarly, $X$-errors forming (products of) star operators commute with the $Z$-stabilizers and thus do not change the logical error class.

After decoding, a correction on the toric code can have three outcomes: (a) all $X$-errors which occurred are successfully removed, or (b) the correction and the error together form an $X$-stabilizer (product of generators), or (c) the correction and the errors complete one of the possible $X_L$ operators, resulting in a logical bit-flip. The toric code can be decoded via MWPM, which is, however, not an ML decoder \cite{dennis_topological_2002}. MWPM decoding on surface and toric codes has been extensively studied \cite{wang_threshold_2010}, yielding a QEC threshold $p_c= 10.3\%$ \rev{for independent bit-flip and phase-flip errors} and a sub-threshold scaling of the logical error rate $p_L \propto p^{(d+1)/2}$. The scaling behavior is identical to the repetition code, as the same number of $X$-errors is needed for the matching decoder to lead to a logical bit-flip.

\subsection{\rev{Phenomenological-noise simulations and logical hazard}}
\label{sec:pheno_methods}

\rev{In contrast to the code-capacity noise model, where errors occur only before decoding, phenomenological noise continuously introduces new data-qubit errors with probability $p$ and syndrome-measurement errors with probability $q$ at each CA step. The decoder must therefore correct errors while new faults continue to occur, using syndrome information that may itself be unreliable.

We quantify decoding performance by the first time $T_F$ at which the residual data-error configuration enters the wrong logical sector, evaluated after each CA update. For SCALA, the corresponding logical failure probability per CA step approaches an approximately stationary value after an initial transient, which we report as the logical error rate $p_L$. For Harrington's decoder, whose failure probability exhibits a more pronounced time dependence, we instead report the mean first-passage lifetime $\langle T_F\rangle$. The precise definitions, numerical estimators, and stationarity checks are provided in App.~\ref{app:hazard}.}

\section{Harrington's decoder}\label{sec:har2d}

Harrington's decoder \cite{harrington_analysis_2004} is a cellular-automaton decoder for the toric code. A cellular automaton (CA) is a discrete model consisting of a grid of cells, each of which can be in one of a finite number of states. An automaton \textit{configuration} is an assignment of states to all cells. The CA evolves a configuration to a new configuration via a synchronous \textit{global update} function. The global update is a composition of a smaller \textit{local rule}, which outputs a cell state in the new configuration based on the states of cells in a finite \textit{neighborhood}, applied to every neighborhood of the current configuration. This local interaction and parallel evolution allows for complex emergent global behavior based on simple underlying local rules. For a global task like error correction some means of communication between distant cells is needed. In the automata we study, this communication is made explicit by introducing signals as part of a cell's state space and signal propagation as part of the local rule. Next, we detail the construction of Harrington's decoder.

\begin{figure*}[hbtp]
    \centering
    \includegraphics[scale=0.255]{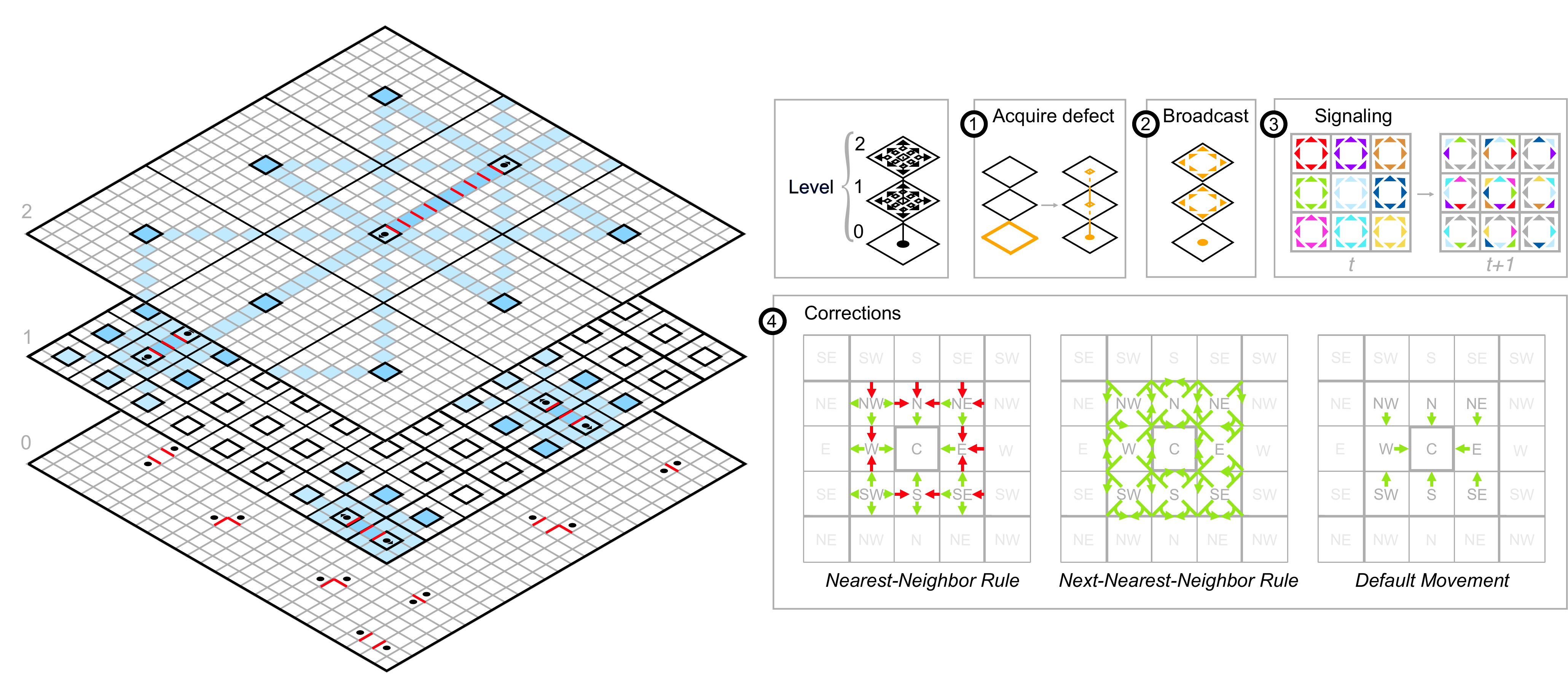}
    \caption{Sketch of Harrington's hierarchical CA decoder. On the left, we show a distance-$27$ toric code from the perspective of the three hierarchy levels Harrington's decoder imposes. Error correction is accomplished by moving defects (black circles) which are close towards each other thereby correcting the errors (red lines) which separate them. At the lowest level (0), one- and some two-qubit errors are corrected within at most two CA steps. Defects which have no other defect in their local neighborhood move towards the center of their colony (group of $Q\times Q$ cells) marked in thick black lines at the next hierarchy level (1). At level-1, defects located at colony centers communicate by exchanging \texttt{CountSignals} (blue color; shade represents amount of signals) via the CA cells in between. These signals are counted for $U$ CA steps, and compared to the threshold values $f_CU,f_NU$ to yield level-1 defects, as explained in detail in the main text. Based on the level-1 defects, the colony center cells calculate level-1 corrections. This correction decision is then propagated via \texttt{FlipSignals} (not shown) to the intermediate cells within the next $Q$ time steps. Similar to level-0, level-1 defects which have no other level-1 neighbor defect are moved to the supercolony center (i.e., the $Q\times Q$ group of level-0 colony centers) which are shown in thick black lines at the next hierarchy level (2). At level-2, the exact same procedure is executed between level-2 defects, communicated by level-2 \texttt{CountSignals} and \texttt{FlipSignals}. In the top center, we show a sketch of a CA cell. A cell contains one syndrome bit at level-0 (black circle) which is connected to one \texttt{DefectCounter} (white square in cell center) per higher level, i.e., for all level $k\geq 1$. Additionally, for every higher level, a cell has eight \texttt{CountSignal} bits (black triangles) each one with attached \texttt{SignalCounter} (white squares), incrementing for every received \texttt{CountSignal} and four \texttt{FlipSignals} (not shown), one for each cardinal direction. Additionally, a cell has memory fields for the working period $U=10$, the linear size of a colony, $Q=3$, the threshold constants $f_C$, $f_N$ and an \texttt{Age} counter which increments each CA step (not shown). Furthermore, for each level a cell has an \texttt{Address} field (not shown), corresponding to the cell's relative position inside its level-$k$ colony. In the boxes labeled (1 --- 4), we subdivide the local rule of Harrington's CA into subrules, as detailed in the main text. In orange, we highlight parts of a cell's memory involved in a substep. In 3), we use color to illustrate the exchange of signals by assigning unique colors to cell signals before the exchange occurs. In 4), green color represents a qubit correction which is executed by the cell from which the arrow emerged if both the cell and its pointed to neighbor cell have a defect present. A red arrow represents a cell performing no action when itself and its corresponding neighbor both have a defect present. For next-nearest neighbors, the green line represents the presence of defects at a cell and its (diagonal) neighbor which causes a correction to the qubit to which the attached arced arrow points. For a detailed description of the local rule we refer to the main text.}
    \label{fig:harrington2d_fig1}
\end{figure*}

\subsection{Construction}\label{sec:har2d_construction}

Harrington's automaton features a hierarchical design. Similar to renormalization group \cite{duclos-cianci_renormalization_2010, rozendaal_analysis_2024} and concatenated decoders, error chains of increasing length are corrected at increasing levels of a hierarchy by coarse-graining the syndrome with increasing granularity. The automaton operates on classical information obtained from syndrome measurements and applies corrections directly to the toric code data qubits in discrete time steps. Inspired by the work of Gacs \cite{gacs_onedimensional_1978, gray_readers_2001,  gacs_reliable_1983, gacs_reliable_2001}, corrections of larger error chains at higher hierarchy levels mimic the corrections at the lowest level via \textit{self-reproduction}. In the following, we explain each of these design aspects in detail.

Since the toric code is a CSS code, phase- and bit-flip errors are corrected by two independent instances of Harrington's decoder, respectively taking $X$- and $Z$-syndromes as input. We focus solely on bit-flip errors which are detected by the plaquette operators.

\subsubsection{Hierarchy}

The hierarchy of the automaton is defined by recursively partitioning the toric-code lattice into functional groups. At the lowest level (level-0), $Q\times Q$ cells are grouped into a \textit{colony}. In this work, we fix $Q=3$ which was found to be optimal in terms of QEC performance \cite{breuckmann_local_2017, harrington_analysis_2004}. Colonies are further grouped into functional groups of size $Q\times Q$ to form a higher-level (level-1) colony. At level 2, $Q\times Q$ level-1 colonies are grouped into a level-2 colony and so on, until the entire toric code lattice is contained within one level-$m$ colony. Due to this recursive partitioning, the only allowed code distances are powers of $Q$, i.e., $3,9,27,81$, and so on. On the left in Fig.~\ref{fig:harrington2d_fig1}, we show an example toric code of distance $d=27$ and its working mechanism on each of its $3$ levels.

For each colony, we choose the center cell to be its representative (thick black lines) at the next higher level. The distance between level-$k$ colony centers, $Q^k$, thereby defines a natural length scale for each level of the hierarchy, corresponding to the length of error chains which can be corrected at level $k$. According to the length scale $Q^k$, we define a corresponding time scale, $U^k$, at which level-$k$ error correction occurs. In this work, we follow \cite{breuckmann_local_2017} and fix $U=10$, thus at level-0 corrections happen every $U^0=1$ time steps, while a correction at level-1 occurs every $U^1=10$ time steps and so on. Next, we discuss the state space of a single CA cell.

\subsubsection{State space}

The smallest unit of a cellular automaton is a cell. As CA cells correspond to stabilizer operators, there are in total $d^2$ cells on a distance-$d$ toric code. The state of a cell, as illustrated in the top center of Fig.~\ref{fig:harrington2d_fig1}, is composed of several classical memory fields. One \texttt{Defect} bit (black circle) is set to the measurement result of the plaquette operator at its location at the beginning of each CA step. The defect bit is counted by $m-1$ \texttt{DefectCounters}, indicated as white squares in the cell's center and attached to the defect bit of level-0. These counters accumulate the defect bits for $U^k$ CA steps and are afterwards reset. Furthermore, a cell has $m$ \texttt{Address} fields (one for each level \rev{$k=0,\ldots,m-1$}) representing a cell's relative position within its level-$k$ colony. The relative positions within a colony correspond to one of four cardinal, four intercardinal and center locations, $\mathcal{L}=\{N,W,E,S,NW,NE,SW,SE,C\}$. Additionally, we include an empty location, $\circ$, which is used as an address for level-$k$ cells that are not part of the level-$k$ hierarchy, i.e., are not center cells of level-$(k-1)$ colonies. We define the set of all locations $\mathcal{L}_{\circ}=\mathcal{L}\cup \{\circ\}$. Moreover, for each level, a cell contains eight \texttt{CountSignal} (black triangles) and four \texttt{FlipSignal} bits (not shown) for communication between level-$k$ representatives. \texttt{CountSignals} at each level are counted by eight \texttt{SignalCounters} for $U^k$ steps after which they are reset. To synchronize local updates of cells, each cell has an \texttt{Age} field, incrementing with each CA step. Finally, a cell stores four additional constants, the working period $U$, the linear size of a colony $Q$ and two so-called threshold values $f_C, f_N \in [0,1]$. These constants play a role in higher-level corrections and are further explained in Sec.~\ref{sec:selfsim}. Data qubits on which the automaton applies corrections are not included in the cell-state space as the syndrome which is relayed to the CA indirectly includes this information. As Ref.~\cite{breuckmann_local_2017} pointed out, when increasing the code distance $d$ the cell-state space, in particular the signal bits and counters, also grow, which breaks strict locality in the sense of constant-size local state space of the automaton.

\subsubsection{Local rule}

The local rule takes the states of CA cells in a \textit{Moore neighborhood}, i.e., a central cell plus its eight closest neighbors, and updates the state of its central cell. Due to the complexity of Harrington's local rule, we split its definition into four sub-rules, which are illustrated in Fig.~\ref{fig:harrington2d_fig1} (1 -- 4). 
\begin{enumerate}
    \item The measurement result of the plaquette stabilizer overwrites the \texttt{Defect} bit and adds its value to all $m-1$ \texttt{DefectCounters} as shown in Fig.~\ref{fig:harrington2d_fig1} (1).
    \item Colony center cells broadcast their defect value to all their \texttt{CountSignal} bits (Fig.~\ref{fig:harrington2d_fig1} (2)).
    \item All cells propagate their signals to all eight neighbors on all levels. Propagation between cells happens at every hierarchy level and is realized by exchanging the corresponding \texttt{CountSignal} and \texttt{FlipSignal} bit values between neighbors, shown in Fig.~\ref{fig:harrington2d_fig1} (3). In the following $Q^k$ time steps, \texttt{FlipSignals} are emitted and \rev{propagate between neighboring cells, including across colony boundaries, to execute the corresponding higher-level correction.}
    \item The decision to apply a correction at any level is based on a) a cell's own relative \texttt{Address} and b) the pattern of \texttt{Defects} in the cell's Moore neighborhood. The specific correction decisions a cell can execute are shown in Fig.~\ref{fig:harrington2d_fig1} (4).
\end{enumerate}

Harrington's original rules prioritize corrections across colony borders. As a consequence, corrections along intercardinal directions across borders are preferred over corrections along cardinal directions within the colony, which can introduce unnecessary bit-flips to the code. Intercardinal corrections within colonies are trivially handled by default movement rules, i.e., situations in which a cell has a defect but its neighbors do not. The authors of Ref.~\cite{breuckmann_local_2017} found that consistently prioritizing cardinal corrections over intercardinal ones yields better performance. To provide a rigorous benchmark, we implement this optimized version of Harrington's local rules. These modified local rules together with a simplified averaging procedure, explained in the following sections, constitute the specific implementation of the Harrington decoder used throughout this work.

In Fig.~\ref{fig:harrington2d_fig1} (4), we show the correction rules based on defects of nearest neighbors (left) and next-nearest neighbors (middle) neighbors and absence of neighbor defects, so-called default movement (right). A cell with a given address (gray label) executes one of eight actions (red and green arrows) or no action based on its own and its neighbors' defect values. A green arrow represents a correction executed by the cell from which the arrow emerges on the qubit to which the arrow points. A red arrow pointing to the center of a cell signifies a cell executes no action. The defect patterns which trigger corrections always consist of two defects, one at the cell from which the arrow emerges and one at the pointed-to neighbor. For next-nearest neighbors, defects must be present at the cell and its (diagonal) neighbor cell, connected by a green diagonal line, in order to cause a correction on the qubit to which the arced arrow points. As an example, a cell at colony address $W$ will flip its western qubit when itself as well as its western neighbor (with colony address $E$) has a defect present while its $W$-neighbor performs no action. Lastly, if a non-center cell has but its neighbors do not have a defect present, the cell pushes its defect towards the center by the default movement rule (Fig.~\ref{fig:harrington2d_fig1} (4) on the right). In order to assure that a cell can at most apply only a single correction to one of its adjacent qubits, a preference scheme is employed. First priority is given to nearest-neighbor corrections, followed by next-nearest neighbor corrections and finally default movement. Furthermore, corrections along the western and southern border have precedence over corrections within a colony. In case of ambiguity, we use the $NESW$ rule, i.e., apply corrections to $N$ before $E$ before $S$ before $W$.

The overall effect these corrections have is that sufficiently isolated errors are corrected immediately, as these lead to neighboring cells having defects that can be handled by corrections along cardinal directions. When two defects meet, they annihilate due to removal of $X$-errors on the data qubits, completion of a stabilizer or of a logical operator. The goal is to eventually annihilate all defects. If there is a chain of errors with defects in two neighboring colonies, the correction rules move those defects to their respective colony centers. Defects at colony centers represent a higher-level syndrome, which is corrected via self-reproduction.

\subsubsection{Self-reproduction}\label{sec:selfsim}

Self-reproduction refers to the emergent behavior where local rules, when iterated across a hierarchical structure, effectively mimic or perform the same function at all levels. The low-level behavior is thereby simulated by the automaton dynamics at different scales. In Harrington's design, the function which is iterated at all levels are the corrections from Fig.~\ref{fig:harrington2d_fig1} (4) mediated by a signaling and counting mechanism described in the following.

Once a colony center cell has a defect present, it first increments all \texttt{DefectCounters} and subsequently broadcasts \texttt{CountSignals} in all eight directions for all levels of the hierarchy, as illustrated by blue color (shade correspond to amount of present signals) to the left of Fig.~\ref{fig:harrington2d_fig1}. \texttt{CountSignals} are propagated, traveling one cell at a time, until reaching their level-$k$ neighbor representative after $Q^{k}$ CA steps. Upon reception, the \texttt{CountSignal} is absorbed by a representative, incrementing its level-$k$ \texttt{SignalCounter} in the corresponding direction. This process continues until the end of a level-$k$ working period, $U^k$. At this point in time, all level-$k$ counters, $c_k^i$ with $i\in\mathcal{L}$ (where $\mathcal{L}$ is the set of all eight neighboring locations plus the center location), are averaged over the working period, $\bar{c}_k^i=c_k^i/U^{k}$ and compared to the threshold values $f_N$ or $f_C$. By this procedure, a local level-$k$ syndrome, i.e., nine defect values within a Moore neighborhood of level-$k$ representatives, is obtained. A cell's own level-$k$ defect is obtained by comparing the averaged level-$k$ \texttt{DefectCounter} value against the threshold $f_C$, $\bar{c}_k^C \geq f_C$, while all averaged level-$k$ \texttt{SignalCounter} values are compared to threshold $f_N$ instead. In Harrington's model, $f_C$ and $f_N$ are hyperparameters. Due to the delay incurred by the distance a signal has to travel until reaching a neighboring representative, $f_N$ is chosen smaller than $f_C$. The values $f_C=9/10$ and $f_N=4/10$ have been numerically found to be optimal for the parameter values $U=10, Q=3$ used in present work. For the numeric optimization, we refer to Ref.~\cite{breuckmann_local_2017}.

A level-$k$ correction is calculated by applying the correction function shown in Fig.~\ref{fig:harrington2d_fig1} (4) to the level-$k$ syndrome (obtained via the averaging procedure described before). The resulting correction decision represents a chain of bit-flips which has to be executed cardinally on all intermediate cells between two level-$k$ representatives. In order to notify the intermediate cells about the correction, another signal, the so-called \texttt{FlipSignal}, is used. A \texttt{FlipSignal} is broadcast at $t=U^k$ in one direction for $Q^k$ time steps, traveling one cell at a time, and reaches the representative neighbor at $t=U^k+Q^k$. Subsequently, all intermediate cells in between the two representatives have an active \texttt{FlipSignal}, thus deciding to flip their qubit in the same direction realizing the desired correction chain. This procedure happens simultaneously at all levels of the hierarchy and is coordinated by the synchronized local \texttt{Age} clocks of each cell. Together, the corrective steps at each hierarchy level enable the decoder to handle error chains at each length scale within their corresponding correction time intervals.

\subsection{Numerical results}

We study the error-correction performance of Harrington's decoder under three noise scenarios: \rev{code-capacity noise, phenomenological noise, and phenomenological noise with additional internal signal errors.} First, we consider code-capacity noise. In this setting, we start with an initially error-free logical reference state. Next, bit-flip errors are independently and randomly applied with probability $p$ on physical qubits. The resulting state is input to the decoder, which runs (noise-free) for as many CA steps as needed until all defects in the error syndrome have been annihilated. Finally, we check if the correction causes a logical bit-flip on any of the two logical qubits. If this happens, we count a logical error. This process is repeated $N$ times. The resulting ratio of logical error counts over $N$ samples gives an estimate of the logical error rate $p_L$.

\rev{Next, we consider phenomenological noise, where independent bit-flip errors occur on data qubits with probability $p$ and syndrome-measurement outcomes are independently flipped with probability $q$ at each CA step. The decoder operates continuously while new errors accumulate, using the noisy syndrome information to determine corrections. We record the first time $T_F$ at which the accumulated data errors and decoder corrections produce a nontrivial logical operator, and use the resulting first-passage statistics to characterize the logical error rate, as described in Sec.~\ref{sec:pheno_methods}.}

The last noise scenario we consider is phenomenological \rev{noise with additional signal errors}. Signal noise occurs in scenarios where the logic of the decoder is subject to internal errors during execution, for instance, due to noisy hardware components. We apply the phenomenological noise model \rev{with equal data and measurement error rates $p=q$} and additionally apply bit-flip noise \rev{with probability $p_{\mathrm{sig}}$ to all \texttt{CountSignal} and \texttt{FlipSignal} bits at all levels after the CA update at the end of each step.}

\subsubsection{Code-capacity noise}

\begin{figure}[hbtp]
    \centering
    \includegraphics[scale=0.7]{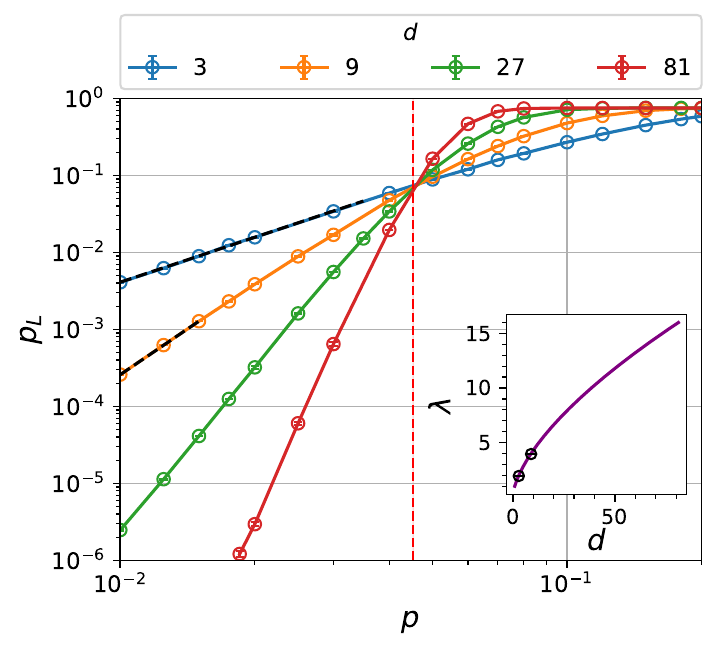}
    \caption{Logical error rate $p_L$ as a function of bit-flip error rate $p$ for Harrington's toric code decoder under code capacity noise for different code distances $d$ (color code). \rev{In the inset, black circles show these fitted exponents with their statistical fit uncertainties, while the purple line shows the expected asymptotic minimum-weight scaling $\lambda(d)=2^{\log_3(d)}\approx d^{0.631}$. For $d=27$ and $81$, the accessible data remain pre-asymptotic and no numerical exponent fit is reported. Fit parameters and fit windows are reported in App.~\ref{app:subthreshold_fits}.}}
    \label{fig:h2d_cc}
\end{figure}

In Fig.~\ref{fig:h2d_cc}, we show our numerical results for Harrington's decoder under code-capacity noise. We perform Monte Carlo simulations with varying number of shots, ranging from $N=10^{4}$ to $10^{6}$ per data point. \rev{Larger sample counts are used at lower physical error rates. A finite-size scaling analysis of the code-capacity data yields a threshold $p_c=0.04523\pm0.00002$ and scaling exponent $\nu=1.997\pm0.020$, as detailed in App.~\ref{app:fss}.} \rev{In the sub-threshold regime, we fit the numerical data to $p_L=A(d)p^{\lambda(d)}$. We report fits only for $d=3$ and $9$, for which the accessible data approach the expected asymptotic minimum-weight scaling; the $d=27$ and $81$ data remain pre-asymptotic. The corresponding fit parameters and fit windows are reported in App.~\ref{app:subthreshold_fits}.} The expected scaling $\lambda(d)=2^{\log_3(d)}\approx d^{0.631}$, where $3$ is the scaling factor of the colony hierarchy, is consistent with the weight of the minimal-error configuration that causes a logical failure. Specifically, these configurations correspond to cases where errors propagate through successive levels of the hierarchy in the most detrimental way, analogous to the 1D failure configurations of a concatenated majority-voting scheme. \rev{For $d=27$ and $81$, the accessible data have not yet reached this asymptotic minimum-weight regime. Targeted tests of the recursive failure configurations nevertheless support the predicted hierarchical mechanism, as detailed in App.~\ref{app:subthreshold_fits}.} These results are similar to the QEC threshold of about $5.2\%$ and same sub-threshold scaling for a (global) renormalization-group decoder \cite{duclos-cianci_renormalization_2010} with block size $3$, as shown in Ref.~\cite{rozendaal_analysis_2024}. \rev{This illustrates the sublinear effective-distance scaling produced by the hierarchical structure of Harrington's decoder and related concatenated block-decoding constructions.}

\subsubsection{Phenomenological noise}

\begin{figure}[hbtp]
    \centering
    \includegraphics[scale=0.7]{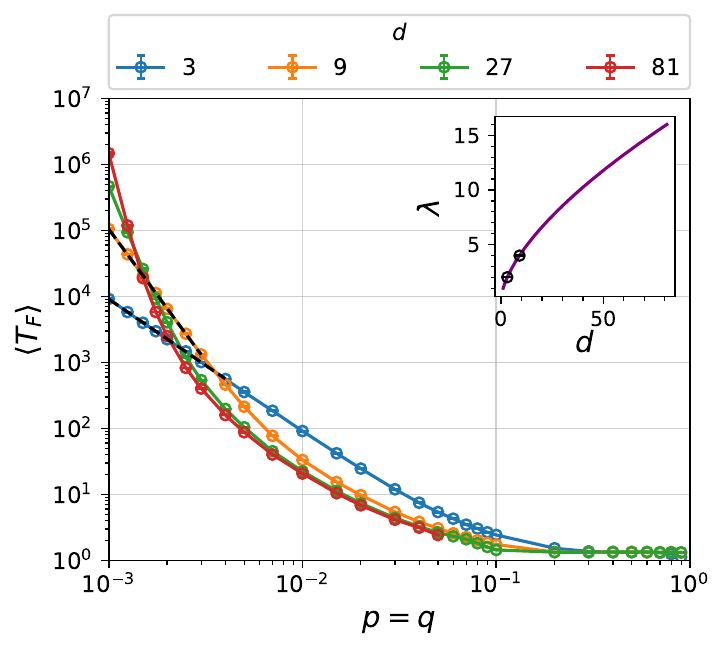}
    \caption{\rev{Average logical lifetime $\langle T_F\rangle$ as a function of equal phenomenological data and measurement error rates $p=q$ for toric codes of distance $d$ (color coded) under correction by Harrington's decoder. Black dashed lines show weighted low-$p$ power-law fits $\langle T_F\rangle=A(d)p^{-\lambda(d)}$ for $d=3$ and $9$. The inset shows the fitted exponents $\lambda(d)$ with their statistical fit uncertainties (black circles), together with the hierarchical scaling prediction $\lambda(d)=2^{\log_3 d}=d^{\log_3 2}$ (purple line). For $d=27$ and $81$, the accessible data remain pre-asymptotic and no numerical exponent fit is reported. Standard errors are shown as black bars.}}
    \label{fig:har2d_pheno}
\end{figure}

In Fig.~\ref{fig:har2d_pheno}, we show numerical results for Harrington's decoder under a phenomenological noise model, i.e., bit-flip errors can occur on data qubits with rate $p$ and on syndrome measurements with rate $q$ before every CA step. \rev{The numerical data use $N=10^3$--$3\times10^6$ independent first-passage trajectories per parameter point.} \rev{We show error-correction performance in terms of $\langle T_F\rangle$ for equal data and measurement error rates, $p=q$; the use of the mean first-passage lifetime for Harrington's decoder is discussed in Sec.~\ref{sec:pheno_methods}, where the first-passage time $T_F$ is introduced. Results for data and measurement noise separately are provided in App.~\ref{app:har2d_pheno}. Harrington proved a non-zero lower bound on the phenomenological threshold \cite{harrington_analysis_2004}, but the systematic drift of the finite-size crossings toward lower error rates (Fig.~\ref{fig:har2d_pheno}) shows that our simulations do not resolve its asymptotic value. In particular, the apparent threshold decreases as larger code distances are considered, so the observed crossings cannot be interpreted as evidence for a nonzero asymptotic threshold.}

\subsubsection{Signal noise}

\begin{figure}[hbtp]
    \centering
    \includegraphics[scale=0.7]{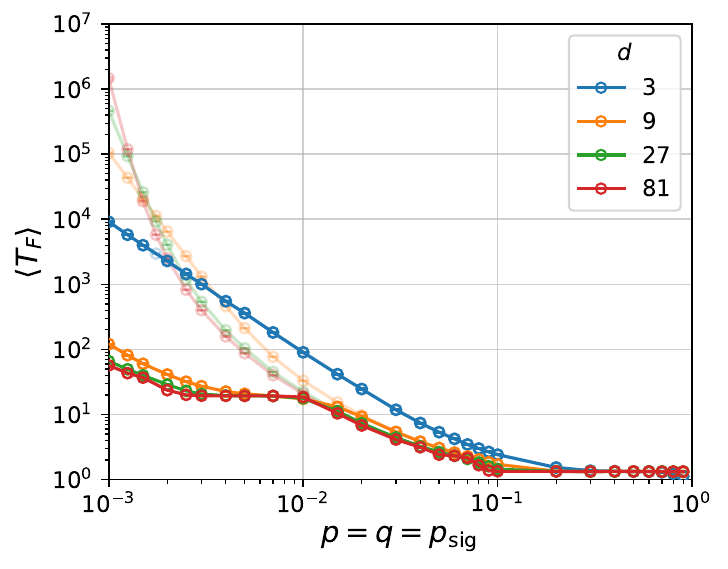}
    \caption{\rev{Average logical lifetime $\langle T_F\rangle$ as a function of equal data, measurement, and internal signal-noise rates, $p=q=p_{\mathrm{sig}}$, for Harrington's decoder on toric codes of distance $d$ (color coded). The faint curves in the background show the corresponding phenomenological-noise baseline with $p=q$ and $p_{\mathrm{sig}}=0$. Internal signal noise strongly degrades the low-error-rate performance and eliminates the improvement with increasing code distance.}}
    \label{fig:H2d_sigs}
\end{figure}

Lastly, we study the QEC performance of Harrington's decoder under signal noise. \rev{After each CA update, we independently flip every \texttt{CountSignal} and \texttt{FlipSignal} bit at every hierarchy level with probability $p_{\mathrm{sig}}$. In Fig.~\ref{fig:H2d_sigs}, we show numerical results for equal data, measurement, and signal error rates, $p=q=p_{\mathrm{sig}}$. The numerical data use $N=10^4$--$10^6$ independent first-passage trajectories per parameter point.} \rev{Signal noise strongly degrades the performance and eliminates the improvement with increasing code distance at sufficiently small error rates.} Harrington's hierarchical CA decoder relies on a robust signaling mechanism. \rev{Our results show that errors in this signaling mechanism compromise the decoder's scaling performance, making reliable internal signaling particularly important for its implementation on noisy hardware.}

\subsection{Other work}

In his original work \cite{harrington_analysis_2004}, Harrington derived a lower bound on the phenomenological-noise threshold of order $10^{-11}$ for $q=0$, while his finite-size numerical results for $d=3,9,27,81$ showed crossings at error rates of order $10^{-3}$, similar to the finite-size crossings observed here. \rev{Breuckmann \textit{et al.}~\cite{breuckmann_local_2017} subsequently estimated} $p_c\approx 2.1\times10^{-3}$ \rev{from an analytic ansatz and numerical fits.} In the code-capacity setting, the same authors find a threshold of about $4\%$ and scaling $\lambda(d)\approx d^{0.631}$, consistent with our results. 

\section{SCALA1D}\label{sec:scala1d}

In this section, we present our \texttt{SCALA} (Signaling Cellular Automaton with Local Attraction) decoders. Motivated by the failure mechanisms identified in Harrington's decoder --- particularly signal-noise amplification and hierarchical error propagation --- we construct \texttt{SCALA} to eliminate these effects by design. \texttt{SCALA1D}, the one-dimensional variant, corrects bit-flip errors on the quantum repetition code, while the two-dimensional version, \texttt{SCALA2D}, corrects errors on the toric code. In two dimensions, we exploit the CSS-nature of the toric code and apply two independent instances of the decoder on the $X$- and on the $Z$-error syndrome, respectively. The two main principles employed in our designs are \textit{Signaling} and \textit{Attraction} of quasi-particles (defects) based on their local environment. The cellular automaton effectively informs defects about the presence of other defects in their vicinity which in turn move towards each other and annihilate upon meeting. Our decoders are tailored to repetition and toric codes with local stabilizers, i.e., localized defects. The measured defects (i.e., the syndrome) is provided as input to the \texttt{SCALA} at the beginning of a CA step, upon which new cell states are calculated and corrective actions on data qubits are applied at the end of the step. This operational principle, also referred to as measurement-and-feedback loop, is a common theme in many cellular automaton decoder proposals such as Refs.~\cite{harrington_analysis_2004, balasubramanian_local_2024, paletta_highperformance_2025, paletta_toric_signal_2026, lang_strictly_2018, herold_cellularautomaton_2015}.

\begin{figure*}[hbtp]
    \centering
    \includegraphics[scale=0.32]{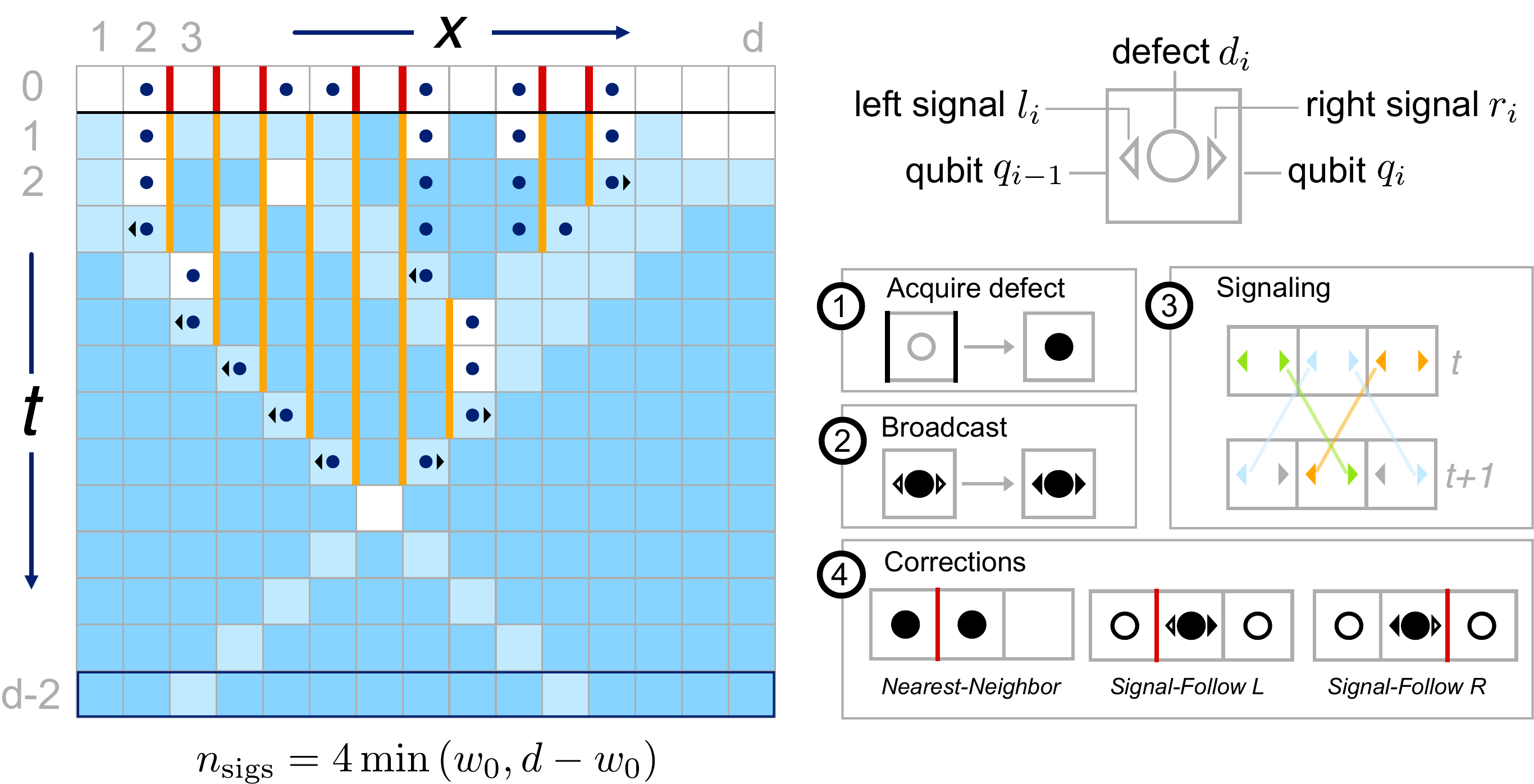}
    \caption{Sketch of the SCALA1D decoder on a distance $d=15$ repetition code with periodic boundary conditions. On the left we show a space-time diagram of an initial ($t=0$) configuration of data qubit errors (thick red lines) with the corresponding syndrome (black dots). Each cell corresponds to a CA cell in space (left to right) and time (top to bottom). The \texttt{SCALA1D} decoder is applied successively for $t=d-2$ steps, leading to the depicted evolution of errors (thick orange lines) and syndrome. Note that errors in orange are ``carried over" from previous time steps while red errors are newly generated due to noise. In the depicted example, errors are only generated at $t=0$ (red lines) and subsequently processed by the automaton under noise-free conditions. In the top right, we show the layout of one CA cell. Besides the defect bit $d_i$ and access to its left and right adjacent qubits $q_{i-1}$ and $q_i$, a cell contains two signal bits $l_i$ and $r_i$, for left and right directions, respectively. In the space-time diagram, a cell which has two active signal bits is colored in dark-blue while cells with one active signal are light-blue. When relevant to understand the dynamics, we additionally show the direction of active signals. At time step $t=d-2$, the number of signals $n_{\text{sigs}}$ is equal to $4\min{(w_0, d-w_0)}$ where $w_0$ is the error weight (i.e., number of errors) at $t=0$, as explained in the text. In the labeled boxes (1---4), we show the local rule divided into four steps, as further detailed in the main text.}
    \label{fig:scala1d_fig1}
\end{figure*}

\subsection{Construction}

\texttt{SCALA1D} is a non-hierarchical cellular automaton decoder for the quantum repetition code. Similar to Harrington's model, CA cells are placed in between data qubits of the repetition code, having access to their corresponding defect measurement information at each discrete time step and the cell states of their left and right neighbors. At the end of each CA step, each cell directly applies its correction to at most one of its two adjacent code qubits.

The state of a CA cell consists of three bits of information --- one \texttt{Defect} bit for the defect value which is overwritten at the beginning of each CA step by the stabilizer measurement at its location, and two \texttt{Signal} bits for the left and right direction. A cell's left and right adjacent qubits, as shown in the top right of Fig.~\ref{fig:scala1d_fig1}, are not part of its internal state but accessible to the cell's corrections. These corrections affect the states of cells indirectly via subsequent syndrome measurement and overwriting of their \texttt{Defect} bits. In Fig.~\ref{fig:scala1d_fig1} (1--4), we illustrate the local rule, divided into four subrules. Applying the local rules synchronously to all neighborhoods yields the global update rule which evolves the current configuration to the next. We summarize this global update step in Alg.~\ref{alg:scala1d_global_rule}. In the global update, we iterate over all $d$ cells in the lattice, choosing the current cell as center cell $C$ with its left ($L$) and right ($R$) neighbor cells. The corresponding defect is extracted from the given input syndrome, which we label $dL, dC, dR$ for left, center and right defects. A cell's left and right signal bits are represented as the attributes $l$ and $r$, e.g., the center cell's left signal bit is accessed by $C.l$. The symbols $\neg$, $\lor$, $\land$ and $\oplus$ represent boolean negation, \texttt{OR}, \texttt{AND} and \texttt{XOR} operations, respectively.

\begin{algorithm}[hbtp]
  \caption{\texttt{SCALA1D} global update}
  \label{alg:scala1d_global_rule}
  \begin{algorithmic}[1]
  \Function{update}{syndrome$[\;]$, cells$[\;]$, qubits$[\;]$, \rev{$t,t_R$}}
      \State $d \gets length(\text{syndrome})$
      \State \rev{$l'[\;],r'[\;]\gets 0$} \Comment{\rev{Next signal state}}

      \Comment{\rev{Determine all signal updates synchronously}}
      \For{$j \gets 1$ to $d$}
          \State $i \gets (j-1)\text{ mod }d$ \Comment{Left neighbor index}
          \State $k \gets (j+1)\text{ mod }d$ \Comment{Right neighbor index}
          \State $L,C,R \gets \mathrm{cells}[i\text{ to }k]$
          \State $dL,dC,dR \gets \mathrm{syndrome}[i\text{ to }k]$
          \State \rev{$l'[j]\gets R.l \lor (dR\land\neg R.l\land\neg R.r)$}
          \State \rev{$r'[j]\gets L.r \lor (dL\land\neg L.l\land\neg L.r)$}
      \EndFor

      \Comment{\rev{Commit the new signal state}}
      \For{$j \gets 1$ to $d$}
          \State \rev{$\mathrm{cells}[j].l\gets l'[j]$}
          \State \rev{$\mathrm{cells}[j].r\gets r'[j]$}
      \EndFor

      \Comment{\rev{Determine corrections from the same syndrome and signal state}}
      \For{$j \gets 1$ to $d$}
          \State $i \gets (j-1)\text{ mod }d$
          \State $k \gets (j+1)\text{ mod }d$
          \State $C \gets \mathrm{cells}[j]$
          \State $dL,dC,dR \gets \mathrm{syndrome}[i\text{ to }k]$

          \If{$dL \land dC$}
              \State $\mathrm{qubits}[j]\gets\mathrm{qubits}[j]\oplus1$
              \Comment{Flip left qubit}
          \ElsIf{$\neg dL \land dC \land \neg dR
                  \land \neg C.l \land C.r$}
              \State $\mathrm{qubits}[j]\gets\mathrm{qubits}[j]\oplus1$
              \Comment{Flip left qubit}
          \ElsIf{$\neg dL \land dC \land \neg dR
                  \land C.l \land \neg C.r$}
              \State $\mathrm{qubits}[k]\gets\mathrm{qubits}[k]\oplus1$
              \Comment{Flip right qubit}
          \EndIf
      \EndFor

      \If{\rev{$t\bmod t_R=0$}}
          \For{$j \gets 1$ to $d$}
              \State \rev{$\mathrm{cells}[j].l,\mathrm{cells}[j].r\gets0,0$}
          \EndFor
      \EndIf
  \EndFunction
  \end{algorithmic}
\end{algorithm}

Next, we discuss the subrules of the local rule, Fig.~\ref{fig:scala1d_fig1} (1--4), in more detail. The first step of the local rule is \textit{defect acquisition} by overwriting a cell's \texttt{Defect} bit with its associated $Z_iZ_{i+1}$ stabilizer. In step 2, a non-trivial defect value is \textit{broadcast} to its left and right \texttt{Signal} bits if and only if both the signals are off. Step 3 is \textit{signal propagation}. Identical to Harrington's construction, left and right signal-bit values are exchanged with neighboring CA cells. Lastly, corrections are applied to adjacent data qubits, depending on the presence of defects in a cell's neighborhood and its own signal values. For clarity, we divide corrections based on 1) defect values and 2) defect and signal values, which we call the \textit{Nearest-Neighbor} and \textit{Signal-Follow} rules, respectively. A Nearest-Neighbor correction takes place if a cell and its left neighbor have a defect present. In this case, the cell decides to flip its left qubit, thereby annihilating the single isolated error on their shared qubit. Clusters of neighboring defects are eroded by this subrule from both sides, two defects per time step. The Signal-Follow subrule, on the other hand, defines corrections of defects which are separated by more than one CA cell. In this case, a cell which has a defect present (while its neighbors have no defect present) decides to move the defect in opposite direction of its received signal, i.e., the defect follows the signal. If a cell received two or no signals, no action is performed.

The local rule of \texttt{SCALA1D} induces a dynamic by which defects are annihilated while bit-flip errors are corrected along the way. An example evolution for an initial error configuration at time $t=0$ under perfect automaton application (code-capacity scenario) is shown in the space-time diagram on the left of Fig.~\ref{fig:scala1d_fig1}. In this diagram, time flows from top to bottom and space from left to right. Each cell corresponds to a CA cell and repetition-code data qubits are represented as vertical edges. The initial configuration of errors is completely removed by time step $t=9$. Light-blue cells represent cells which have a single signal present while dark-blue and white colors represent two and no signals present, respectively. We only indicate the signal direction when relevant to understand the CA dynamics, i.e., when the local rule acts non-trivially. Thick red lines represent bit-flip errors on data qubits due to noise, thick orange lines represent errors which are not created but rather carried over from prior time steps. Black dots represent the syndrome corresponding to the residing error pattern. The equation below the diagram relates the initial error weight $w_0$, i.e., number of errors at time $t=0$, to the number of active signals, $n_{\text{sigs}}$ at the last time step $t=d-2$.

\subsection{Numerical results}

In the following, we provide numerical results on the QEC performance of \texttt{SCALA1D} under different noise scenarios. We compare QEC thresholds to other decoding strategies and reason about the sub-threshold scaling behavior by investigating minimum-weight error configurations that lead to logical failure. We begin our discussion with code-capacity noise.

\subsubsection{Code-capacity noise}

In the code-capacity setting the CA decoder is successively applied to an initial error configuration until all defects are resolved. At the end of this process, we check if a logical error occurred, i.e., if there are more than $(d+1)/2$ errors on the code. We observe that error configurations of $w_0$ errors produce $4 w_0$ signals via the CA dynamics by time step $t=d-2$ if $w_0 < (d+1)/2$ and $4(d-w_0)$ signals if $w_0 \geq (d+1)/2$. The number of signals at this point in time corresponds exactly to the number of bit-flip corrections the decoder applies within $d-2$ time steps. Since the number of signals is limited to $2d$ (as every cell has only two signal bits), all initial error configurations with error weight $w<(d+1)/2$ are corrected to the error-free state and for all other error configurations the decoder applies bit-flips to qubits not afflicted by errors, completing the code's logical $X$-operator. Thereby, \texttt{SCALA1D} finds the same solution as global majority vote and thus reproduces the maximum-likelihood decoding rule for the repetition code. 

To get a better intuition for the relation between initial error weight $w_0$ and number of signals, $n_{\text{sigs}}$, consider a single error cluster of weight $l$, causing two defects a distance $l$ apart. With every time step, the cells at which the defects are located each broadcast two signals (one to the left, one to the right). At $t=l$, there are $4l$ signals in the system of which two arrive at the cells with the defects. In the next time step, these defects start moving towards each other, each with unit speed. After $\lceil l/2 \rceil$ more steps, the defects are annihilated. During defect movement, no further broadcasting takes place as the intermediate cells all have signals present. After the defects are annihilated no further signals are produced. Assuming every possible cluster is eroded by $t=d-2$, the number of signals remains $4l$. If we have a cluster with $l\geq (d+1)/2$, the defects attract each other along the shorter path, leading to annihilation by time step $t=d-w_0$. The relation between $w_0$ and $n_{\text{sigs}}$ not only applies to single clusters but also to more complicated error configurations, e.g., the one shown in the space-time diagram of Fig.~\ref{fig:scala1d_fig1}. Next, we show that all defects have been removed within at most $t=d-2$ by considering the worst-case error configuration.

\begin{figure}[hbtp]
    \centering
    \includegraphics[scale=0.3]{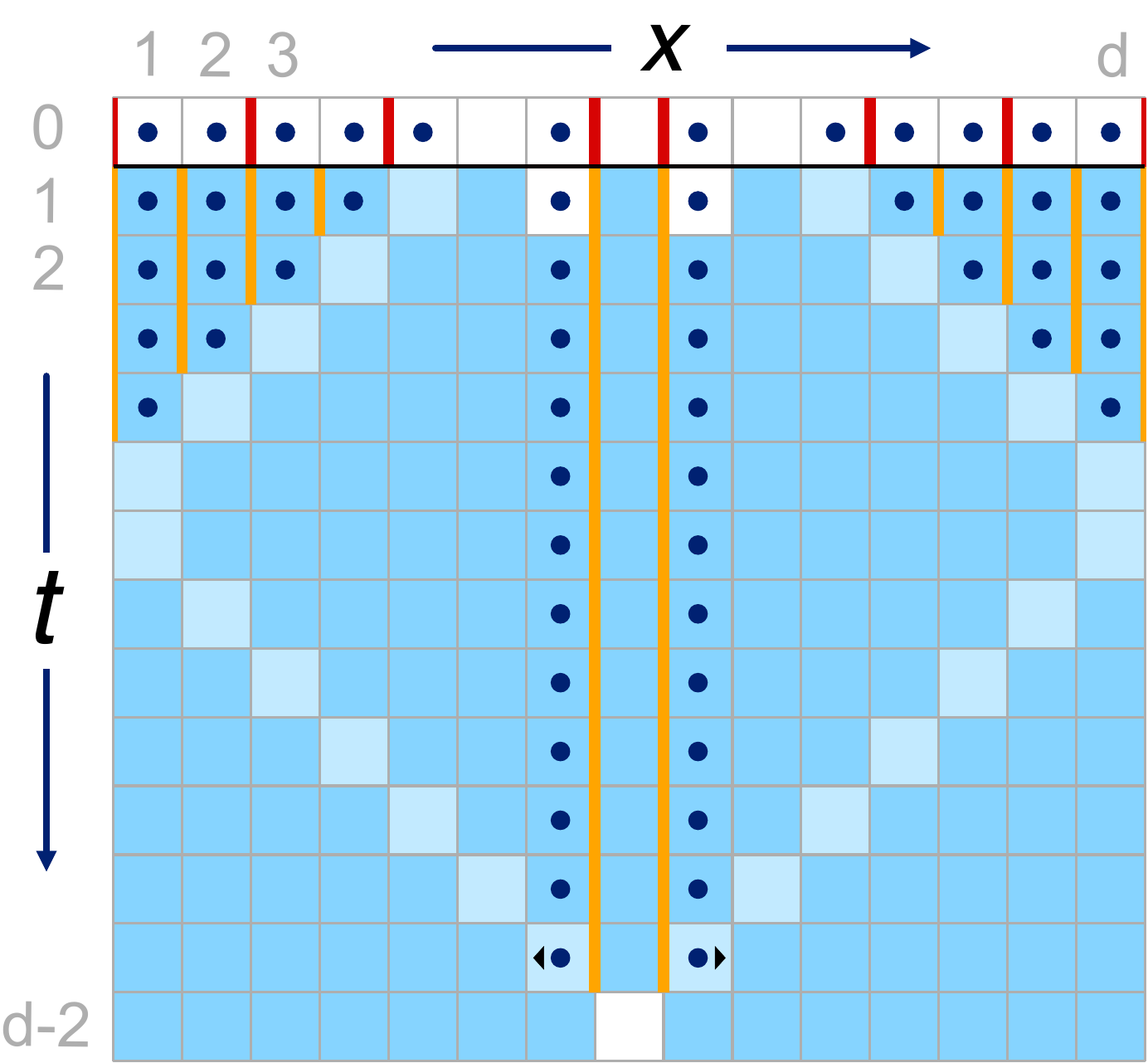}
    \caption{Space-time diagram of the evolution of an initial configuration of errors (thick red lines) causing an associated measured syndrome (black circles) under \texttt{SCALA1D} dynamics for a quantum repetition code of distance $d$. Orange edges correspond to errors carried over from the previous time step. White-colored cells have no active signal, light-blue ones have a single (left or right) signal and dark-blue cells have both left and right signal present. We show the direction of the signal for light-blue cells only when relevant to understand the CA dynamics. For the initial error configuration at $t=0$, it takes exactly $t=d-2$ time steps until removal, corresponding to the longest observed erosion time for any possible error configuration.}
    \label{fig:scala1d_t_most}
\end{figure}

For all code distances $d$, there exists a translation-invariant syndrome pattern which is eroded in exactly $t=d-2$ CA steps, the longest duration for the removal of any error configuration. The pattern, as shown in Fig.~\ref{fig:scala1d_t_most}, is caused by $w_0=(d-1)/2$ errors, such that there exists a cluster of two errors, causing two defects a distance 2 apart, surrounded by a cluster of $d-5$ defects whose endpoints are a distance 2 away from the inner defects. During evolution, the removal of the inner error cluster is blocked as its endpoints receive signals from both directions and thus trigger no correction. Only when the large syndrome cluster is removed (by nearest-neighbor correction steps) and the last residual signal emitted from the large syndrome cluster moved past the end points of the inner defects at time step $t=d-5$, the central cluster is allowed to be removed within another 3 steps. This entire process takes exactly $t=d-2$ steps for arbitrary $d$.

\begin{figure}[hbtp]
    \centering
    \includegraphics[scale=0.7]{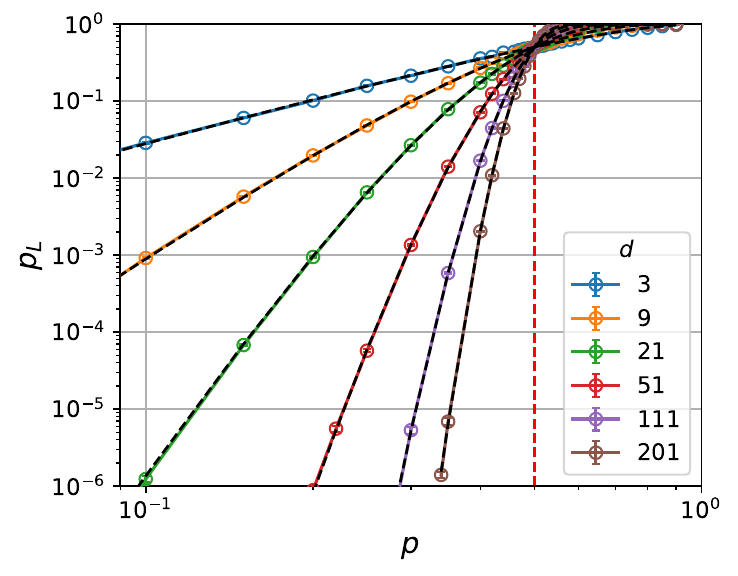}
    \caption{Logical error rate $p_L$ as a function of data qubit error rate $p$ under code capacity noise. Performance for \texttt{SCALA1D} on repetition codes corresponds to the solid lines. Black dashed lines correspond to the analytical performance of global majority vote (Eq.~\ref{eq:pL_ML} and Fig.~\ref{fig:rep_mwpm}). \rev{The dashed red line marks the exact threshold} $p_c=1/2$. Each data point was obtained from $10^4$ to $10^{6}$ Monte Carlo shots until statistical uncertainty is small relative to the measured values. Standard errors are represented by black error bars.}
    \label{fig:scala1d_code_capa}
\end{figure}

In Fig.~\ref{fig:scala1d_code_capa}, we show numerical logical error rates $p_L$ as a function of data qubit error rate $p$ in the code-capacity setting. Solid (and colored) lines correspond to performance under \texttt{SCALA1D} corrections, black dashed lines correspond to performance under global majority voting (Eq.~\ref{eq:pL_ML}). The coincidence of the data with analytical results numerically underpins the equivalence of the two decoding strategies, as discussed in the previous paragraphs. \rev{The dashed red line marks the exact threshold $p_c=1/2$. As a numerical consistency check, finite-size scaling of the Monte Carlo data gives $p_c=0.50006\pm0.00011$ and $\nu=2.006\pm0.010$, consistent with the exact majority-voting transition; details are given in App.~\ref{app:fss_scala1d}.}

\subsubsection{Phenomenological noise}

Next, we analyze the QEC performance of \texttt{SCALA1D} under phenomenological noise. In this setting, bit-flip errors occur on data qubits with rate $p$ and on syndrome measurements with rate $q$ at the beginning of every CA step. While in the code-capacity setting a logical-error check is performed after $t=d-2$ CA steps, in the phenomenological setting the check is performed after every CA step. Exactly this property allows for error configurations of weight $w<(d+1)/2$ to lead to logical failure, since it is not guaranteed that \texttt{SCALA1D} monotonically shrinks the number of errors during its evolution. In fact, the example in the space-time diagram of Fig.~\ref{fig:scala1d_fig1} illustrates this behavior. An initial configuration of error weight $w_0=7$ on a $d=15$ code grows by time step $t=2$ to $w=8$ through the \texttt{SCALA1D} dynamics, thereby constituting a logical error at this time step, although $w_0=7<(15+1)/2=8$. Next, we identify error configurations of minimal weight that lead to logical failure. Throughout this work, ‘minimal-weight’ refers to dominant error configurations inferred from the observed scaling behavior, rather than a formal proof of optimal decoding.

Consider a configuration of initial error weight $w_0=kn_e$, consisting of $k$ clusters of each $n_e$ data qubit errors which are separated by $n_e-1$ qubits that have no errors. After $n_e-1$ time steps of \texttt{SCALA1D} evolution, the first signals arrive, causing defect attraction via the intermediate error-free qubits. After $n_e-1$ more time steps, these intermediate qubits now all have errors due to successive execution of the Signal-Follow rule, connecting the initially separate error clusters into one large cluster of $w = kn_e + (k-1)(n_e-1) = 2w_0 - (n_e+k) + 1$ errors. In order to obtain the largest possible error weight \rev{$w_{\max}$} for a given initial weight $w_0$, we have to minimize $k+n_e$. Thus, $w_{\max}(w_0)=2w_0-f(w_0)+1$ with $f(w_0)=\min\{k+n_e\mid k,n_e\in\mathbb{Z}^+,\;kn_e=w_0\}$. \rev{For odd code distance $d$, the corresponding discrete minimum initial
weight predicted by this failure mechanism is therefore
\begin{equation}
    w_0(d)
    =
    \min\left\{
        w\in\mathbb{Z}^{+}
        \;\middle|\;
        w_{\max}(w) \geq \frac{d+1}{2}
    \right\}.
    \label{eq:scala_discrete_w0}
\end{equation}
This integer-valued expression gives the discrete weight predicted by this failure mechanism and is used below when comparing to the numerical effective distances.} By the inequality of arithmetic and geometric mean
\rev{\cite{euclid_elements_1956, hardy_inequalities_1952}} we have
$(n_e+k)/2 \geq \sqrt{n_ek}$, which yields the lower bound
\begin{equation}
    w_0(d) \geq \frac{(1+\sqrt{d})^2}{4}.
    \label{eq:scala_w0_bound}
\end{equation}
\rev{Equation~\ref{eq:scala_w0_bound} gives a closed-form lower bound on the discrete prediction in Eq.~\ref{eq:scala_discrete_w0}.} Thus, the initial error weight $w_0$ necessary to cause logical failure scales asymptotically linearly with $d$.

Next, we consider an error process of minimum weight leading to logical failure under measurement noise. Consider an error configuration of $w=(d+1)/2 + 1$ measurement errors corresponding to a cluster of $(d+1)/2 + 1$ defects. Since the defects are direct neighbors, \texttt{SCALA1D} executes the Nearest-Neighbor subrule on all but one cells with defects, leading to $(d+1)/2$ data errors after correction which constitutes a logical error, detected at the end of the step. \rev{Therefore, if this process is of minimum weight, we expect the stationary logical error rate under pure measurement noise to scale as $p_L\propto q^{(d+3)/2}$. Since this weight grows faster with $d$ than the data-error failure weight $w_0(d)$ derived above, the latter provides the leading candidate failure mechanism when data and measurement errors occur with equal rates, $p=q$. Mixed data- and measurement-error processes could in principle lead to lower-weight failures; however, the numerical effective exponents for $p=q$ approach the predicted $w_0(d)$ staircase, as shown below, providing numerical evidence that the data-error mechanism determines the leading low-error scaling.}

\begin{figure}[hbtp]
    \centering
    \includegraphics[width=\columnwidth]{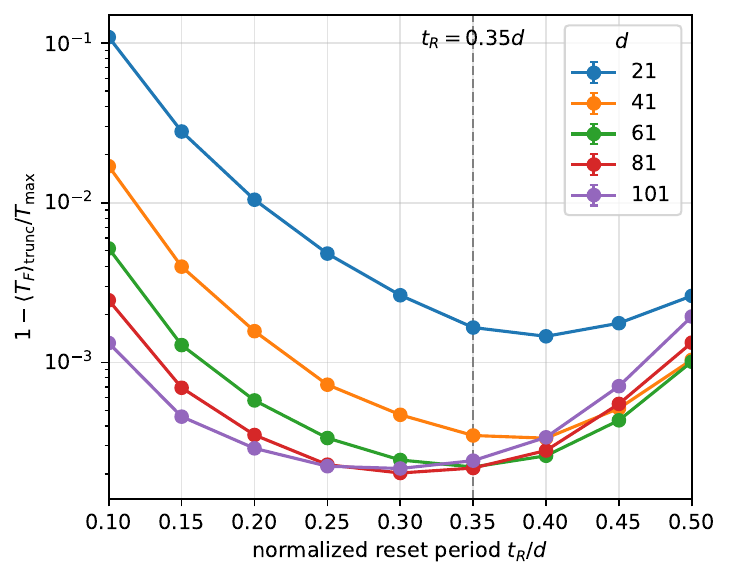}
    \caption{\rev{Dependence of the \texttt{SCALA1D} phenomenological-noise performance on the reset period at fixed equal data-qubit and measurement error rates $p=q=0.014$. All simulations are truncated at $T_{\max}=120d$; we show $1-\langle T_F\rangle_{\mathrm{trunc}}/T_{\max}$ as a function of the normalized reset period $t_R/d$. Smaller values correspond to longer observed lifetimes within this common simulation window. The dashed line marks the fixed choice $t_R=0.35d$ used in the simulations.}}
    \label{fig:scala1d_reset_time}
\end{figure}

\rev{Under phenomenological noise,} signal bits have to be reset periodically to prevent pile-up of signals from past (already eroded) errors over time. The signal reset time $t_R$ \rev{is} a hyperparameter in our model. \rev{The CA update rules remain strictly local, while the reset schedule introduces an externally chosen global control parameter.} As two defects on a periodic one-dimensional lattice are at most \rev{$\lfloor(d-1)/2\rfloor$} cells apart, \rev{we restrict $t_R$ to this range.} \rev{To choose a fixed reset schedule, we vary $t_R/d$ at $p=q=0.014$ in Fig.~\ref{fig:scala1d_reset_time}. Each simulation is truncated at $T_{\max}=120d$, and we plot $1-\langle T_F\rangle_{\mathrm{trunc}}/T_{\max}$, where $\langle T_F\rangle_{\mathrm{trunc}}$ denotes the mean first-passage lifetime after truncation at $T_{\max}$. Smaller values correspond to longer observed lifetimes. For all investigated distances, the performance exhibits a broad minimum around $t_R/d\simeq0.3$--$0.4$, with only weak distance dependence. We therefore use the predetermined schedule $t_R=0.35d$ for all phenomenological-noise simulations.}

\begin{figure}[hbtp]
    \centering
    \includegraphics[width=\columnwidth]{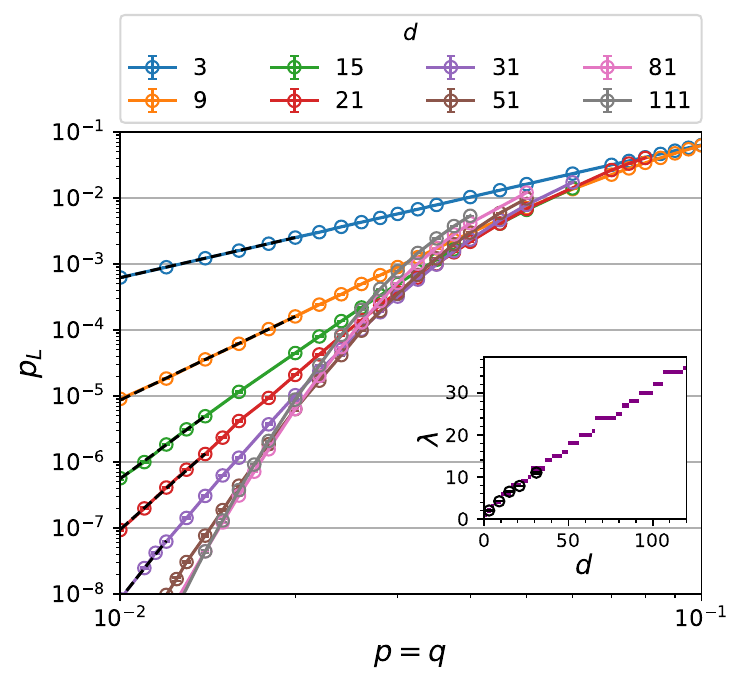}
    \caption{\rev{Stationary logical error rate $p_L$ for \texttt{SCALA1D} under phenomenological noise with equal data-qubit and measurement error rates $p=q$, for different code distances $d$. Black dashed lines show power-law fits $p_L=A(d)p^{\lambda(d)}$ to the statistically resolved low-$p$ data. The inset shows the fitted exponents $\lambda(d)$ together with the discrete failure-weight prediction $w_0(d)$ of Eq.~\ref{eq:scala_discrete_w0} (purple segments).}}
    \label{fig:scala1d_pheno}
\end{figure}

\rev{We estimate the stationary logical error rate $p_L$ as described in App.~\ref{app:hazard}, with representative hazards shown in App.~\ref{app:hazard_stationarity}. Figure~\ref{fig:scala1d_pheno} shows $p_L$ for equal data and measurement error rates $p=q$. We fit the statistically resolved low-$p$ data to $p_L=A(d)p^{\lambda(d)}$ and compare $\lambda(d)$ with the discrete data-error failure weight $w_0(d)$ of Eq.~\ref{eq:scala_discrete_w0}. The smaller-distance results approach the predicted staircase, supporting this as the leading low-error failure mechanism for $p=q$; at larger distances, the accessible effective exponents remain below the prediction.}

\rev{The finite-size crossings of the phenomenological $p_L$ curves drift systematically toward lower physical error rates as $d$ increases. In App.~\ref{app:scala_screening}, we identify screening of the signal-mediated defect attraction at finite error density as a possible dynamical explanation for this behavior. Together, these observations suggest a vanishing asymptotic phenomenological threshold for the present \texttt{SCALA1D} signaling rule. Nevertheless, $p_L$ remains strongly suppressed with increasing code distance over the investigated finite-size regime, demonstrating that the decoder can provide substantial protection against continuously occurring data and measurement errors even without an established asymptotic threshold. Importantly, this suppression is achieved with a constant number of classical state bits per cell and simple local update rules, making \texttt{SCALA1D} an attractive architecture for finite-size quantum error correction.}

\subsubsection{Signal noise}

\begin{figure}[hbtp]
    \centering
    \includegraphics[width=\columnwidth]{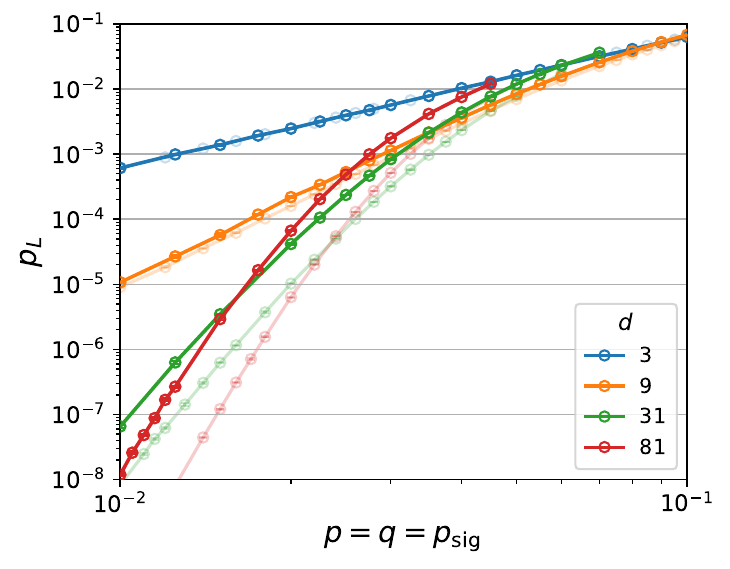}
    \caption{\rev{Stationary logical error rate $p_L$ of \texttt{SCALA1D} in the presence of signal noise. Solid curves correspond to equal data-qubit, measurement, and signal error rates, $p=q=p_{\mathrm{sig}}$, while the low-opacity curves show the corresponding phenomenological-noise results with $p=q$ and $p_{\mathrm{sig}}=0$.}}
    \label{fig:scala1d_sig}
\end{figure}

Similar to our study of Harrington's decoder, we analyze how the decoder performance is affected by signal noise. Additional to data-qubit noise with rate $p$, we subject all signal bits at the end of each CA step to bit-flip errors with rate $p_{\text{sig}}$. \rev{In Fig.~\ref{fig:scala1d_sig}, we compare the stationary logical error rate for $p=q=p_{\text{sig}}$ with the corresponding phenomenological-noise results for $p=q$ and $p_{\text{sig}}=0$. Both data sets use the same fixed reset schedule $t_R=0.35d$ and the stationary-hazard procedure described in Sec.~\ref{sec:pheno_methods}. The addition of signal noise increases $p_L$, while the strong distance-dependent suppression remains qualitatively intact over the investigated range.}

Signal errors can only lead to logical failure in conjunction with either data qubit or measurement errors. This is the case since the Signal-Follow subrule is only triggered when a defect is present. Moreover, isolated data errors are corrected by the Nearest-Neighbor subrule, ignoring the presence of any signals. Thus, only for an error cluster of size at least $2$, signal noise can lead to data qubit errors, specifically when a signal error affects one of the endpoints (isolated defects) of the cluster causing a correction in opposing direction of the cluster via the Signal-Follow subrule. In this case, a cluster of size $w$ grows to $w+1$ due to the occurrence of two data and one signal error with occurrence probability $\mathbb{P}(w\rightarrow w+1) \propto p^wp_{\text{sig}}$. When we choose $p=p_{\text{sig}}$, then $\mathbb{P}(w\rightarrow w+1) \propto p^{w+1}$, the same probability as the occurrence of a $(w+1)$-cluster under pure data noise. Thus, this mechanism does not reduce the minimum total fault order required for logical failure, but increases the number of contributing error configurations.

\subsubsection{Other work}

Recently, a \rev{related} CA decoder design for the quantum repetition code~\cite{paletta_highperformance_2025} has been proposed that also corrects bit-flip errors dynamically via a signal-based automaton. Our \texttt{SCALA1D} design differs in \rev{several} major ways. First, the cell state space of our decoder is much smaller and the local rule less complex. \texttt{SCALA1D} requires only three bits (defect, left and right signal) while the design of Ref.~\cite{paletta_highperformance_2025} involves several types of signals traveling at different speeds and a stack as part of each cell's state. The simplicity of \texttt{SCALA1D} makes the design amenable to future implementation as a quantum cellular automaton. Second, whereas the design of Ref.~\cite{paletta_highperformance_2025} is known to not be a maximum-likelihood decoder in the code capacity scenario, we provide strong numerical evidence (Fig.~\ref{fig:scala1d_code_capa}) that ours is. Third, whereas in Ref.~\cite{paletta_highperformance_2025} all signals are removed (``uncomputed") by faster moving anti-signals, we use periodic resets of the signal bits. The uncomputation requires additional overhead in terms of CA steps which \texttt{SCALA1D} avoids.

\rev{Finally, the two designs behave differently under phenomenological noise. Ref.~\cite{paletta_highperformance_2025} reports numerical evidence for exponential logical-error suppression with increasing system size, whereas the present \texttt{SCALA1D} rule exhibits the screening and finite-size crossing drift discussed above. This comparison shows that different signal-mediated local decoders can exhibit substantially different finite-size behavior under phenomenological noise. The origin of this difference remains an open question. Two potentially relevant ingredients are the asymmetric signaling dynamics of Ref.~\cite{paletta_highperformance_2025} and its continuous removal of obsolete signals through uncomputation, in contrast to the symmetric signaling and periodic resets used in \texttt{SCALA1D}. Determining how these ingredients affect screening and its associated characteristic length scale while retaining the simplicity of \texttt{SCALA1D} is an important direction for future work.}

\begin{figure*}[hbtp]
    \centering
    \includegraphics[scale=0.25]{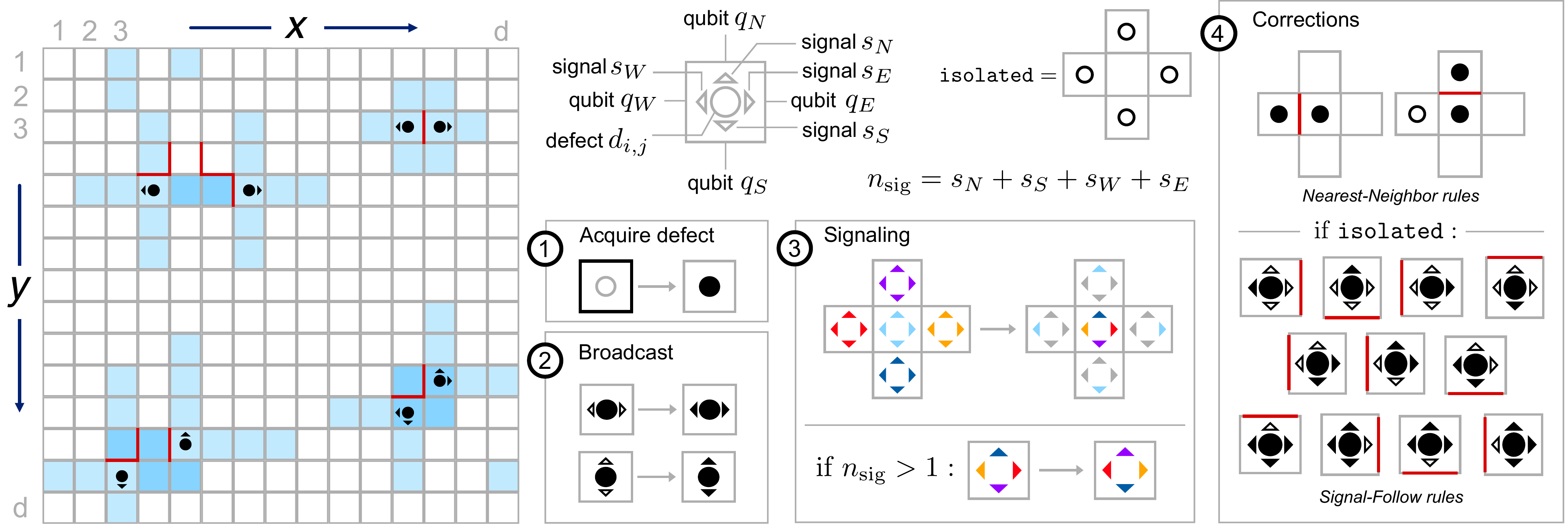}
    \caption{Graphical representation of \texttt{SCALA2D} on the 2D toric code correcting bit-flip errors. A cellular automaton cell, corresponding to a cell of the grid on the left and shown in detail in the top center, holds classical bits for signals in four directions, $s_N,s_E,s_S,s_W$, and a defect bit $d_i$ storing the measurement result of its associated $Z$-type stabilizer. A cell can apply one bit-flip to any of its four adjacent qubits within one CA step. For convenience, we also define $n_{\text{sig}}$ as the number of all active signals in a cell and the boolean variable \texttt{isolated} which is on when all four neighbors of a cell have no defect present. These definitions are used in the description of the local rule (1--4). During one CA step, \texttt{SCALA2D} alters both its defect as well as its signal bits and might apply a correction to one of its qubits. The local rule (1--4) of our decoder is split into four sub-rules, similar to \texttt{SCALA1D}. We elaborate on these rules in the main text. On the left, we show an example error configuration (red edges) resulting in the error syndrome (black dots) on the CA cells (faces). In response, the CA cells emit signals (blue shades) in all four directions. Light-blue cells have one active signal while dark-blue ones contain two active signals. We indicate the directions of signals when relevant to understand which correction the local rule applies in the next time step.}
    \label{fig:scala2d_fig1}
\end{figure*}

\section{SCALA2D}\label{sec:scala2d}

In this section, we now extend \texttt{SCALA1D} to two dimensions to protect the toric code. Both the toric and repetition codes are defined by local stabilizers that indicate the endpoints of error chains, making them amenable to correction through local attraction and annihilation of defects. We begin by discussing the decoder construction.

\subsection{Construction}

We start constructing \texttt{SCALA2D} by placing independent \texttt{SCALA1D} decoders on each of the $d$ rows and $d$ columns of the toric code lattice shown on the left in Fig.~\ref{fig:scala2d_fig1}. The cells of the lattice correspond to the position of plaquette operators which provide the input syndrome to the decoder correcting bit-flip errors. Phase-flip errors are corrected analogously via the star-stabilizers by an independent \texttt{SCALA2D} decoder. A CA cell, shown in the top center of Fig.~\ref{fig:scala2d_fig1}, has four \texttt{Signal} bits, $s_N,s_E,s_S,s_W$, one for each cardinal direction, a \texttt{Defect} bit, $d_i$, as well as access to its four adjacent toric-code data qubits. Similar to \texttt{SCALA1D}, we divide the local rule into subrules for clarity. 

After syndrome collection, the defect bit is overwritten by the measurement result of a cell's associated plaquette operator (1) at the beginning of a CA step. Together with the subrules for broadcasting (2) and signalling (3) these steps correspond exactly to two independent \texttt{SCALA1D} decoders, one oriented horizontally and one vertically. However, the problem with independent \texttt{SCALA1D} decoders is that there is no communication between them. For example a weight-2 error on the $N$ and $W$ data qubits of one cell cannot be removed without communication. To mitigate this issue, we modify the signaling process. If a cell has no defect present and receives signals from horizontal and vertical directions at the same time, the cell \textit{reflects} those signals back to their emitters. As a consequence, cells with defects which are not in the same row or column are able to communicate. We achieve signal reflection by the additional subrule shown in the bottom of box (3). This subrule swaps northern with southern and western with eastern signals if at least two of these signals are non-trivial and the cell has no defect present. 

The last step of the local rule consists of applying recovery operations to the toric-code data qubits. Similar to \texttt{SCALA1D}, we subdivide correction operations into Nearest-Neighbor and Signal-Follow corrections. During one CA step, a cell can at most apply a correction to a single qubit. If a cell and its western neighbor both have defects present, the cell flips its western qubit. If a cell and its northern neighbor have a defect but its western neighbor has no defect present, the cell flips its northern qubit. These rules constitute the Nearest-Neighbor corrections. A Signal-Follow correction, on the other hand, is only executed when a cell has a defect present but all its neighbors have no defect (we call this the \texttt{isolated} condition) and the cell's signal bits match a certain pattern. We group the signal-bit patterns into three subgroups, distinguished by the number of active, i.e., non-trivial signals ($n_{\text{sig}}$). If a cell has one or three active signals, it flips its qubit in opposite direction of a received (non-blocking) signal. A \textit{blocking} signal refers to two signals received from opposite sites at the same time. \textit{Non-blocking} signals are thus received only from a single direction. For the case $n_{\text{sig}}=2$ (central row of Signal-Follow subrule in Fig.~\ref{fig:scala2d_fig1} (4)), corrections are only applied when one of the shown three patterns is matched. The Signal-Follow subrules mirror the Nearest-Neighbor subrules using signals instead of neighbor defects and excluding actions for blocking situations. This construction method is identical to the Signal-Follow corrections of \texttt{SCALA1D}. Next, we investigate the QEC performance of \texttt{SCALA2D}.

\subsection{Numerical results}

In the following, we study the performance of our \texttt{SCALA2D} decoder on the 2D toric code. We show numerical simulation results under code-capacity, phenomenological and signal noise, starting with the code-capacity setting.

\subsubsection{Code-capacity noise}

Under code-capacity bit-flip noise, the decoder is applied on the error syndrome of an initial error configuration for $d^2$ time steps, corresponding to one complete ramp-up and ramp-down of the signal reset time $t_R$ as described below. \rev{A failure is recorded if the residual error contains a logical $X$ operator on either encoded logical qubit or remains unresolved with a nonzero syndrome.}

The decoder runtime of $d^2$ steps in the code-capacity setting is determined by the signal-reset scheme we employ. Since below the QEC threshold (corresponding to the correctable phase) smaller error clusters are exponentially more likely than larger ones \cite{sykes_percolation_1976}, it is beneficial to correct smaller clusters first. \rev{We reinforce this behavior by gradually increasing the signal reset time $t_R$ from $1$ to $d$ and subsequently decreasing it from $d-1$ to $1$. The maximum $t_R=d$ allows even maximally separated defects to communicate. The complete schedule takes
$\sum_{k=1}^{d}k+\sum_{k=1}^{d-1}k=d^2$ CA steps. While the CA update rules are strictly local, this predetermined, distance-dependent reset schedule introduces a global timescale without requiring non-local communication during runtime.}

\begin{figure}[hbtp]
    \centering
    \includegraphics[scale=0.7]{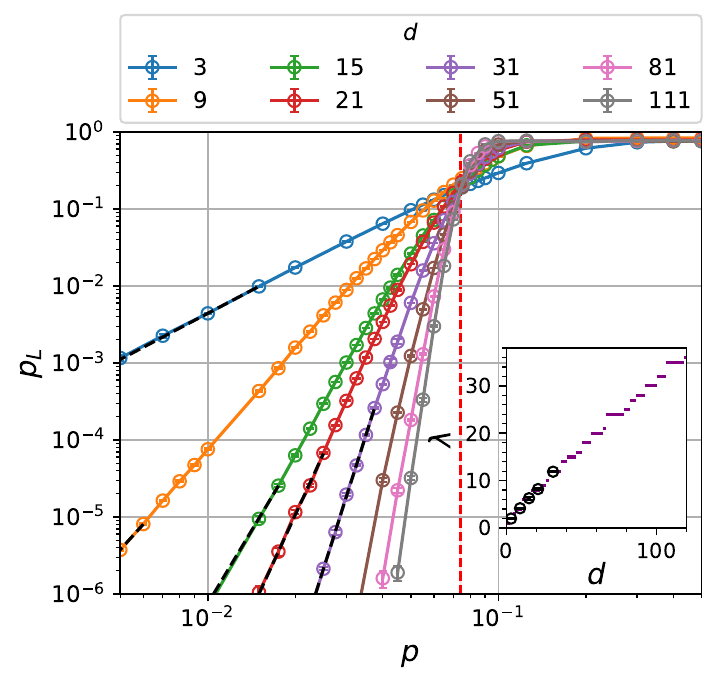}
    \caption{Logical error rate $p_L$ as a function of bit-flip error rate $p$ for a toric code of distance $d$ (color coded) under \texttt{SCALA2D} correction. \rev{The red dashed line and shaded region indicate the code-capacity threshold estimate $p_c=0.0740(5)$.} Black dashed lines correspond to fits of the function $f(d,p)=A(d)p^{\lambda(d)}$, \rev{with the fitted exponents $\lambda(d)$ shown as black circles in the inset and the discrete prediction $w_0(d)$ of Eq.~\ref{eq:scala_discrete_w0} as purple segments. The lower bound $w_0(d)\geq(1+\sqrt{d})^2/4$ grows asymptotically as $d/4$. Black error bars show statistical standard errors.}}
    \label{fig:scala2d_cc}
\end{figure}

In Fig.~\ref{fig:scala2d_cc}, we show numerical results for logical error rates $p_L$ as a function of bit-flip error rate $p$ in the code-capacity model for different code distances $d$ under \texttt{SCALA2D} corrections. In dashed black lines, we provide fits to the function $f(d,p)=A(d)p^{\lambda(d)}$ with $\lambda(d)$ shown in the inset; \rev{the fit details are reported in App.~\ref{app:subthreshold_fits}. The fitted exponents approach the discrete prediction $w_0(d)$ of Eq.~\ref{eq:scala_discrete_w0} where the low-$p$ regime is statistically resolved.} We can explain this similarity in performance by noting that the \rev{failure configurations underlying $w_0(d)$ for} \texttt{SCALA1D} also lead to logical errors under \texttt{SCALA2D} correction in the code-capacity setting.

\rev{The large-distance pairwise crossings yield the code-capacity threshold estimate $p_c=0.0740(5)$, where the uncertainty accounts for their residual finite-size variation; details are given in App.~\ref{app:scala2d_crossings}.}

For comparison, the threshold of the toric and surface code under minimum-weight perfect matching (MWPM) decoding in the code-capacity setting is about $10.3\%$ \cite{fowler_practical_2012}, while optimal (maximum-likelihood) decoding yields a threshold of approximately $10.9\%$ \cite{nishimori_statistical_2001, bravyi_efficient_2014, chubb_general_2021}, corresponding to the critical point of the associated 2D random-bond Ising model. \rev{The \texttt{SCALA2D} threshold is lower than these global-decoder thresholds while using local CA update rules supplemented by the reset schedule.} In the next section, we analyze the QEC performance under phenomenological noise.

\subsubsection{Phenomenological noise}

In the phenomenological setting, errors can occur with every CA step. We thus check after every time step if a logical error has occurred \rev{by applying an offline PyMatching \cite{higgott_pymatching_2022} MWPM decoder to the noise-free syndrome of the residual data-error configuration. If no logical error is detected, the \texttt{SCALA2D} evolution continues from the residual configuration.}

\begin{figure}[t]
    \centering
    \includegraphics[width=\columnwidth]{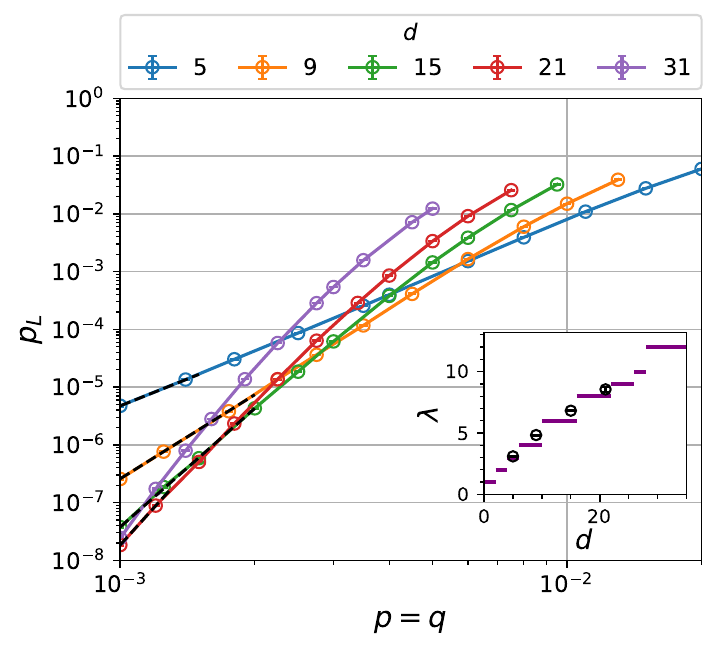}
    \caption{\rev{Stationary logical error rate $p_L$ for equal data and measurement error rates $p=q$ under \texttt{SCALA2D} correction with $t_R=w_0(d)$. For the shown distances $d\leq31$, the finite-size crossings remain within $p\geq10^{-3}$ and shift toward lower physical and logical error rates with increasing $d$. Black error bars show statistical standard errors.}}
    \label{fig:scala2d_pheno}
\end{figure}

\rev{For continuous phenomenological noise, we use the fixed reset period $t_R=w_0(d)$ of Eq.~\eqref{eq:scala_discrete_w0}. Figure~\ref{fig:scala2d_pheno} shows the stationary logical error rate $p_L$ for $p=q$. The finite-size crossings shift toward lower physical and logical error rates with increasing distance. We show $d\leq31$, for which the crossings remain within the simulated range $p\geq10^{-3}$. This drift, together with the finite-density screening mechanism demonstrated for \texttt{SCALA1D} in App.~\ref{app:scala_screening}, provides evidence against a nonzero asymptotic  threshold against phenomenological noise in the thermodynamical limit. The same mechanism applies to one-dimensional error chains embedded in \texttt{SCALA2D}, which use the same signal-mediated attraction. Nevertheless, at $p=q\simeq10^{-3}$, \texttt{SCALA2D} reaches substantially lower $p_L$ at smaller distances than the corresponding $1/\langle T_F\rangle$ for Harrington's decoder. Thus, despite Harrington's nonzero phenomenological threshold, \texttt{SCALA2D} provides stronger error suppression over the finite-size regime studied here with a simpler local architecture.}

\subsubsection{Signal noise}

\begin{figure}[hbtp]
    \centering
    \includegraphics[scale=0.7]{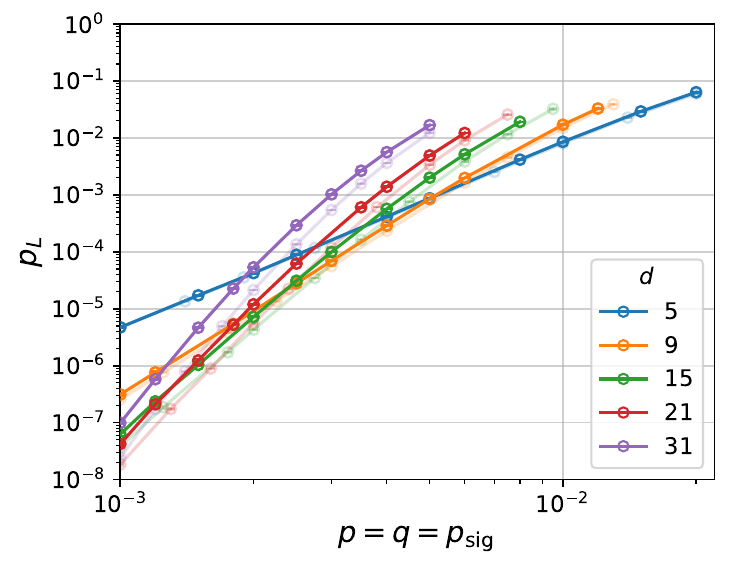}
    \caption{\rev{Stationary logical error rate $p_L$ of \texttt{SCALA2D} with signal noise. Solid curves show $p=q=p_{\mathrm{sig}}$, while low-opacity curves show the corresponding $p=q$, $p_{\mathrm{sig}}=0$ baseline. All simulations use $t_R=w_0(d)$. Black error bars show statistical standard errors.}}
    \label{fig:scala2d_sig}
\end{figure}

\rev{In addition to data-qubit and measurement noise with rates $p$ and $q$, respectively, we apply independent bit-flip noise with rate $p_{\mathrm{sig}}$ to each signal bit after every CA step. Figure~\ref{fig:scala2d_sig} compares matched data, measurement, and signal noise, $p=q=p_{\mathrm{sig}}$, with the phenomenological-noise baseline $p=q$, $p_{\mathrm{sig}}=0$. Signal noise increases $p_L$ primarily by an overall performance offset, while the distance-dependent scaling remains approximately unchanged within the investigated finite-size regime. Thus, \texttt{SCALA2D} remains functional even when its internal signaling dynamics are noisy.}

\subsubsection{Other work}

\rev{To place \texttt{SCALA2D} in the broader context of local decoders, Table~\ref{tab:toric_decoder_comparison} compares representative constructions in terms of local resources, code-capacity and phenomenological thresholds, and subthreshold scaling.}

\begin{table*}[t]
\centering
\scriptsize
\renewcommand{\arraystretch}{1.08}
\setlength{\tabcolsep}{3pt}
\rev{
\begin{tabular*}{\textwidth}{@{\extracolsep{\fill}}lccccc@{}}
\hline\hline
\textbf{Decoder} & \textbf{Local resources} & \textbf{Robust?}
& \textbf{$p_c^{\rm CC}$} & \textbf{$p_c^{\rm pheno}$}
& \textbf{Subthreshold scaling} \\
\hline
Harrington~\cite{harrington_analysis_2004} (2004), \cite{breuckmann_local_2017} (2017)
& $O(\log d)$ & $\times$
& $4.523(2)\%$ & $0.133\%^{a}$
& $\lambda(d)\propto d^{\log_3 2}$ \\

Herold et al.~\cite{herold_cellularautomaton_2015} (2015), \cite{herold_cellular_2017} (2017)
& polylog. & ?
& $\sim8.2\%$ & $\sim0.17\%$
& ? \\

Balasubramanian et al.~\cite{balasubramanian_local_2024} (2024)
& hierarchical & $\checkmark$
& ? & $\lesssim0.01\%^{b}$
& $\lambda(d)=O(d^\alpha)$ \\

Lake~\cite{lake_fast_2025,lake2025localactiveerrorcorrection} (2025)
& $O(\mathrm{polylog}\,d)$ & ?
& $\sim7.3\%$ & $\sim1.5\%$
& $\lambda(d)=O(d^\alpha)$ \\

Paletta et al.~\cite{paletta_toric_signal_2026} (2026)
& $O(\log d)$ & ?
& ? & $\sim0.68\%^{c}$
& $\lambda(d)=O(d^\alpha)$ \\

\hline
\texttt{SCALA2D}
& $O(1)$ & $\checkmark$
& $7.40(5)\%$ & --$^{d}$
& $\lambda(d)\propto d$ \\
\hline\hline
\end{tabular*}

\caption{Comparison of local decoders for the 2D toric code. ``Robust?'' denotes demonstrated robustness to noise in the internal signaling or local decoder operations; ``CC'' denotes code capacity, ``pheno'' denotes phenomenological noise; A question mark ``?'' denotes that the corresponding quantity or property was not studied. For $\lambda(d)=O(d^\alpha)$, $0<\alpha<1$. $^{a}$Ref.~\cite{breuckmann_local_2017} obtains $p_c=(0.133\pm0.001)\%$ from a fit to a hierarchy-motivated memory-time ansatz; the finite-size curves do not exhibit a common crossing, so the quoted fit uncertainty does not account for systematic uncertainty in the threshold estimate. $^{b}$Circuit-level noise, which includes faults in the local decoder operations~\cite{balasubramanian_local_2024}. $^{c}$Ref.~\cite{paletta_toric_signal_2026} obtains $\sim0.68\%$ from a finite-size fit, but argues against interpreting it as a genuine asymptotic phenomenological threshold because recurrent noise can inhibit long-range signal propagation. $^{d}$For \texttt{SCALA2D}, the downward crossing drift and screening mechanism provide numerical evidence against a nonzero asymptotic phenomenological threshold.}
\label{tab:toric_decoder_comparison}
}
\end{table*}

\rev{Recent local decoders illustrate the trade-off between asymptotic fault tolerance and local computational resources. Balasubramanian \textit{et al.}~\cite{balasubramanian_local_2024} construct a hierarchical local automaton that is fault-tolerant under circuit-level noise, while Lake~\cite{lake2025localactiveerrorcorrection} obtains fault-tolerant local active error correction by simulating confinement using $O(\mathrm{polylog}\,d)$ local classical resources. The latter achieves stretched-exponential memory lifetime and a numerical phenomenological threshold of approximately $1.5\%$. These constructions demonstrate that scalable local decoding is possible when additional scale-dependent structure or resources are introduced.}

\rev{The signal-rule decoder of Ref.~\cite{paletta_toric_signal_2026} provides a particularly close comparison with \texttt{SCALA2D}, as it likewise uses local signaling to mediate defect attraction. It shows numerical evidence for exponential suppression under phenomenological noise, with a fitted critical error rate of approximately $0.68\%$ and near-linear effective-distance scaling over the investigated system sizes. Thus, alternative signal-mediated dynamics can substantially extend the finite-size regime of logical-error suppression. \texttt{SCALA}, in turn, emphasizes minimal local resources and robustness: its local state consists of a constant number of binary signaling and defect variables per lattice site, independent of system size and its dynamics remain functional under internal signal noise. This constant local memory requirement is complemented by a globally coordinated reset mechanism whose period depends on the system size. 
The combination of local update rules and globally coordinated resets makes \texttt{SCALA} a potential candidate for implementation using quantum cellular automata (QCA) on programmable quantum processors. Recent experimental demonstrations of QCA dynamics in dual-species Rydberg-atom arrays \cite{white_qca_2026}, together with theoretical proposals for globally controlled QCA architectures \cite{cesa_qca_2026} and fermionic quantum processors \cite{calliari_fermionic_2026}, motivate exploring such implementations. However, realizing the required local syndrome-dependent feedback remains an open challenge.
Identifying which local ingredients control the characteristic length scale associated with screening while preserving the simplicity and robustness of \texttt{SCALA} is an important direction for future work.}

\rev{Other distributed real-time decoder architectures have been developed with explicit hardware implementation in mind, including proposals for hardware-efficient decoding~\cite{holmes_nisq_2020,ueno_qecool_2022,ravi_better_2023}; we leave a hardware-level comparison with \texttt{SCALA} to future work.}

Finally, the CA decoder of Ref.~\cite{herold_cellularautomaton_2015} has served as initial inspiration for our work, especially the idea of mutual attraction of defects via a force conveyed by signals. \rev{Its reported code-capacity threshold of approximately $8.2\%$ exceeds that of \texttt{SCALA2D}, but its field-based dynamics require substantially larger local resources. \texttt{SCALA} instead prioritizes minimal local state, simple signaling dynamics, and robustness to internal signal noise.}

\section{Conclusion and Outlook}\label{sec:conc}

This work analyses the performance of different types of cellular-automaton decoders for quantum error correction. More specifically, we investigate \rev{the trade-offs between} a non-hierarchical, locally interacting and parallel architecture \rev{and} a hierarchical, multi-scale one. By comparing the hierarchical Harrington decoder \cite{harrington_analysis_2004} to our novel non-hierarchical CA design, \texttt{SCALA}, on the quantum repetition and toric codes, \rev{we find that the non-hierarchical construction combines strong code-capacity performance with low local computational overhead and robustness to noise within the decoder dynamics. In particular, \texttt{SCALA2D} achieves a code-capacity threshold of $p_c\simeq7.40(5)\%$ and subthreshold scaling $\lambda(d)$ that grows approximately linearly with code distance.}

We emphasise, however, that while the local update rules of \texttt{SCALA} are strictly local and translation-invariant, the protocol includes a reset time, $t_R$, that scales with the lattice size. \rev{This system-size dependence enters through a global reset schedule rather than through growing computational resources of individual cells. Such global control can be comparatively inexpensive on architectures supporting parallel operations, while a fully local implementation could instead generate the reset timing internally at the cost of additional local resources.}

\rev{The principal architectural trade-offs and attractive features of our non-hierarchical CA design are summarized as follows.}

\begin{enumerate}
    \item Hierarchical designs impose a recursive structure on the decoding process similar to renormalization-group and concatenated decoders. \rev{For Harrington's decoder, this structure leads to the subthreshold scaling $\lambda(d)\propto d^{\log_3 2}\approx d^{0.631}$ in $p_L=A(d)p^{\lambda(d)}$. In contrast, \texttt{SCALA} exhibits $\lambda(d)$ consistent with linear scaling in $d$ under code-capacity noise and over the accessible finite-size regime under phenomenological noise.}

    \item The hierarchy imposes constraints on the allowed code distances. For Harrington's model, only code distances $d=3^k$ for $k\in\mathbb{N}^+$ are allowed. \texttt{SCALA}, on the other hand, is applicable to any code distance.

    \item Harrington's decoder requires the memory per CA cell to scale with the code distance, which increases the complexity and resource requirements for larger quantum codes. In contrast, our \texttt{SCALA} designs \rev{retain constant local memory per CA cell, with the remaining system-size dependence entering through the global reset schedule}. This makes \texttt{SCALA} a modular local design, simplifying hardware implementation for large-scale systems.

    \item Harrington's automaton is highly \rev{sensitive} to signal noise. \rev{Our designs, on the other hand, remain functional under the signal-noise models considered here, with signal noise primarily producing an offset in logical-error performance while leaving the distance-dependent scaling approximately unchanged over the investigated system sizes.}
\end{enumerate}

\rev{The phenomenological-noise results identify an important limitation of the present \texttt{SCALA} signaling rule. Despite the approximately linear finite-size scaling of $\lambda(d)$, the crossings shift systematically toward lower physical error rates with increasing distance. In \texttt{SCALA1D}, we trace this behavior to screening of the signal-mediated defect attraction at finite error density, as continuously generated defects interfere with long-range signaling. The same one-dimensional mechanism can occur along error chains embedded in \texttt{SCALA2D}, providing evidence against a nonzero asymptotic phenomenological threshold for the present signaling rule. This asymptotic limitation does not preclude strong finite-size performance: over the investigated regime and system sizes, \texttt{SCALA2D} reaches substantially lower logical error rates at smaller distances than Harrington's decoder. Understanding how signaling rules control screening and its characteristic length scale while retaining constant local resources, simple dynamics, and robustness to internal noise is therefore a central direction for future work.}

The design of \texttt{SCALA} makes it particularly well-suited for real-time decoding. Here, one can envision a hybrid approach where \texttt{SCALA} runs continuously during a quantum computation to rapidly clear low-weight errors, complemented by a slower conventional decoder that handles higher-weight errors in larger periods, thereby avoiding the backlog problem \cite{terhal_quantum_2015}. The simple logic of \texttt{SCALA} allows for implementation on classical hardware like field-programmable gate arrays (FPGAs) or application-specific integrated circuits (ASICs) similar to Refs.~\cite{ravi_better_2023,ueno_qecool_2022,holmes_nisq_2020,barber_realtime_2025}.

\rev{The combination of simple local rules, constant local memory, and robustness to internal signal noise also motivates a quantum-cellular-automaton (QCA) realization of \texttt{SCALA} \cite{arrighi_overview_2019,wintermantel_unitary_2020}. Recent experimental demonstrations of QCA dynamics in dual-species Rydberg-atom arrays \cite{white_qca_2026} further illustrate the level of local interactions and global control now achievable on programmable quantum hardware. Replacing the classical CA dynamics by local quantum dynamical maps could yield a measurement-free QEC protocol operating directly on quantum degrees of freedom \cite{guedes_quantum_2024, day_faulttolerant_2024, butt_demonstration_2025}. Such an approach is particularly attractive for platforms where measurements are costly but controlled operations can be applied efficiently and in parallel \cite{evered_highfidelity_2023,skoric_parallel_2023}. Constructing such a QCA and determining whether the signal-noise robustness observed here extends to circuit-level errors in the corresponding quantum dynamics constitutes an important next step.}

\rev{Whereas the two-dimensional analysis in this work focuses on the toric code with periodic boundaries, we have also found that the \texttt{SCALA1D} dynamics can be adapted to open boundaries using reflecting signals while retaining its code-capacity performance. Extending this construction to two-dimensional planar geometries and establishing suitable boundary rules for \texttt{SCALA2D} is an important direction toward experimentally relevant surface-code implementations.}

\rev{Lastly, an important open question is whether the local-attraction principle underlying \texttt{SCALA} can be generalized beyond regular lattices to other decoding graphs. Extending the signaling and correction dynamics to variable nodes coupled to more than two checks would provide a route toward testing this architecture on quantum low-density parity-check (LDPC) codes \cite{tillich_quantum_2014}.}

\section{Code availability}

All cellular automaton decoders studied in this work are available in a public repository \cite{winter_scala_2026}. The repository contains modular implementations of all models and scripts for reproducing the numerical analysis.

\section{Acknowledgements}

We gratefully acknowledge support by the ERC Starting Grant QNets through Grant No. 804247. MM acknowledges support by the BMFTR project MUNIQC-ATOMS, the European Union’s Horizon Europe research and innovation programme under grant agreement No 101114305 (“MILLENION-SGA1” EU Project). Furthermore, we acknowledge support by the BMFTR project NeuQuant and the DFG Schwerpunkt Programm Quantum Software, Algorithms and Systems (https://www.spp2514.kit.edu). This research is also part of the Munich Quantum Valley (K-8), which is supported by the Bavarian state government with funds from the Hightech Agenda Bayern Plus. The authors gratefully acknowledge funding by the Deutsche Forschungsgemeinschaft (DFG, German Research Foundation) through Grant No. 449905436, and under Germany’s Excellence Strategy ‘Cluster of Excellence Matter and Light for Quantum Computing (ML4Q) EXC 2004/1’ 390534769. T.L.M.G.~also acknowledges the support of the Simons Foundation (grant number 1023171, RC), the Brazilian National Council for Scientific and Technological Development (CNPq, grant number 315081/2025-2) and the Financiadora de Estudos e Projetos (grant 1699/24 IIP-FINEP). We acknowledge computing time provided at the NHR Center NHR4CES at RWTH Aachen University (Project No. p0020074). This is funded by the Federal Ministry of Education and Research and the state governments participating on the basis of the resolutions of the GWK for national high performance computing at universities.

\bibliography{refs} 
\bibliographystyle{apsrev4-2}

\appendix

\section{\rev{Phenomenological-noise methodology}}
\label{app:hazard}

\rev{In contrast to the code-capacity noise model, where errors occur only before decoding, phenomenological noise continuously introduces new data-qubit errors with probability $p$ and syndrome-measurement errors with probability $q$ at each CA step. The decoder must therefore correct errors while new faults continue to occur, using syndrome information that may itself be unreliable. The precise noise implementation is specified in the corresponding decoder sections below. Under phenomenological noise, we treat logical failure as a first-passage process. After each CA update, we determine whether the physical data-error configuration has entered the wrong logical sector. Measurement errors affect only the syndrome supplied to the CA decoder and are not included in this check. For the repetition code, the logical sector is determined by the exact majority-vote criterion; for the toric code, we compute the noise-free syndrome of the residual data-error configuration and apply an offline minimum-weight perfect-matching decoder. The first time step at which this check identifies a logical failure defines the first-passage time $T_F$.}

\rev{The corresponding discrete first-passage hazard,
\begin{equation}
    h(t)=\Pr(T_F=t\mid T_F\geq t),
    \label{eq:hazard}
\end{equation}
is the probability of failing at time $t$ conditioned on having survived up to that time, following the standard first-passage formulation \cite{redner2001firstpassage}. For \texttt{SCALA}, $h(t)$ approaches an approximately time-independent value after an initial transient (as verified numerically in App.~\ref{app:hazard_stationarity}), allowing us to define the stationary logical error rate per CA step as
\begin{equation}
    p_L\equiv h_\infty.
    \label{eq:stationary_pL}
\end{equation}
To estimate $h_\infty$, we simulate an ensemble of independent trajectories, each initialized without logical errors and evolved until its first logical failure. Once a trajectory fails, it is terminated and does not contribute further to the statistics. After an initial burn-in interval, we divide the subsequent evolution into two consecutive time windows, during which we independently estimate the logical failure probability per CA step. The purpose of using two windows is to check for systematic drift in the estimated failure rate between two successive post-transient intervals: trajectories surviving the first window continue into the second, while those that have already failed are excluded. For \texttt{SCALA1D}, the predetermined reset period is $t_R=\operatorname{round}~(0.35d)$ and the burn-in and each measurement window have duration $B=10t_R$; for \texttt{SCALA2D}, we use $B=20t_R$. Thus, $0\leq t<B$ is discarded and the intervals $B\leq t<2B$ and $2B\leq t<3B$ are used for measurement.}

\rev{In each window $i$, we record the number of first logical failures $F_i$ and the total exposure $E_i$, obtained by summing the number of trajectories at risk at the beginning of each CA step. The logical failure rate in window $i$ and its standard error are
\begin{equation}
    h_i=\frac{F_i}{E_i},
    \qquad
    \sigma_{h_i}\simeq\frac{\sqrt{F_i}}{E_i},
\end{equation}
where the uncertainty follows from the standard deviation $\sigma_{F_i}$ of failure count $F_i$ via the Poisson counting approximation $\sigma_{F_i}\simeq\sqrt{F_i}$. We accept the two windows as approximately stationary if each contains at least $100$ first failures, corresponding to a counting uncertainty of at most approximately $10\%$, and if the relative drift between their hazard estimates satisfies
\begin{equation}
    \frac{|h_1-h_2|}{(h_1+h_2)/2}\leq 0.20.
\end{equation}
If these criteria are satisfied, we combine the failure counts and total exposure from both windows to obtain
\begin{equation}
    p_L=\frac{F_1+F_2}{E_1+E_2},
    \qquad
    \sigma_{p_L}\simeq\frac{\sqrt{F_1+F_2}}{E_1+E_2}.
\end{equation}
If either the minimum-statistics requirement or the stationarity test is not satisfied, the corresponding parameter point is marked unresolved and excluded from subsequent fits. Representative time-dependent hazard curves supporting the existence of an approximately stationary regime are shown in App.~\ref{app:hazard_stationarity}.}

\rev{Harrington's decoder exhibits longer distance-dependent relaxation and persistent temporal modulation of its hazard (App.~\ref{app:hazard_stationarity}). We therefore characterize its performance by the mean first-passage lifetime $\langle T_F\rangle$ rather than a stationary logical error rate. For a time-independent discrete hazard, the first-passage distribution is geometric, yielding $\langle T_F\rangle=1/h_\infty=1/p_L$. Consequently, $1/p_L$ approximates the mean first-passage lifetime of \texttt{SCALA} when its initial transient is short compared with $\langle T_F\rangle$. The inverse mean lifetime $1/\langle T_F\rangle$ reported for Harrington should not, however, be interpreted as a stationary hazard.}

\rev{Unless stated otherwise, all Monte Carlo trajectories are statistically independent. Error bars for probabilities estimated from $N$ independent samples denote the binomial standard error,
\begin{equation}
    \sigma_{\hat p}=\sqrt{\frac{\hat p(1-\hat p)}{N}},
\end{equation}
while error bars for $\langle T_F\rangle$ denote the standard error of the sample mean. Sample-count ranges are given with the corresponding numerical results.}

\section{\rev{Additional phenomenological-noise results}}
\label{app:hazard_stationarity}

\subsection{\rev{First-passage hazard and stationarity}}

\begin{figure*}[hbtp]

    \centering
    \includegraphics[width=\textwidth]{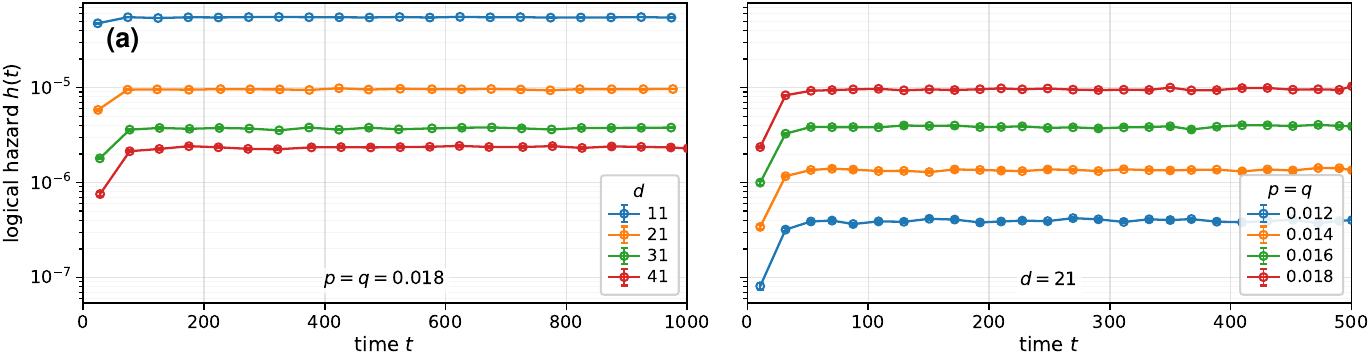}
    \vspace{0.5em}
    \includegraphics[width=\textwidth]{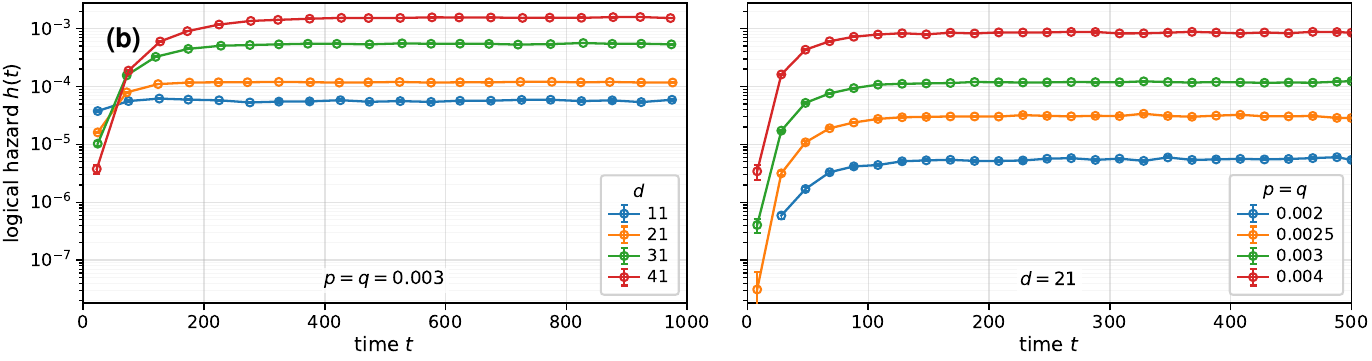}
    \vspace{0.5em}
    \includegraphics[width=\textwidth]{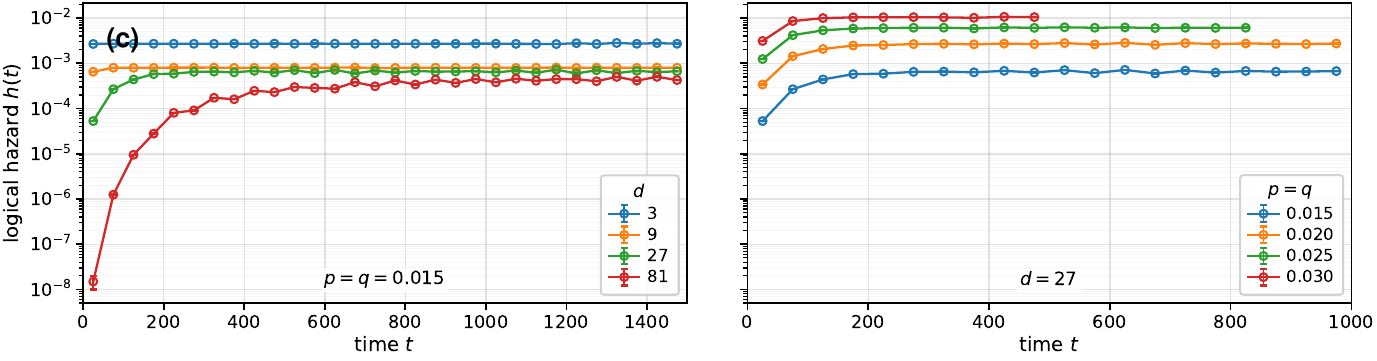}
    \vspace{0.5em}
    \includegraphics[width=\textwidth]{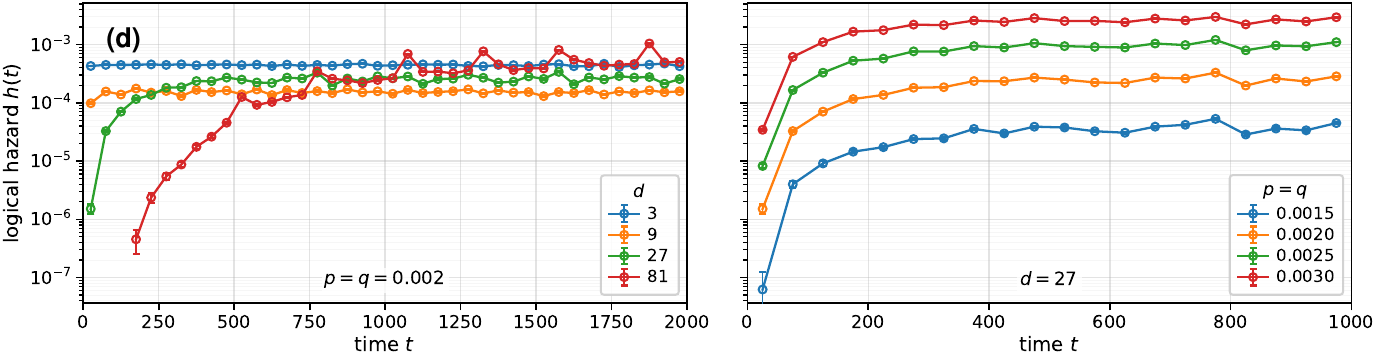}

    \caption{\rev{Representative first-passage hazard $h(t)$ under phenomenological noise for (a) \texttt{SCALA1D}, (b) \texttt{SCALA2D}, (c) Harrington1D, and (d) Harrington2D. In each row, the left panel varies the code distance $d$ at fixed $p=q$, while the right panel varies $p=q$ at fixed $d$. The representative error rates are chosen separately for each decoder. The \texttt{SCALA} traces approach an approximately time-independent post-transient regime, whereas Harrington's decoder exhibits longer distance-dependent relaxation and residual temporal modulation.}}

    \label{fig:hazard_stationarity}
\end{figure*}

\rev{Figure~\ref{fig:hazard_stationarity} shows representative time-dependent first-passage hazards $h(t)$, defined in Eq.~\ref{eq:hazard}, for \texttt{SCALA1D}, \texttt{SCALA2D}, Harrington1D, and Harrington2D. The left panels vary the code distance at fixed physical error rate, while the right panels vary the physical error rate at fixed distance.}

\rev{For both \texttt{SCALA1D} and \texttt{SCALA2D}, the hazard approaches a clear plateau after an initial transient. This behavior is visible both when varying the code distance at fixed $p=q$ and when varying $p=q$ at fixed distance. The onset of the plateau therefore provides an operational separation between the initial relaxation and a post-transient regime in which $h(t)$ fluctuates around an approximately time-independent value $h_\infty$. The transient duration varies with the simulation parameters, but remains short compared with the time interval over which the plateau is resolved in these representative traces. This behavior underlies the stationary-hazard estimator used for \texttt{SCALA} in App.~\ref{app:hazard}.}

\rev{The Harrington traces show qualitatively different temporal behavior. In both one and two dimensions, the relaxation becomes progressively slower with increasing code distance, as is particularly visible for the largest distances in Fig.~\ref{fig:hazard_stationarity}(c,d). This is consistent with the hierarchical dynamics, in which corrections at successively higher levels operate on increasingly long timescales. Moreover, the late-time hazard does not exhibit the same clean plateau as for \texttt{SCALA}: residual temporal modulation remains visible after the initial relaxation, most clearly for Harrington2D. Consequently, an estimate of $h_\infty$ would depend on the chosen averaging window, consistent with our use of $\langle T_F\rangle$ to characterize Harrington's performance (App.~\ref{app:hazard}).}

\subsection{\rev{Phenomenological-noise performance of Harrington's decoder}}
\label{app:har2d_pheno}

\begin{figure*}[hbtp]
    \centering
    \includegraphics[scale=0.7]{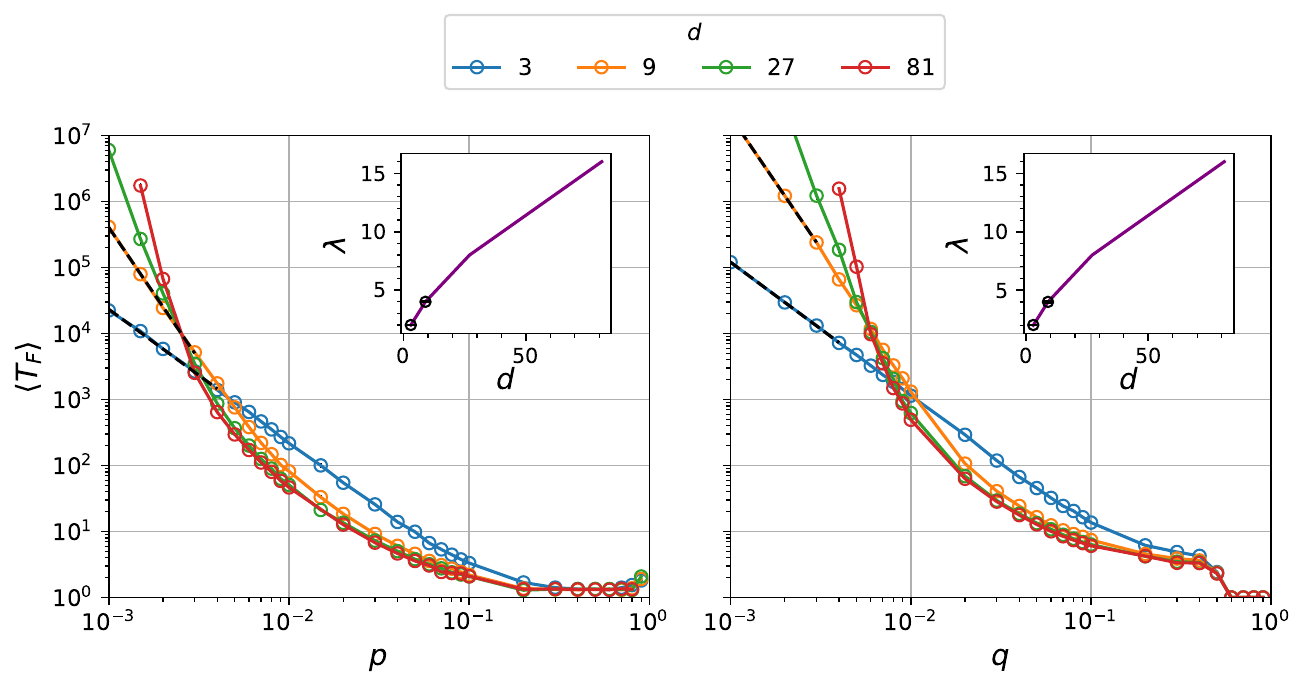}
    \caption{\rev{Average logical lifetime $\langle T_F\rangle$ as a function of phenomenological data-qubit error rate $p$ with $q=0$ (left) and measurement error rate $q$ with $p=0$ (right) for Harrington's toric-code decoder. Black dashed lines show weighted low-error fits $\langle T_F\rangle=A(d)r^{-\lambda(d)}$, with $r=p$ or $q$, for $d=3$ and $9$. The corresponding fitted exponents are shown as black circles in the insets, with error bars denoting $1\sigma$ fit uncertainties, together with the hierarchical prediction $\lambda(d)=d^{\log_3 2}\approx d^{0.631}$ (purple). For $d=27$ and $81$, the accessible data remain pre-asymptotic and no numerical exponent fit is reported. Error bars on $\langle T_F\rangle$ denote standard errors of the sample mean.}}
    \label{fig:har2d_pheno_separate}
\end{figure*}

\rev{Figure~\ref{fig:har2d_pheno_separate} shows the separate response of Harrington's decoder to data-qubit noise ($q=0$) and measurement noise ($p=0$). The numerical data use $N=10^3$--$3\times10^6$ independent first-passage trajectories per parameter point. In both noise channels, increasing the code distance strongly increases the logical lifetime at sufficiently small error rates. However, the finite-size crossings occur at error rates of order $10^{-3}$ and shift toward lower physical error rates with increasing distance. The accessible system sizes therefore do not provide a common finite-size crossing from which an asymptotic phenomenological threshold could be directly extracted.}

\rev{At low error rates, the smaller-distance data approach the hierarchical power-law scaling $\langle T_F\rangle\propto r^{-\lambda(d)}$ expected from the recursive structure of Harrington's decoder. We therefore fit only $d=3$ and $9$, where this regime is numerically resolved. The larger-distance data remain pre-asymptotic over the accessible error-rate range, so fitting them would not provide a reliable estimate of the asymptotic exponent. The resulting $d=3$ and $9$ exponents are consistent with the hierarchical prediction $\lambda(d)=d^{\log_3 2}$ within the numerically accessible regime.}

\section{One-dimensional reduction of Harrington's decoder}\label{app:har1d}

\rev{As discussed in the main text,} the error-correction performance of Harrington's decoder scales in the sub-threshold regime as $p^{\lambda(d)}$ with $\lambda(d)=\rev{d^{\log_3 2}\approx d^{0.631}}$ for toric code distance $d$. In this section, we show that this scaling can be explained solely by error processes and decoder corrections confined to a one-dimensional cross section of the toric code lattice. In this case, the minimum bit-flip logical operator corresponds to the logical operator of the repetition code, thus error correction on this cross section is identical to protecting a 1D repetition code.

In the following, we reduce Harrington's decoder to one dimension, inducing an emergent concatenated majority voting on qubit errors via the CA dynamics. Afterwards, we study how logical errors arise from data qubit, measurement and signal error processes of minimum qubit support (error weight). The results of this section not only explain the scaling behavior of Harrington's decoder in two dimensions but also highlight \rev{the consequences of its hierarchical design discussed in the main text}. We begin by discussing the reduction of Harrington's decoder to 1D.

\subsection{Construction}

\begin{figure*}[hbtp]
    \centering
    \includegraphics[scale=0.34]{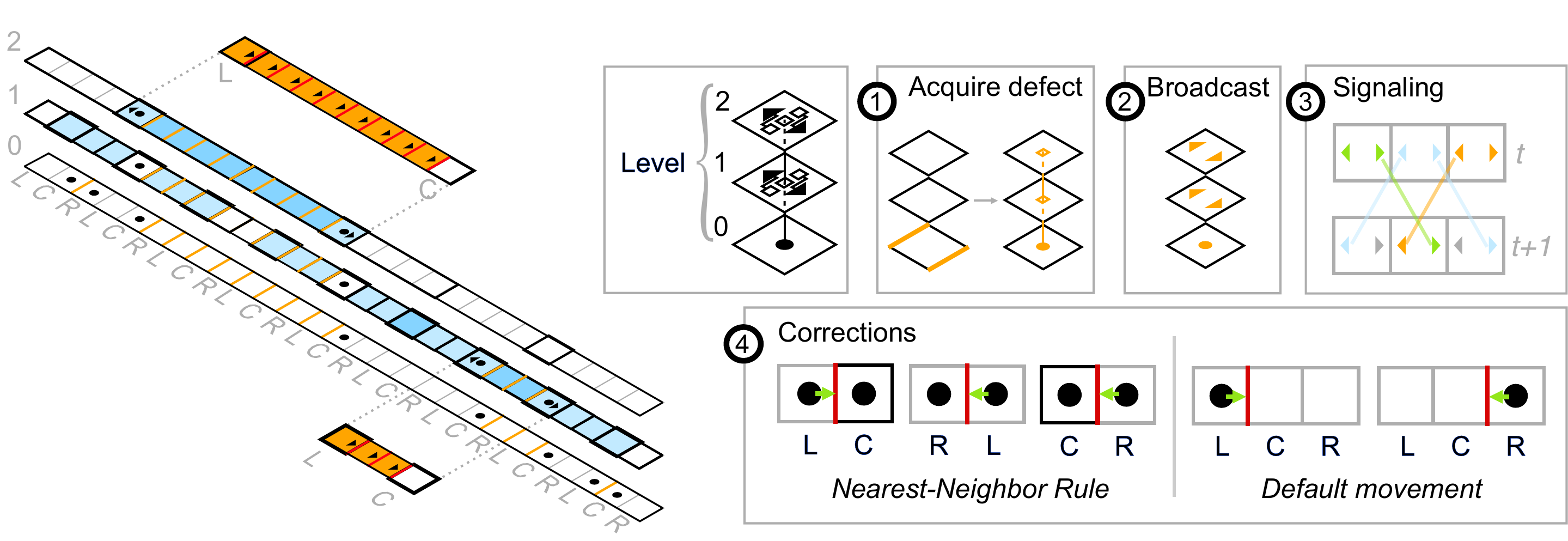}
    \caption{Sketch of Harrington's decoder confined to one dimension. On the left, we show a distance-$27$ repetition code from the perspective of the three hierarchy levels Harrington's decoder imposes. Error correction is accomplished by moving close-by defects (black circles) towards each other, thereby correcting the errors (red and orange lines) which separate them. At the lowest level (0), one- and some two-qubit errors are corrected within at most two time steps, while defects that have no other defect in their neighborhood are moved towards their colony center (marked in black thick lines at the next level). At the next level (1), \texttt{CountSignals} (colored in blue; shade corresponds to amount of signals) are exchanged between colony center cells and counted upon reception for $U$ time steps after which they are averaged and compared to the threshold values $f_C, f_N$ to yield level-1 defects\rev{.} Based on the level-1 defects, the colony center cells calculate level-1 corrections which are subsequently propagated via \texttt{FlipSignals} (orange) to all intermediate cells between the cell and its neighboring center cells within the next $Q$ time steps. Similar to level-0, level-1 defects which have no other level-1 neighbor defect are moved to the super-colony center, shown in thick black lines at level-2. At level-2, the exact same procedure is executed between level-2 defects, communicated by level-2 \texttt{CountSignals}, counted over $U^2$ time steps and finally corrected via level-2 \texttt{FlipSignals}. In the top center of the right-hand diagrams, we show a sketch of a CA cell. A cell contains one \texttt{Defect} bit at level-0, which is counted by \texttt{DefectCounters} (white squares) at levels $k\geq 1$. For each level $k\geq 1$, a cell has two \texttt{CountSignals} (black triangles), two \texttt{SignalCounters} (attached white squares) and two \texttt{FlipSignals} (not shown). Additionally, a cell stores working period $U$, linear colony size $Q=3$, threshold values $f_C, f_N$, \rev{$m-1$ \texttt{Age} counters}, as well as \rev{$m$ \texttt{Address} fields} which hold the relative address of a cell within its level-$k$ colony. In the boxes labeled (1---4), we show the local rule, divided into subrules, \rev{as detailed below}. In orange, we highlight parts of a cell’s memory involved in a substep. In 3), we use color to illustrate the exchange of signals by assigning unique colors to cell signals before the exchange occurs. In 4), a green arrow represents a correction resulting in a bit-flip (red line) applied to a data qubit.}
    \label{fig:harrington1d_fig1}
\end{figure*}

The dynamics of the one-dimensional automaton, as sketched on the left of Fig.~\ref{fig:harrington1d_fig1}, protect the repetition code against bit-flip errors. We modify the local rules of \rev{the two-dimensional construction of Sec.~\ref{sec:har2d_construction}} to match the one-dimensional scenario, shown in Fig.~\ref{fig:harrington1d_fig1} (1---4). First, the information about the possible presence of a defect is acquired by measuring the $Z_iZ_{i+1}$ stabilizer of the repetition code (1), while incrementing the \texttt{DefectCounters} at all levels $k\geq 1$. \rev{At each hierarchy level, representative cells broadcast (2) their instantaneous defect state to the left and right via \texttt{CountSignals}, which are subsequently propagated through the intermediate cells.} These signals propagate (3) to the left and right by swapping with their neighbors. In one dimension, a colony reduces to three cells with addresses $L,C,R$, corresponding to \textit{left}, \textit{center} and \textit{right} relative positions. The correction rules (4) in one dimension simplify significantly. Neighboring defects are annihilated directly (Nearest-Neighbor subrule) and defects with no matching partner are moved to the colony center (Default movement), analogous to the two-dimensional CA. Movement across colony borders is prioritized via a deterministic tie-breaking rule: if an $L$-cell and its neighboring $R$-cell both contain defects, the $L$-cell executes a flip while the $R$-cell performs no action. A CA cell (Fig.~\ref{fig:harrington1d_fig1} top center) needs only $m-1$ left and right \texttt{CountSignals} (black triangles) and \texttt{FlipSignals} (not shown), $m-1$ left and right \texttt{SignalCounters} (white squares attached to triangles), one \texttt{Defect} bit and $m-1$ associated \texttt{DefectCounters} (white squares attached to defect bit), as well as $m$ \texttt{Address} fields taking one of the values $\mathcal{L}_\circ^{(1D)} = \{L,R,C,\circ\}$ (not shown) where $m$ is the largest hierarchy level. Finally, each cell has \rev{$m-1$ \texttt{Age} counters, one for each higher hierarchy level,} and storage for the constants $U,Q,f_C,f_N$ (not shown), similar to the two-dimensional automaton. In Alg.~\ref{alg:har1d_global_rule}, we present pseudo-code describing one global CA update step. \rev{The signal-update, reset, and correction subroutines are specified explicitly in Alg.~\ref{alg:har1d_subroutines}.} The symbols $\lor, \land, \oplus$ represent logical disjunction (OR), conjunction (AND) and exclusive disjunction (XOR). For simplicity, we introduce the constants $U,Q,f_C,f_N,d,m$ in the global scope. The memory of a cell is accessed via dot notation, e.g., $c_j.FlipSig[k][L]$ corresponds to the left \texttt{FlipSignal} bit at level-$k$ of the cell at location $j$. \rev{For the hierarchical construction considered here we take $d=3^m$, such that $m=\log_3 d$ is an integer. We denote the level-$k$ \texttt{DefectCounter} by $\mathrm{Count}[k][C]$ and the left and right \texttt{SignalCounters} by $\mathrm{Count}[k][L]$ and $\mathrm{Count}[k][R]$, respectively. Signal propagation is synchronous: primed signal variables denote the state to be committed simultaneously after all cells have been evaluated. At the beginning of each update, \rev{each} level-$k$ clock is incremented modulo $U^k+Q^k$. A level-$k$ correction is decided when $\mathrm{Age}[k]=U^k-1$. Its \texttt{FlipSignal} subsequently propagates along the correction path, and cells carrying the signal execute the corresponding physical correction when $\mathrm{Age}[k]=U^k+Q^k-1$. At this point the level-$k$ signals and counters are reset; the age clock itself wraps to zero on the following update.}

\begin{algorithm}[hbtp]
  \caption{Harrington1D global update}
  \label{alg:har1d_global_rule}
  \begin{algorithmic}[1]
  \Procedure{update}{syndrome$[\;]$, cells$[\;]$, qubits$[\;]$}

        \State $U \gets 10$ \Comment{Working period}
        \State $Q \gets 3$ \Comment{Colony size}
        \State $f_C \gets 9/10$ \Comment{Center defect threshold}
        \State $f_N \gets 4/10$ \Comment{Neighbor defect threshold}
        \State $d \gets$ \Call{length}{syndrome}
        \State $m \gets \log_3(d)$ \Comment{Max. hierarchy level}

        \Comment{\rev{Update hierarchy signals and counters}}
        \For{$k \gets 1$ to $m-1$}
            \State \rev{\Call{UpdateSignals}{$k$, syndrome, cells}}
        \EndFor

        \Comment{\rev{Determine corrections, highest level first}}
        \For{$j \gets 1$ to $d$}
            \State $c_j \gets \mathrm{cells}[j]$
            \State \rev{$\mathrm{dir}\gets\circ$}

            \For{\rev{$k \gets m-1$ downto $1$}}
                \If{\rev{$c_j.\mathrm{Age}[k]=U^k-1
                    \land c_j.\mathrm{Addr}[k]\neq\circ$}}
                    \State $c \gets c_j.\mathrm{Count}[k][C]\geq f_CU^k$
                    \State $l \gets c_j.\mathrm{Count}[k][L]\geq f_NU^k$
                    \State $r \gets c_j.\mathrm{Count}[k][R]\geq f_NU^k$
                    \State \rev{$\mathrm{move}\gets
                        \Call{correct}{l,c,r,c_j.\mathrm{Addr}[k]}$}
                    \If{\rev{$\mathrm{move}\neq\circ$}}
                        \State \rev{$c_j.\mathrm{FlipSig}[k][\mathrm{move}]
                            \gets1$}
                    \EndIf
                \EndIf

                \If{\rev{$c_j.\mathrm{Age}[k]=U^k+Q^k-1$}}
                    \State \rev{$f_L\gets c_j.\mathrm{FlipSig}[k][L]$}
                    \State \rev{$f_R\gets c_j.\mathrm{FlipSig}[k][R]$}
                    \State \rev{\Call{ResetLevel}{$c_j,k$}}
                    \If{\rev{$f_L$}}
                        \State $\mathrm{dir}\gets L$
                        \State \textbf{break}
                    \ElsIf{\rev{$f_R$}}
                        \State $\mathrm{dir}\gets R$
                        \State \textbf{break}
                    \EndIf
                \EndIf
            \EndFor

            \If{\rev{$\mathrm{dir}=\circ$}}
                \State $l,c,r\gets
                    \mathrm{syndrome}[(j-1)\text{ to }(j+1)\bmod d]$
                \State $\mathrm{dir}\gets
                    \Call{correct}{l,c,r,c_j.\mathrm{Addr}[0]}$
            \EndIf

            \If{$\mathrm{dir}=L$}
                \State $\mathrm{qubits}[j]\gets\mathrm{qubits}[j]\oplus1$
            \ElsIf{$\mathrm{dir}=R$}
                \State $\mathrm{qubits}[(j+1)\bmod d]\gets
                    \mathrm{qubits}[(j+1)\bmod d]\oplus1$
            \EndIf
        \EndFor
  \EndProcedure
  \end{algorithmic}
\end{algorithm}

\begin{algorithm}[hbtp]
  \caption{\rev{Harrington1D update subroutines}}
  \label{alg:har1d_subroutines}
  \begin{algorithmic}[1]

  \Function{\rev{UpdateSignals}}{$k$, syndrome$[\;]$, cells$[\;]$}
      \State \rev{$d\gets\Call{length}{syndrome}$}
      \State \rev{$\mathrm{CountSig}'[\;],\mathrm{FlipSig}'[\;]\gets0$}

      \For{\rev{$j\gets1$ to $d$}}
          \State \rev{$c_j\gets\mathrm{cells}[j]$}
          \State \rev{$c_j.\mathrm{Age}[k]\gets (c_j.\mathrm{Age}[k]+1)\bmod(U^k+Q^k)$}
          \State \rev{$j_L\gets(j-1)\bmod d,\quad j_R\gets(j+1)\bmod d$}

          \State \rev{$s_L\gets\mathrm{cells}[j_R].\mathrm{CountSig}[k][L]$}
          \State \rev{$s_R\gets\mathrm{cells}[j_L].\mathrm{CountSig}[k][R]$}
          \State \rev{$f_L\gets\mathrm{cells}[j_R].\mathrm{FlipSig}[k][L]$}
          \State \rev{$f_R\gets\mathrm{cells}[j_L].\mathrm{FlipSig}[k][R]$}

          \If{\rev{$c_j.\mathrm{Addr}[k]\neq\circ$}}
              \State \rev{$\mathrm{CountSig}'[j][k][:]\gets
                  \mathrm{syndrome}[j]$}
              \State \rev{$\mathrm{FlipSig}'[j][k][:]
                  \gets c_j.\mathrm{FlipSig}[k][:]$}
              \State \rev{$c_j.\mathrm{Count}[k][C]\gets
                  c_j.\mathrm{Count}[k][C]+\mathrm{syndrome}[j]$}
              \State \rev{$c_j.\mathrm{Count}[k][L]\gets
                  c_j.\mathrm{Count}[k][L]+s_R$}
              \State \rev{$c_j.\mathrm{Count}[k][R]\gets
                  c_j.\mathrm{Count}[k][R]+s_L$}
          \Else
              \State \rev{$\mathrm{CountSig}'[j][k][L]\gets s_L$}
              \State \rev{$\mathrm{CountSig}'[j][k][R]\gets s_R$}
              \State \rev{$\mathrm{FlipSig}'[j][k][L]\gets f_L$}
              \State \rev{$\mathrm{FlipSig}'[j][k][R]\gets f_R$}
          \EndIf
      \EndFor

      \For{\rev{$j\gets1$ to $d$}}
          \State \rev{$\mathrm{cells}[j].\mathrm{CountSig}[k][:]
              \gets\mathrm{CountSig}'[j][k][:]$}
          \State \rev{$\mathrm{cells}[j].\mathrm{FlipSig}[k][:]
              \gets\mathrm{FlipSig}'[j][k][:]$}
      \EndFor
  \EndFunction

  \State
  \Function{\rev{ResetLevel}}{$c_j,k$}
      \State \rev{$c_j.\mathrm{Count}[k][:]\gets0$}
      \State \rev{$c_j.\mathrm{CountSig}[k][:]\gets0$}
      \State \rev{$c_j.\mathrm{FlipSig}[k][:]\gets0$}
  \EndFunction

  \State
  \Function{correct}{$l,c,r,\mathrm{addr}$}
      \If{\rev{$\neg c\lor\mathrm{addr}=C$}}
          \State \Return $\circ$
      \ElsIf{$\mathrm{addr}=L$}
          \If{$l$}
              \State \Return $L$
          \Else
              \State \Return $R$
          \EndIf
      \ElsIf{$\mathrm{addr}=R$}
          \If{\rev{$r$}}
              \State \Return $\circ$
          \Else
              \State \Return $L$
          \EndIf
      \EndIf
      \State \Return $\circ$
  \EndFunction

  \end{algorithmic}
\end{algorithm}

\subsection{Analysis of logical errors}

\begin{figure}[htb]
    \centering
    \includegraphics[scale=0.2]{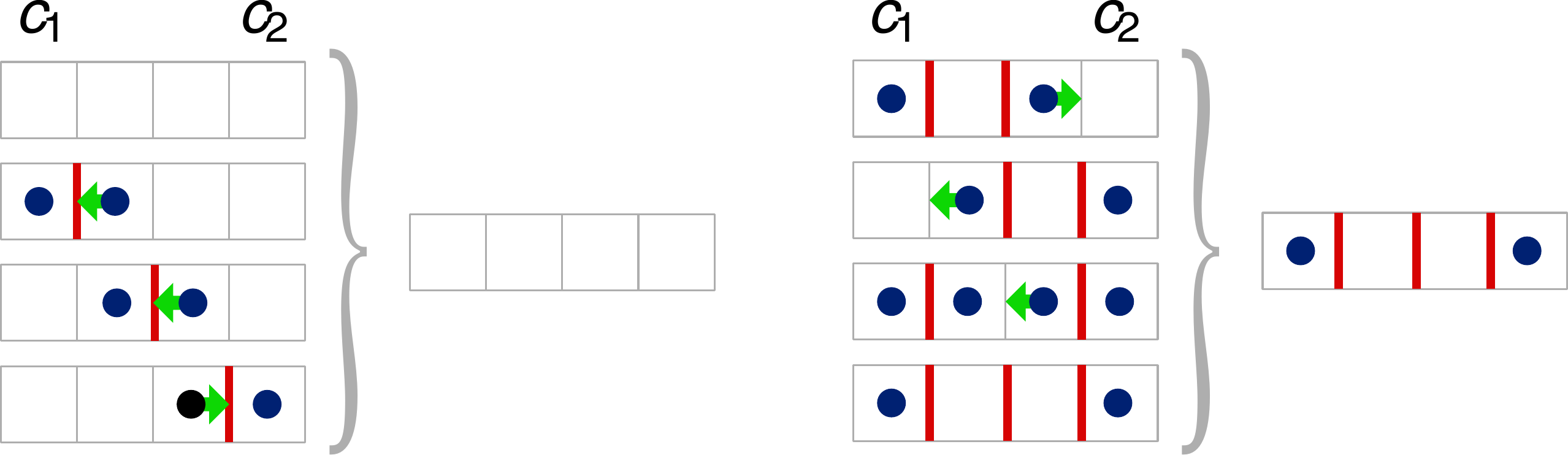}
    \caption{All possible error configurations of qubits within a block. Data-qubit errors (red bars) generate a syndrome (black circles) between two colony centers $C_1$ and $C_2$. The green arrows indicate the local update prescribed by the correction rule. Configurations with zero or one error (left) are corrected to the trivial configurations with no errors between the centers, whereas configurations with two or three errors (right) are completed to the nontrivial configuration with three errors between the centers.}
    \label{fig:LL_maj}
\end{figure}

The local rule of Harrington's automaton is designed in such a way that any isolated error on one of the data qubits is removed within one time step. When two errors occur on the data qubits exactly between two colony centers, the decoder adds an error to the third qubit. We will refer to a group of qubits between two colony centers as (level-0) \textit{block}. The decoder action on all error configurations of a block can be seen in Fig.~\ref{fig:LL_maj}. These corrections effectively induce a majority vote on the data qubit errors of a block. If all qubits within a block have an error, we say a \textit{block error} occurred. Block errors cause defects at their enclosing colony centers, which are counted over the working period $U$ and subsequently corrected at the next level. A single block error is then corrected after emission of \texttt{FlipSignals} and subsequent $Q$ steps of travel time. If, however, two block errors are present between two neighboring level-1 colony centers for at least $f_C U$ time steps, the level-1 correction decides to create a new block error, similar to the action at level-0. This process is applied \rev{recursively} at each level resulting in an effective concatenated majority vote.

\begin{figure}[!hbtp]
    \centering
    \includegraphics[scale=0.7]{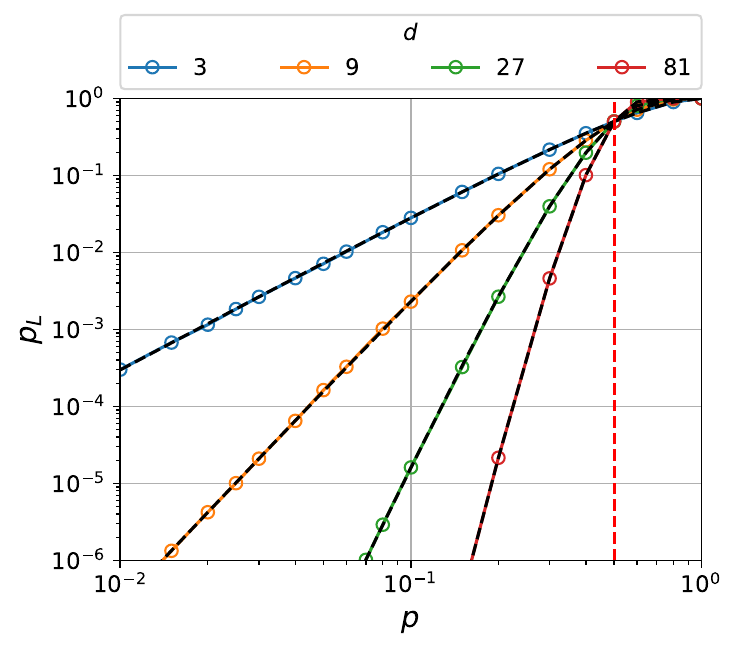}
    \caption{Logical error rate $p_L$ as \rev{a} function of data qubit error rate $p$ for Harrington's decoder confined to one dimension under code capacity noise on distance-$d$ repetition codes. The red dashed line marks the QEC threshold at $p_c=1/2$. The dashed black lines correspond to the \rev{analytical} performance of \rev{global} concatenated majority voting, $p_L=p_{\mathrm{maj}}^m(p)$.}
    \label{fig:har1d_cc}
\end{figure}

In the code-capacity setting, we recover the same performance as concatenated majority voting on the repetition code, as shown in Fig.~\ref{fig:har1d_cc}. \rev{The numerical data use $N=10^5$--$10^9$ independent Monte Carlo samples per parameter point, depending on the logical error rate.} The black dashed lines correspond to (global) concatenated voting, obeying $p_L = p_{\mathrm{maj}}^m(p)$, where $p_{\mathrm{maj}}(p) = 3p^2(1-p)+p^3$ and $m=\log_3(d)$ for code distance $d$ and data qubit error rate $p$, while the solid colored lines correspond to the performance of Harrington's decoder in one dimension. 

\rev{To quantitatively determine the code-capacity threshold, we additionally perform a finite-size scaling analysis using the ansatz
\begin{equation}
    p_L = F\left[(p-p_c)d^{1/\nu}\right],
\end{equation}
where the scaling function $F$ is approximated by a cubic polynomial. The resulting data collapse is shown in Fig.~\ref{fig:har1d_cc_fss}, with the fit parameters and uncertainties reported in the caption.

\begin{figure}[!hbtp]
    \centering
    \includegraphics[scale=0.7]{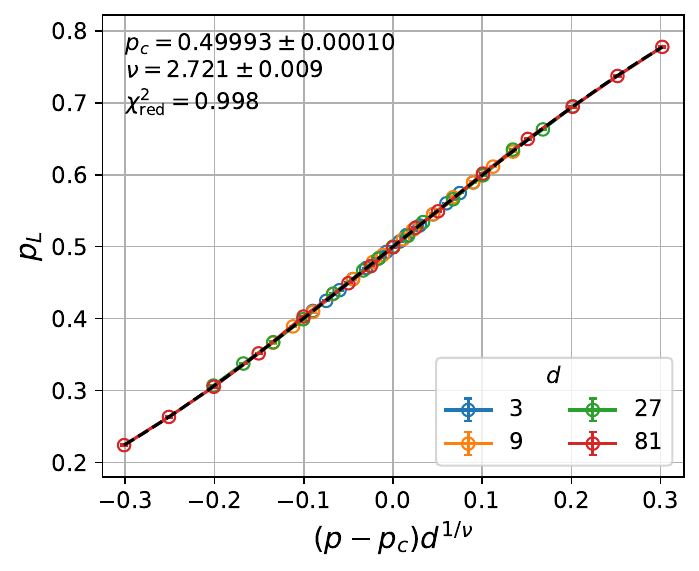}
    \caption{\rev{Finite-size scaling collapse of the Harrington1D code-capacity data shown in Fig.~\ref{fig:har1d_cc}, using the scaling variable $(p-p_c)d^{1/\nu}$. The fit includes distances $d=3,9,27,81$ and physical error rates $p\in[0.44,0.56]$. The dashed black line shows the fitted cubic approximation to the scaling function $F$. A weighted least-squares fit using the Monte Carlo standard errors gives $p_c=0.49993\pm0.00010$ and $\nu=2.721\pm0.009$, with $\chi^2_{\mathrm{red}}=0.998$. Error bars show the Monte Carlo standard errors.}}
    \label{fig:har1d_cc_fss}
\end{figure} }

\begin{figure*}[hbtp]
    \centering
    \includegraphics[scale=0.7]{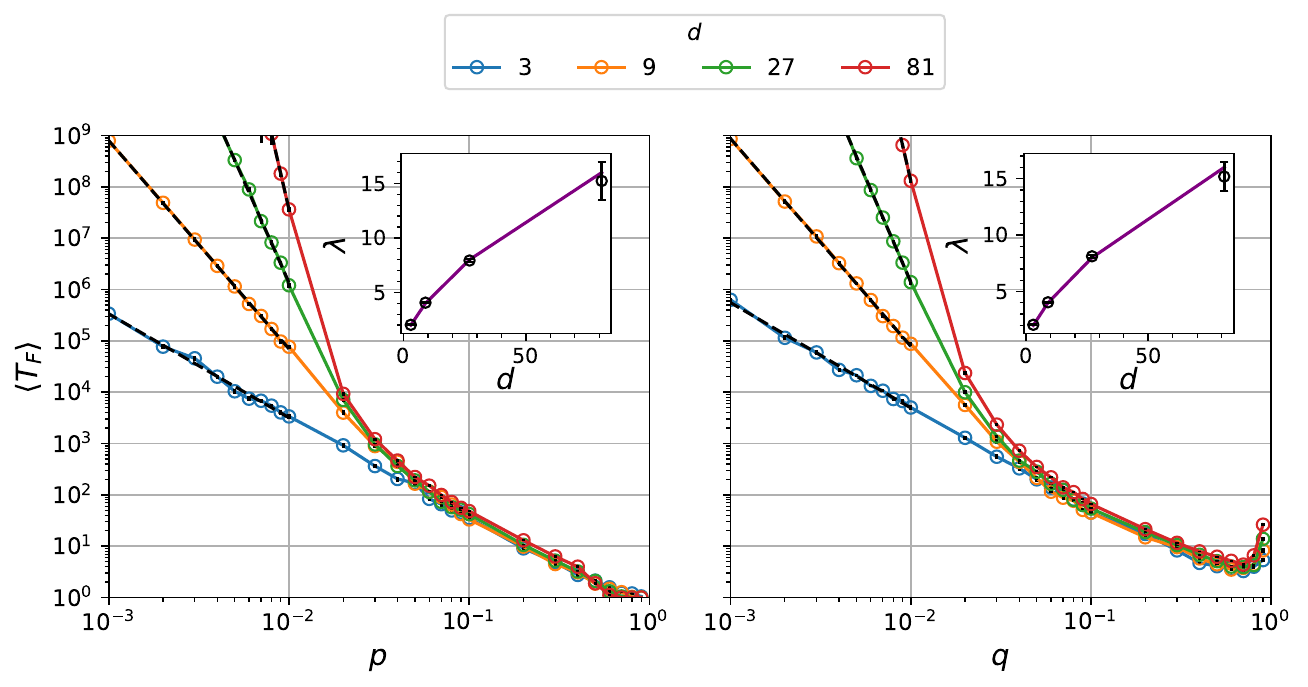}
    \caption{Logical lifetime $\langle T_F \rangle$ as a function of data qubit error rate $p$ (left) and measurement error rate $q$ (right) with $q=0$ ($p=0$), respectively, for the one-dimensional Harrington decoder under phenomenological noise. \rev{The dashed black lines show weighted fits of $\langle T_F\rangle=A(d)r^{-\lambda(d)}$, with $r=p$ ($q$) in the left (right) panel. The fitted exponents $\lambda(d)$ are shown in the insets, with error bars denoting $1\sigma$ fit uncertainties. The purple lines indicate the hierarchical prediction $\lambda(d)=d^{\log_3 2}\approx d^{0.631}$.}}
    \label{fig:har1d_pheno}
\end{figure*}

In the phenomenological setting, we obtain the numerical performance shown in Fig.~\ref{fig:har1d_pheno} for pure data qubit noise (left) and measurement noise (right). \rev{The numerical data use $N=10^3$--$10^6$ independent first-passage trajectories per parameter point. In the low-error regime, we fit the mean logical lifetime to $\langle T_F\rangle=A(d)r^{-\lambda(d)}$, with $r=p$ or $q$, using weighted least squares and the Monte Carlo standard errors of $\langle T_F\rangle$ as statistical uncertainties. The fitted exponents are shown in the insets, with error bars denoting $1\sigma$ fit uncertainties.} In the following, we analyze relevant error mechanisms \rev{underlying the observed subthreshold scaling}. We start with data qubit errors.

\subsubsection{Data errors}\label{har1ddata}

\begin{figure}[hbtp]
    \centering
    \includegraphics[scale=0.65]{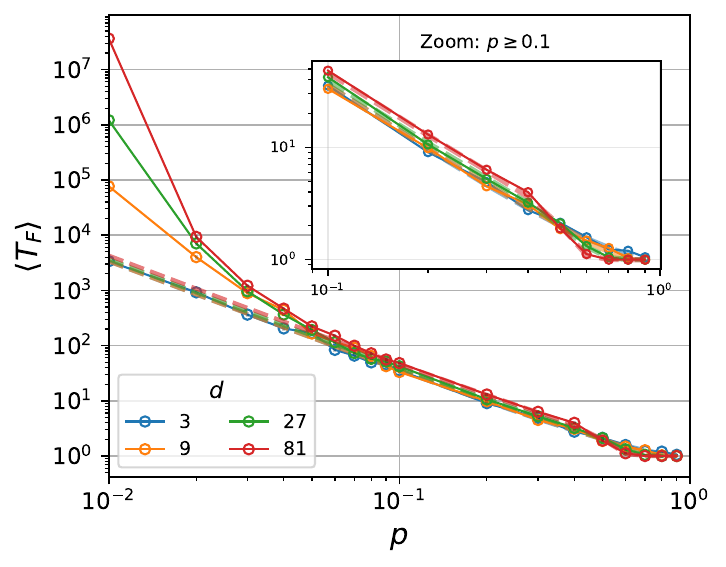}
    \caption{Average logical lifetime $\langle T_F \rangle$ data from Fig.~\ref{fig:har1d_pheno} for data errors. In colored dashed lines we show analytical results obtained from Eq.~\ref{eq:T_ana} considering only the lowest level of the error correction hierarchy. In the inset, we zoom in to the regime of large data qubit error rates, $p\geq 0.1$, to illustrate that the analytical results closely follow the numerical ones.}
    \label{fig:har1d_pheno_lvl1}
\end{figure}

When physical qubits suffer independent bit-flip errors with probability $p$, a block error occurs by the corrections (Fig.~\ref{fig:LL_maj}) with block-flip error probability $p_b=3p^2(1-p)+p^3$. For $n=3^m$ qubits in a (concatenated) repetition code, there are $n_b=n/3$ level-1 blocks, which are independently corrected. Thus, starting with an initial state without errors, a logical error occurs when at least $\lceil n_b/2 \rceil$ ($n$ odd) block errors have accumulated. This accumulation process can be modeled as a Markov chain of $n_b$ Bernoulli random variables with bit-flip probability $p_b$. The average logical lifetime corresponds to the average hitting time of the set of states which have more ones (block errors) than zeros (error-free blocks). These states are absorbing states at which the Markov process terminates once hit. The resulting absorbing Markov chain is described by transition matrix $P$, which has the structure,

\begin{equation}
    P= 
    \begin{pmatrix}
        W & R\\
        0 & I
    \end{pmatrix},
\end{equation}

where $W$ represents the transitions between transient states (states of more zeros than ones), $R$ the transitions from transient to absorbing states and $I$ and $0$ represent the identity matrix and the matrix of all zeros, since absorbing states never return to any other state. The average hitting time, i.e., logical lifetime $\langle T_{1} \rangle$ with respect only to level-1 block corrections can then be calculated as

\begin{equation}\label{eq:T_ana}
    \langle T_{1} \rangle = \pi_0 (\mathbb{I}-W)^{-1}\mathbf{1},
\end{equation}

where $\pi_0$ is the \rev{initial-state distribution, which we take to be concentrated on the error-free all-zero state}.~$\mathbf{1}$ is a vector of ones. A detailed derivation of this equation~ can be found in App.~\ref{app:T_proof}. Although Eq.~\ref{eq:T_ana} allows the exact calculation of logical \rev{lifetimes}, its computation \rev{quickly becomes computationally infeasible as} the transition matrix $P$ grows exponentially in the number of blocks $n_b$ and therefore the transient block $W$ also grows exponentially. In App.~\ref{app:markov}, we exploit permutation symmetry of blocks in the Markov process to reduce this description to a matrix with size linear in $n$. This simplification allows us to calculate hitting times for larger systems as well. We show the resulting performance curves as black dashed lines over the numerical results for data bit-flip noise in Fig.~\ref{fig:har1d_pheno_lvl1}. In the regime of large data error rates, $p\geq 0.1$, the analytical curves  closely follow the numerical ones. For lattice size $n=3$ the analytical curve matches the numerical data for all $p$. This is not very surprising, as for $n=3$ there is only one level of error correction, thus no other correction processes have influence on the performance. For the other code distances, the higher hierarchy levels have a chance to correct errors at time steps $t=U^k+Q^k\rev{-1}$ which become dominant contributions for low error rates of $p<10^{-1}$.

To model the logical lifetime for small physical error rates we have to consider the effect of higher levels in the error correction hierarchy. We focus on code distance $d=9$, for which there are two levels. The level-0 correction occurs every CA step while the level-1 correction \rev{operates with period $\tau=U+Q$}. If the level-1 correction works perfectly, i.e., only considers the state at time step $t=\rev{\tau}$ after the last level-0 correction to vote on the majority state of the three blocks, only two outcomes can occur at this time step. Either all errors are successfully removed and the CA configuration is effectively reset to its initial state of all-zeros or a logical error occurs, in which case the Markov process terminates. We can model this perfect reset mechanism using the law of total expectation,

\begin{equation}
    \mathbb{E}(X) = \sum_i \mathbb{E}(X|A_i)\mathbb{P}(A_i),
\end{equation}

where $\{A_i\}$ is a finite partition of sample space, and $X$ a random variable over sample space, $\mathbb{P}(A_i)$ the sum of probabilities of all atomic events in $A_i$ and $\mathbb{E}(X|A_i)$ the expectation value of random variable $X$ constrained to subset $A_i$. In our case, we have the random variable $T$ over the sample space of all logical lifetimes from 1 to infinity and we partition the lifetimes into times before the level-1 correction takes place, i.e., $t<\tau=U+Q$, and after, i.e., $t\geq \tau$. The corresponding subsets are denoted by $A_{t<\tau}$ and $A_{t\geq\tau}$, respectively. The cumulative probabilities are denoted by $\mathbb{P}_{t<\tau}$ and $\mathbb{P}_{t\geq\tau}$. Thus, we obtain for the average lifetime $\langle T_2 \rangle$ for $d=9$ considering both level-0 and level-1 corrections together by
 
\begin{equation}
    \label{eq:L9LawOfExp}
    \begin{split}
        \langle T_2 \rangle = \mathbb{E}(T) &= \mathbb{P}_{t < \tau} \mathbb{E}(T|A_{t< \tau}) + (1-\mathbb{P}_{t< \tau}) \mathbb{E}(T|A_{t\geq\tau})\\
                      &= \mathbb{P}_{t<\tau} \mathbb{E}(T|A_{t<\tau}) + (1-\mathbb{P}_{t<\tau}) (\tau + \mathbb{E}(T))\\
                      &= \mathbb{E}(T|A_{t<\tau}) + \tau (1-\mathbb{P}_{t<\tau})/\mathbb{P}_{t<\tau}.
    \end{split}
\end{equation}

The second line in Eq.~\ref{eq:L9LawOfExp} follows from resetting the system to the state it was in at time $t=0$ if no logical error occurred. Thus, for the lifetime after the reset we have $\mathbb{E}(T|A_{t\geq\tau})=\tau+\mathbb{E}(T)$. From line 2 to 3, we first subtract $(1-\mathbb{P}_{t<\tau})\mathbb{E}(T)$ from both sides and then divide by $\mathbb{P}_{t<\tau}$. The cumulative probability $\mathbb{P}_{t<\tau}$ and expectation value $\mathbb{E}(T|A_{t<\tau})$ can be obtained from the Markov chain transition matrix $P$ by

\begin{equation}
    \mathbb{P}_{t<\tau}
    =
    \rev{\sum_{t=1}^{\tau-1} \pi_0 W^{t-1} R \mathbf{1}},
\end{equation}
and 
\begin{equation}
    \mathbb{E}(T|A_{t<\tau})=
    \rev{\frac{1}{\mathbb{P}_{t<\tau}}}
    \sum_{t=1}^{\tau-1} t\,\rev{\pi_0 W^{t-1}R\mathbf{1}}.
\end{equation}

\begin{figure}[hbtp]
    \centering
    \includegraphics[scale=0.7]{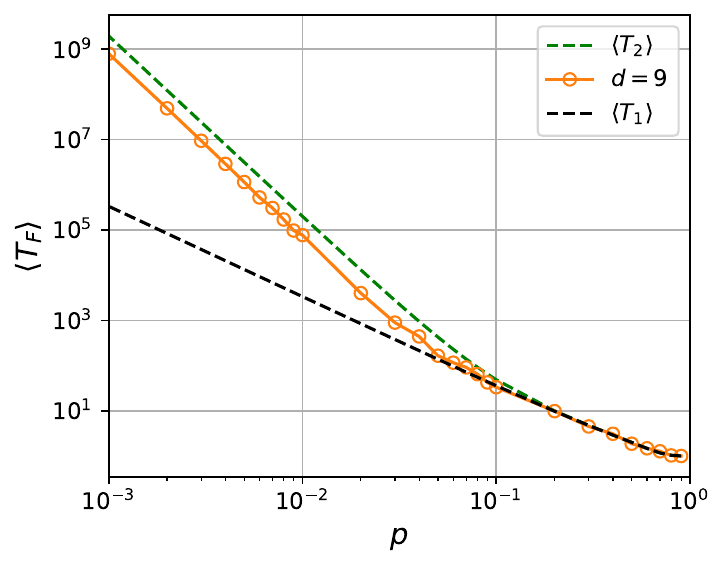}
    \caption{Average logical lifetime from Fig.~\ref{fig:har1d_pheno} for data noise and $d=9$. \rev{The black dashed line shows the analytical result $\langle T_1\rangle$ obtained from Eq.~\ref{eq:T_ana} when only the lowest level of the error-correction hierarchy is included, while the green dashed line shows $\langle T_2\rangle$ from Eq.~\ref{eq:L9LawOfExp}, including the idealized level-1 reset.} In the regime of large data qubit error rates, $p>10^{-1}$, the \rev{level-1 analytical result closely matches} the numerical \rev{data}.}
    \label{fig:har1d_pheno_lvl2}
\end{figure}

In Fig.~\ref{fig:har1d_pheno_lvl2}, we show the numerical lifetime for code distance $d=9$ together with the analytical lifetimes considering only level-1 corrections, $\langle T_1 \rangle$ (black dashed line) and analytical lifetime considering level-0 and level-1 corrections, $\langle T_2 \rangle$ (green dashed line). The analytical results clearly bound the numerical performance. In the regime $p<10^{-1}$, the numerical performance scales in the same way as $\langle T_2 \rangle$, only offset by a constant factor. The reason for this is that in practice, higher-level corrections are not perfect. Due to the CA dynamics, we find additional minimum-weight space-time error configurations which lead to logical errors. Consequently, \rev{additional minimum-weight configurations can change the prefactor while preserving the same leading power-law scaling, consistent with} our observations of Fig.~\ref{fig:har1d_pheno_lvl2}.

Although it was straightforward to calculate average logical lifetimes for $d=9$, for larger code distances this analysis becomes more involved. At the highest level, the system is always reset under perfect voting, at intermediate levels, only a partial reset occurs. This partial reset cannot be directly modeled via the law of total expectation. Nonetheless, \rev{the low-$p$ scaling observed in Fig.~\ref{fig:har1d_pheno} is consistent with $\langle T_F\rangle\propto p^{-2^m}$ and therefore with the hierarchical minimum-weight mechanism described above. In particular, the fitted exponents approach the prediction $2^m=d^{\log_3 2}$ as the low-error regime is resolved.} This supports the interpretation that imperfect voting primarily changes the number of failure-inducing configurations rather than the leading power-law exponent.

\subsubsection{Measurement errors}\label{har1dmeas}

For Harrington's two-dimensional decoder and for our one-dimensional reduction we observe \rev{compatible low-error scaling exponents} of the logical lifetime $\langle T_F \rangle$ under data-qubit and measurement noise. \rev{This similarity can be understood from the fact that} single measurement errors are converted to data errors via the default movement rule, pushing defects to colony centers. This process only occurs when the measurement error occurs on either an $L$ or an $R$ cell, but not on a $C$ cell \rev{where no correction is performed}. Thus, most measurement errors directly cause a data error, \rev{providing a mechanism consistent with the similar low-error scaling observed for pure data noise.} When a measurement error occurs on a $C$ cell, no data error is generated, \rev{which reduces the effective rate at which measurement errors induce data errors and is consistent with the observed offset relative to pure data noise.}

\subsubsection{Signal errors}\label{har1dsig}

In Harrington's decoder, two types of signals are employed. \texttt{CountSignals} are transmitted at each time step between colony representatives to exchange information about the presence of defects at their neighbors. \texttt{FlipSignals} are emitted at the end of a level-$k$ work period, $t=U^k$, and travel for $Q^k$ time steps to a level-$k$ neighbor representative\rev{, which they reach at $t=U^k+Q^k$}. At $t=U^k+Q^k-1$, \rev{the cells along the correction chain} which \rev{have} a \texttt{FlipSignal} present \rev{flip the corresponding data qubit}.

\begin{figure}[hbtp]
    \centering
    \includegraphics[scale=0.3]{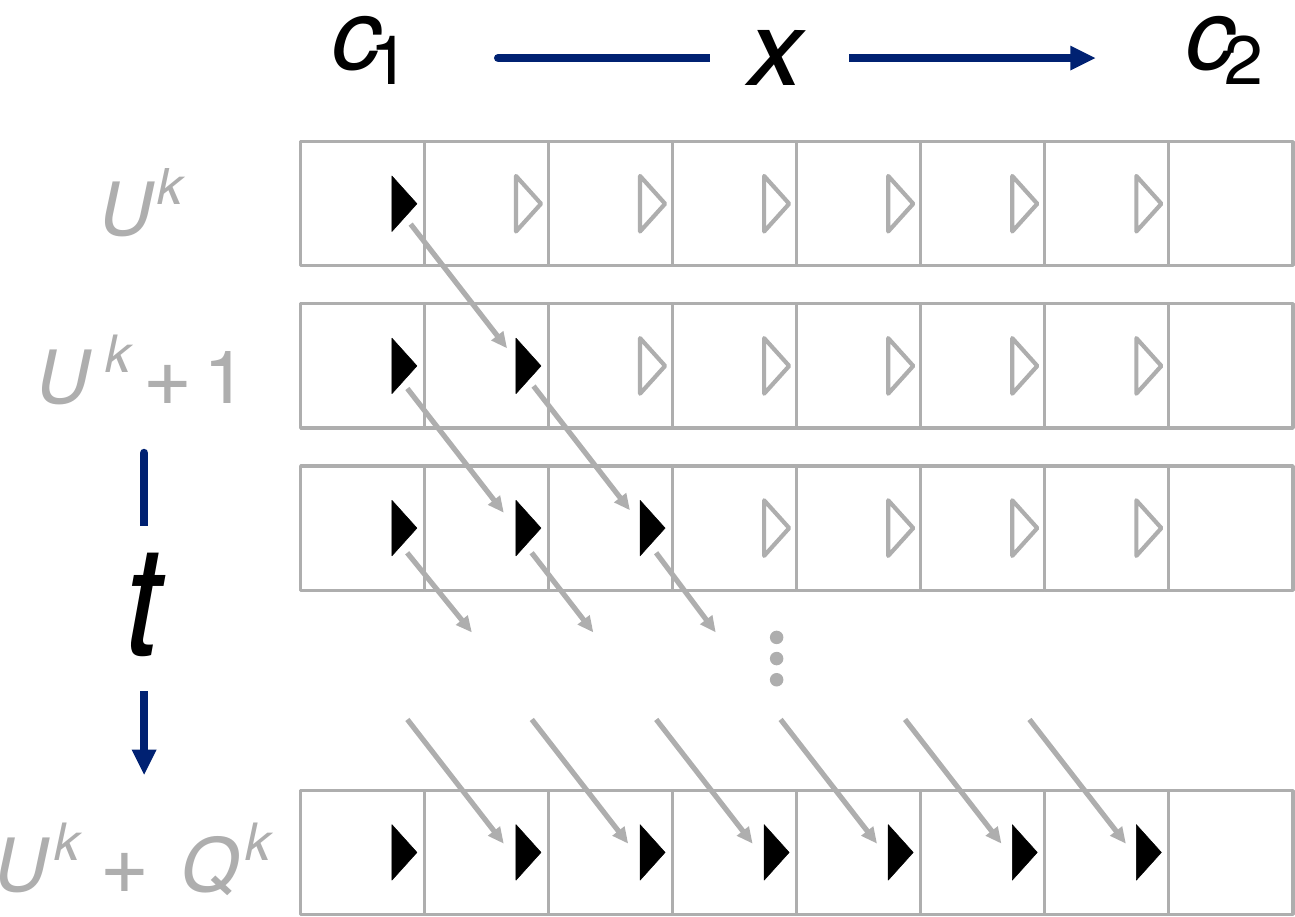}
    \caption{Illustration of \texttt{FlipSignal} propagation and replication between two colony centers. At time step $t=U^k$, cell $C_1$ decides to send a \texttt{FlipSignal} to its neighboring center cell $C_2$ by activating its right \texttt{FlipSignal} bit (blue filled triangles). All intermediate cells propagate this signal to the right by synchronously overwriting their right \texttt{FlipSignal} bit (gray filled triangles) according to the state of their left neighbor's right \texttt{FlipSignal} (gray arrows). Thus, at $t=U^k+1$, the bit \rev{has propagated} one cell to the right. Note that the blue source signal is never overwritten but only copied, i.e., replicated by its right neighbor. This process continues for $Q^k-1$ time steps, \rev{after which} cell $C_1$ and all intermediate cells in the direction of $C_2$ have an active \texttt{FlipSignal} present and decide to flip their right qubit\rev{ at $t=U^k+Q^k-1$; the signal reaches $C_2$ at $t=U^k+Q^k$}. \texttt{FlipSignal} noise is applied at the end of a CA step to all \texttt{FlipSignal} bits.}
    \label{fig:har1d_flipSig_propScheme}
\end{figure}

Pure \texttt{CountSignal} errors can never lead to logical errors, since \rev{a defect must be present for a cell to make an error-correction decision, and defects cannot be induced by} \texttt{CountSignals}. \texttt{FlipSignals}, on the other hand, can cause logical errors after conversion to data qubit errors. \texttt{FlipSignal} errors are a very destructive source of error, as they are ``replicated'' from the source center cell during propagation \rev{by the mechanism that notifies} all intermediate cells about a planned correction chain. Thus, a single \texttt{FlipSignal} error at time step $t=U^k$ can cause an error chain of length $Q^k$ at time $t=U^k+Q^k-1$. We illustrate this process in Fig.~\ref{fig:har1d_flipSig_propScheme}. 

\begin{figure}[hbtp]
    \centering
    \includegraphics[scale=0.7]{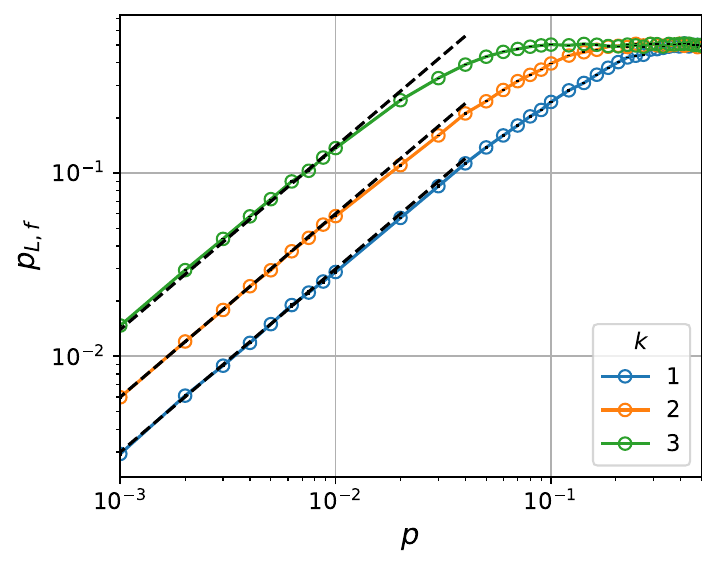}
    \caption{Occurrence probability $p_{L,f}$ of at least $\lceil Q^k/2 \rceil$ data errors due to \texttt{FlipSignal} errors for different hierarchy levels $k$ (color code), as explained in the text. For larger hierarchy levels, we observe larger occurrence probability. The dashed lines correspond \rev{to the} functions $h(k,p)=\lceil Q^k/2 \rceil p$.}
    \label{fig:flipSig_num}
\end{figure}

If an erroneous correction chain is realized with \rev{at least} $\lceil Q^k/2 \rceil$ data errors by \texttt{FlipSignal} errors, it will be completed to a full error chain of $Q^k$ data errors within the next $U^k$ CA steps via lower-level correction processes. We denote the probability of such a correction event by $p_{L,f}$ and simulate noisy \texttt{FlipSignal} propagation numerically for three different hierarchy levels $k$ and corresponding parameters $U^k, Q^k$ with $U=10$ and $Q=3$, shown in Fig.~\ref{fig:flipSig_num}. \rev{The occurrence probabilities are estimated from $N=10^4$--$10^6$ independent Monte Carlo samples per parameter point.} We can see that higher levels in the error correction hierarchy are more sensitive to \texttt{FlipSignal} errors, leading to \rev{a larger occurrence probability of long error chains}. In leading order, a single bit-flip on the source \texttt{FlipSignal} before time step $t=\lceil Q^k/2 \rceil$ will \rev{generate such an erroneous correction chain, giving} $p_{L,f} \approx \lceil Q^k/2 \rceil p$, as shown by fits in dashed lines in Fig.~\ref{fig:flipSig_num}. As error chains along straight one-dimensional lines represent minimum-weight errors causing logical errors on the toric code, \rev{the same mechanism is present in Harrington's two-dimensional decoder, providing an explanation for} the strong detrimental effect under signal noise observed in Fig.~\ref{fig:H2d_sigs}. The rapid decrease in $\langle T_F \rangle$ with increasing code distance $d$ is attributed to \rev{the increasing length of erroneous correction chains generated at higher hierarchy levels} under \texttt{FlipSignal} noise.

\section{Average logical lifetime of an absorbing Markov process}\label{app:T_proof}

In this section, we derive expressions for the average logical lifetime (mean first-passage time) in absorbing Markov processes and relate them to the scaling behavior observed in the main text. For a general introduction to absorbing Markov chains and first-passage times, see, e.g., standard references such as \cite{norris_markov_1998,levin_markov_2017}. In our setting, transient states correspond to configurations without logical error, while absorbing states represent logical failure. The resulting hitting time therefore models the logical lifetime of the encoded information. In the simplest case, an absorbing Markov chain has a single absorbing and a single transient state with transient-to-absorbing-state transition probability $p$ and transient-to-transient-state transition probability $q=1-p$. For the corresponding absorbing Markov chain transition matrix $P$ we have

\begin{equation}
    P = 
    \begin{pmatrix}
        q & p \\
        0 & 1
    \end{pmatrix}.
\end{equation}

As the transition probabilities to transient and absorbing states are constant, the probability of \rev{first reaching} the absorbing state at time $t$, given the initial state is transient is obtained by

\begin{equation}\label{eq:P_1}
    \mathbb{P}(t)=q^{t-1}p.
\end{equation}

The expected time until absorption can be calculated from the general formula for expectation values,

\begin{align}\label{eq:E_1}
    \mathbb{E}(T) &=  \sum_{t=1}^{\infty} t q^{t-1} p 
                = p \frac{d}{dq} \sum_{t=0}^{\infty} q^t \\
                &= p \frac{d}{dq} \frac{1}{1-q}
                = \frac{p}{(1-q)^2} = 1/p,
\end{align}

where $T$ is a random variable over all absorption times $t$ and we use the geometric series as $q<1$. In the main text, we observe \rev{in the approximately geometric regime} that the logical failure probability scales with physical error rate $r$ as $p_L\sim r^{\lambda(d)}$, where $\lambda(d)$ defines an effective distance. Combining this with the above equations, we obtain $\langle T_F \rangle \sim r^{-\lambda(d)}$, \rev{consistent with} the observed scaling of the logical lifetime. Eqs.~\ref{eq:P_1} and \ref{eq:E_1} are exactly the probability mass function and mean of the geometric distribution. The geometric distribution models a stochastic Bernoulli process, i.e., repeated coin flips with time-independent probability $p$ of hitting the absorbing state and probability \rev{$p=1-q$} of hitting the transient state. In the context of decoding, this corresponds to a regime in which logical failures occur as rare and approximately independent events, with an effective failure probability per time step $p_L$. In this regime, the logical lifetime is therefore given by $\langle T_F \rangle = 1/p_L$. Next, we generalize the geometric series to multiple transient and absorbing states. 

\rev{To avoid confusion with the colony-size parameter $Q$ used for Harrington's decoder, we denote the transient block of the transition matrix by $W$.}

The transition matrix of an absorbing Markov process, $P$, can always be written in block-matrix form,
\begin{equation}
    P =
    \begin{pmatrix}
        W & R \\
        0 & \mathbb{I}
    \end{pmatrix},
\end{equation}
with square matrix $W$ describing transient-to-transient-state transitions and matrix $R$ describing transient-to-absorbing-state transitions, identity matrix $\mathbb{I}$ and the matrix of all zeros $0$. In this setting, transient states correspond to different error configurations of the system, while matrix $W$ captures their stochastic evolution under noise and decoding. The matrix $R$ describes transitions from these configurations to logical failure. Since $P$ is a stochastic matrix, its rows must sum to unity. Thus, for $\dim{W}=d_W \times d_W$ and $\dim{R}=d_W \times d_R$ we have,
\begin{align}
    \begin{pmatrix}
        W & R
    \end{pmatrix}\mathbf{1}_{d_W+d_R} &= \mathbf{1}_{d_W} \\
    W\mathbf{1}_{d_W} + R\mathbf{1}_{d_R} &= \mathbf{1}_{d_W} \\
    R\mathbf{1}_{d_R} &= (\mathbb{I}-W)\mathbf{1}_{d_W},
\end{align}
where $\mathbf{1}_{d_x}$ is the $d_x$-element column vector of ones. Thus, the matrix relation $R\mathbf{1} = (\mathbb{I}-W)\mathbf{1}$ plays the role of the scalar identity $q=1-p$ in the geometric process, relating transition probabilities to absorption probabilities.

Next, we define the probability vector whose $k^{th}$ element is
\begin{equation}
    \mathbb{P}_k(t) = (W^{t-1}R\mathbf{1})_k.
\end{equation}
Each element of vector $\vec{\mathbb{P}}(t)$ describes the probability of the Markov chain starting in the transient state $k$ and transitioning to one of the absorbing states for the first time at time step $t$. In analogy with the geometric series, we obtain the \rev{vector of expectation values $\vec{\mathbb{E}}(T)$, whose $k^{th}$ element $\mathbb{E}_k(T)$ is} the average hitting time of the Markov chain started in transient state $k$,
\begin{align}\label{eq:E_k}
    \rev{\vec{\mathbb{E}}(T)}
    &= \rev{\sum_{t=1}^\infty t\,\vec{\mathbb{P}}(t)}
     = \sum_{t=1}^\infty t W^{t-1} R \mathbf{1} \\
    &= (\underbrace{\mathbb{I}+2W+3W^2+\ldots}_{S})R\mathbf{1} \\
    &= (\mathbb{I}-W)^{-2}R\mathbf{1} \\
    &= (\mathbb{I}-W)^{-2}(\mathbb{I}-W)\mathbf{1}\\
    &= (\mathbb{I}-W)^{-1}\mathbf{1}.
\end{align}
This expression shows that the logical lifetime is governed by the spectral properties of the transient transition matrix $W$. In particular, when \rev{the spectral radius $\rho(W)$} approaches unity, the expected hitting time becomes large, corresponding to long-lived logical information. To obtain the equality $S=(\mathbb{I}-W)^{-2}$ used in the second line of Eq.~\ref{eq:E_k}, we calculate,
\begin{align*}
    S &= \mathbb{I} + 2W + 3W^2 + \ldots\\
    WS &= W + 2W^2 + 3W^3 + \ldots\\
    S - WS &= (\mathbb{I} - W)S \\
    &= \mathbb{I} + 2W + 3W^2 + 4W^3+\ldots - W - 2W^2 - 3W^3 - \ldots \\
    &= \mathbb{I} + W + W^2 + \ldots \\
    &= (\mathbb{I}-W)^{-1},
\end{align*}
where we use the geometric series for square matrix $W$ with spectral radius $\rho(W)<1$. The spectral radius is defined as the maximal absolute eigenvalue of a matrix $X$, i.e., $\rho(X)=\rev{\max_i |\lambda_i|}$. Thus, we identify
\begin{equation}
    S = (\mathbb{I}-W)^{-2}.
\end{equation}
The geometric description assumes approximately independent failure events \rev{with a time-independent failure probability}. While the decoder dynamics can introduce temporal correlations, \rev{the general absorbing-Markov-chain result derived above does not require this approximation.} In all simulations under phenomenological noise, we choose an error-free state at $t=0$. Thus, the corresponding Markov chain starts with an initial (Kronecker)-delta distribution \rev{$(\pi_0)_i=\delta_{i,0}$}, i.e., a single peak at index 0 corresponding to the all-0 state. Hence, by Eq.~\ref{eq:E_k}, we obtain our final result for the average logical lifetime $\langle T_F \rangle$,
\begin{equation}\label{eq:T_analytic}
    \langle T_F \rangle = \mathbb{E}_{\pi_0}(T) = \pi_0 (\mathbb{I}-W)^{-1} \mathbf{1}.
\end{equation}

Note, in Markov-chain theory, state distribution vectors are \rev{represented as row vectors and multiplied from the left by} the transition matrix by convention.

\section{Efficient representation of an absorbing Markov process} \label{app:markov}

In the following, we present a method to obtain a more efficient matrix representation of the Markov process which describes $n$ independent Bernoulli variables, all starting in the 0-state and experiencing bit-flips with probability $p$. In particular, we describe how permutation symmetry of bit-strings, i.e., the (microscopic) states of the Markov process, can be used to reduce the dimension of its transition matrix $P$ from \rev{$2^n$ microscopic states to $n+1$ macroscopic states} where $n$ is the number of bits in a bit-string. The new transition matrix, $P^\prime$ describes transitions between (macroscopic) states corresponding to subsets of states with the same number of bits with value $1$. The reduced transition matrix remains an exact description of the dynamics.

For a single Bernoulli random variable with bit-flip probability $p$ and, respectively, probability $q=1-p$ to remain in its state, we have transition matrix $M$,
\begin{equation}
    M = 
    \begin{pmatrix}
        q & p\\
        p & q
    \end{pmatrix}.
\end{equation}
For $n$ independent such processes, we obtain the transition matrix as tensor product of $n$ single-state matrices,
\begin{equation}
    P = M^{\otimes n}.
\end{equation}
As $\dim{M}=2 \times 2$, the transition matrix $P$ grows exponentially in $n$, $\dim{P}=2^n\times 2^n$. Next, we use the structure of the state space of $P$ to find a more concise matrix $P^\prime$.
Notice that, due to the construction of $P$, the transition probabilities $P_{ij}$ are invariant under permutations of bits, i.e., $P$ commutes with the matrix representation of the entire symmetric group $S_{n}$ \rev{acting on the $n$ bit positions} and therefore partitions the state space into equivalence classes containing all states of the same number of ones. The action of $P$ on an arbitrary distribution $\pi$ is to effectively apply probabilistic transitions from any element of class $k$ ($k$ ones, $n-k$ zeros) in $\pi$ to any other class $l$ ($l$ ones, $n-l$ zeros). Notice that, although the probability to transition from one state in $k$ to two different states in $l$ might have different probability, the \rev{total probability of transitioning to the class $l$, obtained by summing over all states in that class,} is the same for every state in $k$. Thereby, we can express $P$ as a new matrix $P^\prime$, entirely in terms of transitions between  equivalence classes. 

In order to calculate the elements of $P^\prime$, we consider transitions from a state of $k$ ones to a state of $l$ ones. We start with the case in which $k=l$. Starting from a particular bit-string of $k$ ones, we reach another bit-string of $l=k$ ones by \textit{redistributing} the ones in the original string. To redistribute a single one, we need at least two bit-flips, one to swap one of the initial ones to a zero and one to flip an initial zero to a one. For this process, we can select one of the ${k\choose{1}}$ initial ones and one of the ${n-k\choose{1}}$ initial zeros to produce a different bit-string with exactly $l=k$ ones. Thus, for redistributing a single one we obtain overall $\mathbb{P}_1={k\choose{1}}{n-k\choose{1}}p^2 q^{n-2}$. Next, we consider redistributing two initial ones. In this case, we have $k\choose{2}$ possibilities to choose two initial ones and $n-k\choose{2}$ possibilities to choose two initial zeros. Thus, we have probability $\mathbb{P}_2 = {k\choose{2}}{n-k\choose{2}}p^4 q^{n-4}$ to obtain a new configuration of $k$ ones by four independent bit-flips. At most \rev{$\min(k,n-k)$ ones can be redistributed, since the same number of initial zeros must be flipped}. Summing up all those probabilities then yields the overall probability to remain in the class of states with $k$ ones after one step of the Markov process, i.e., element of the transition matrix 
\begin{equation}
    P^\prime_{k,k}=\sum_{j=0}^{\rev{\min(k,n-k)}} {k\choose{j}} {n-k\choose{j}} p^{2j} q^{n-2j}. 
\end{equation}
Note, we included $j=0$, as the transition to the same state is also a valid transition to stay in class $k$.

Next, we consider the case $k<l$. Following the same logic as before, we start with a specific initial configuration of $k$ ones and $n-k$ zeros for which we want to obtain the probability to transition to any state with $l$ ones via a single Markov step. In order to generate a state of $l$ ones, we need to flip $\Delta = l-k$ zeros in the initial configuration. There are exactly ${n-k\choose{\Delta}}$ ways to choose such zeros, i.e., probability $\mathbb{P}_\Delta={n-k \choose{\Delta}}p^\Delta q^{n-\Delta}$ for this process to occur. Next, we consider flipping one of the initial ones and $\Delta + 1$ \rev{initial} zeros, generating a new bit-string of $l$ ones different from all prior generated strings. For this process, we have ${k\choose{1}}$ ways to choose the initial one and ${n-k \choose{\Delta+1}}$ ways to choose $\Delta+1$ initial zeros, yielding the probability $\mathbb{P}_{\Delta+1}={k\choose{1}}{n-k\choose{\Delta+1}} p^{\Delta+2} q^{n-\Delta-2}$. In general, we have for \rev{$j\leq\min(k,n-l)$} that $\mathbb{P}_{\Delta+j}={k\choose{j}}{n-k\choose{\Delta+j}} p^{\Delta + 2j} q^{n-\Delta - 2j}$ and thus in general for $k<l$,
\begin{equation}
    P^\prime_{k,l}=\sum_{j=0}^{\rev{\min(k,n-l)}} {k\choose{j}} {n-k\choose{\Delta + j}} p^{\Delta + 2j} q^{n-\Delta-2j}. 
\end{equation}

Finally, we consider the case $k>l$. In this case, we have to flip at least $\Delta = k-l$ initial ones to obtain a configuration of $l$ ones, i.e., $\mathbb{P}^\prime_\Delta = {k\choose{\Delta}}p^{\Delta}q^{n-\Delta}$. Next, we flip $\Delta+1$ initial ones and one initial zero to obtain $l$ ones which happens with probability $\mathbb{P}^\prime_{\Delta+1}={k\choose{\Delta+1}}{n-k\choose{1}}p^{\Delta+2} q^{n-\Delta-2}$. Thus, in general for \rev{$j \leq \min(l,n-k)$} we have $\mathbb{P}^\prime_{\Delta+j} = {k\choose{\Delta+j}}{n-k\choose{j}}p^{\Delta+2j}q^{n-\Delta-2j}$, yielding the overall probability
\begin{equation}
    P^\prime_{k,l} = \sum_{j=0}^{\rev{\min(l,n-k)}} {k\choose{\Delta+j}} {n-k\choose{j}} p^{\Delta+2j} q^{n-\Delta-2j},
\end{equation}
for $k>l$, which completes our analysis. In the main text, we calculate these matrix elements at runtime for each pair of values of $n,p$ and subsequently obtain \rev{the corresponding exact mean first-passage times} by Eq.~\ref{eq:T_analytic}.

\section{Minimum-weight error configuration for \texttt{SCALA1D} for code distance $d=7$} \label{app:scala1d_d7_minweight}

In this section, we describe a finite-size error mechanism in \texttt{SCALA1D} under phenomenological data-qubit noise that leads to a logical failure at code distance $d=7$ with total weight $w=3$, below the minimum weight predicted by the \rev{large-distance scaling} in the main text.

\begin{figure}[hbtp]
    \centering
    \includegraphics[scale=0.65]{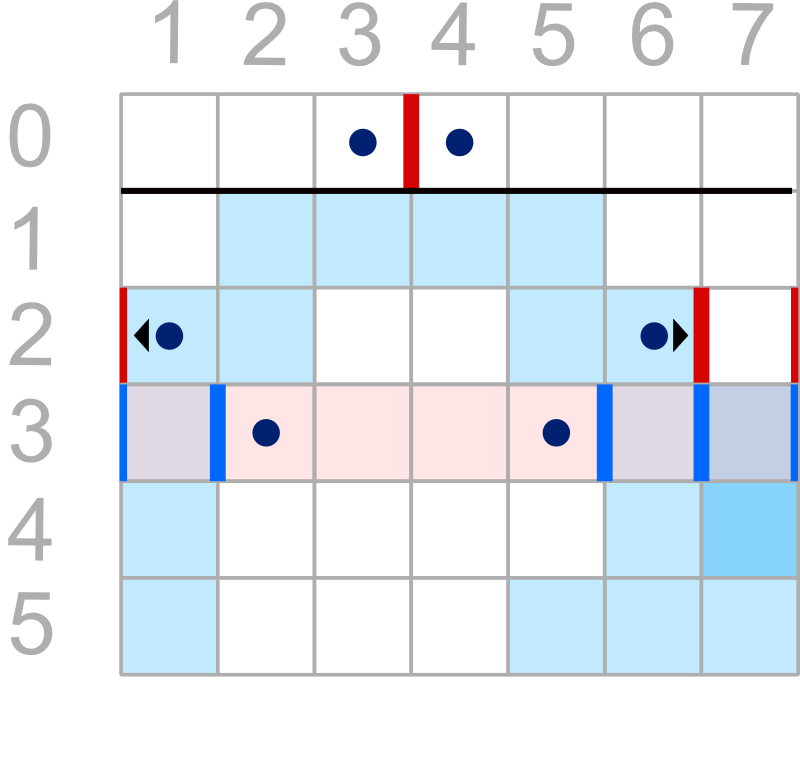}
    \caption{Illustration of a finite-size error process in \texttt{SCALA1D} for code distance $d=7$ under phenomenological noise leading to a logical error (shaded red) at time step 3 consisting of four errors (blue bars) induced by \rev{a total of} $w=3$ errors (red bars at $t=0$ and $t=2$). Signals (blue squares) generated by locally seeded defects (dots) propagate in opposite directions and meet after wrapping around the periodic lattice, triggering a correction chain for \rev{the two-qubit error configuration present at $t=2$} that results in a logical failure. This process constitutes a finite-size effect and produces a lower-weight logical error than predicted by the asymptotic minimal-weight analysis.}
    \label{fig:scala1d_d=7_min_w}
\end{figure}

The process is illustrated in Fig.~\ref{fig:scala1d_d=7_min_w}. For a configuration of three data-qubit errors, the decoder generates a set of signals which propagate along the one-dimensional lattice according to the local signaling rules. Due to the periodic boundary conditions, these signals travel in opposite directions and meet only after wrapping around the lattice. Upon meeting, the greedy signal-follow subrule connects the corresponding defects and produces a nontrivial error chain, resulting in a logical failure.

The periodic wrap-around trajectory therefore provides a \rev{finite-size mechanism for the anomalously low minimum weight observed at $d=7$}. \rev{For the larger code distances investigated, we do not observe a corresponding deviation from the large-distance minimum-weight scaling described in the main text. These numerical results support interpreting the $d=7$ wrap-around process as a finite-size effect rather than a change in the observed large-$d$ scaling.}

\section{\rev{Phenomenological crossing drift and screening in \texttt{SCALA}}}
\label{app:scala_screening}

\rev{The low-$p$ effective-distance scaling discussed in Sec.~\ref{sec:scala1d} characterizes the suppression of rare logical failures at fixed finite distance, but does not by itself establish a nonzero phenomenological threshold. To probe the large-distance behavior independently, we first extract pairwise crossings of the stationary logical-error-rate curves of \texttt{SCALA1D}.}

\begin{figure}[hbtp]
    \centering
    \includegraphics[width=\columnwidth]{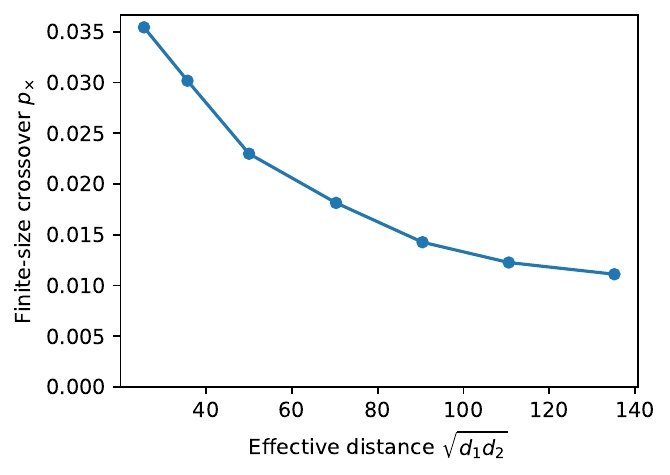}
    \caption{\rev{Pairwise finite-size crossover points $p_\times$ of the \texttt{SCALA1D} phenomenological logical-error-rate curves as a function of the effective pair distance $d_{\mathrm{eff}}=\sqrt{d_1d_2}$. The crossover points move systematically toward lower physical error rates with increasing distance, with no indication of convergence to a common nonzero value over the investigated range.}}
    \label{fig:scala1d_pheno_crossings}
\end{figure}

\rev{As shown in Fig.~\ref{fig:scala1d_pheno_crossings}, the pairwise crossings shift systematically toward lower physical error rates as the distance increases, with no indication of convergence to a common nonzero value over the investigated range. This finite-size drift is conceptually distinct from the low-$p$ failure-weight scaling: the latter describes how rapidly $p_L$ decreases with physical error rate at fixed $d$, whereas the crossing drift probes whether increasing $d$ continues to improve the decoder at a fixed nonzero physical error rate.}

\rev{A mechanism consistent with this drift is screening of the signal-mediated attraction. For an isolated pair of defects, signals can in principle propagate over a large distance and bias the local dynamics so that the pair approaches and annihilates. Under phenomenological noise at fixed $p=q>0$, however, the decoder operates in a finite-density background of continuously generated defects and their signals. Signals associated with different defects overlap and compete, so the directional information associated with any particular defect pair becomes progressively less effective with increasing separation. The result is an effective interaction range set density of background defects and their signals rather than by the total system size.}

\rev{We measure this effect directly in \texttt{SCALA1D} by preparing a tagged pair of defects with initial separation $r_0$ in a phenomenological-noise background and measuring the change in their separation over one reset period $t_R$, conditioned on both tagged defects surviving until the end of that period. Negative $\langle\Delta R\rangle_{\mathrm{surv}}$ therefore corresponds to an effective attraction between the tagged defects.}

\begin{figure}[hbtp]
    \centering
    \includegraphics[width=\columnwidth]{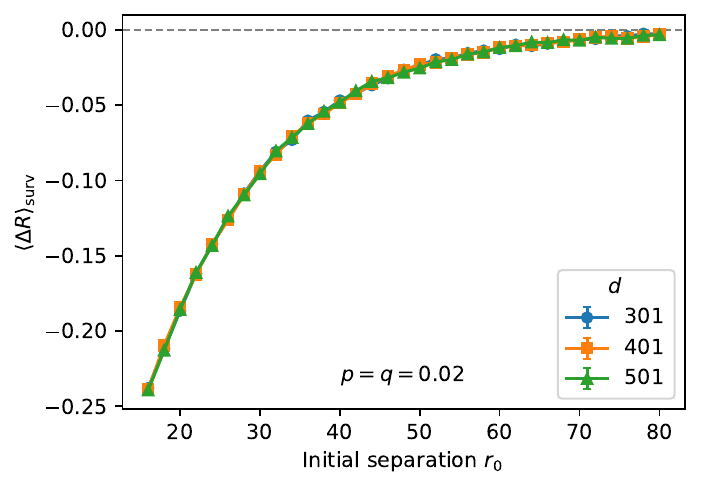}
    \caption{\rev{Screening of the signal-mediated attraction in \texttt{SCALA1D} at $p=q=0.02$. We show the mean change $\langle\Delta R\rangle_{\mathrm{surv}}$ in the separation of a surviving tagged defect pair as a function of its initial separation $r_0$. Negative values indicate effective attraction. The attraction decreases with separation and approaches zero, while the results for $d=301$, $401$, and $501$ nearly collapse. The interaction range is therefore controlled by the finite-density background rather than by the total code distance over the investigated range.}}
    \label{fig:scala1d_screening}
\end{figure}

\begin{table*}[hbtp]
    \centering
    \caption{\rev{Sub-threshold power-law fits $p_L=A(d)p^{\lambda(d)}$ for the code-capacity data of Harrington's two-dimensional decoder and \texttt{SCALA2D}. The quoted uncertainties are statistical fit uncertainties. No fit is reported when the available data do not sufficiently resolve the asymptotic sub-threshold regime.}}
    \label{tab:code_capacity_tail_fits}
    \begin{tabular}{lcccccc}
        \hline\hline
        Decoder & $d$ & fit interval & $A(d)$ & $\lambda(d)$ & $N_{\mathrm{fit}}$ & $\chi^2_{\mathrm{red}}$ \\
        \hline
        Harrington2D & 3  & $[0.0100,0.0350]$ & $3.0769\times10^{1}$ & $1.9378\pm0.0084$ & $9$ & $1.092$ \\
                     & 9  & $[0.0025,0.0150]$ & $1.9975\times10^{4}$ & $3.9443\pm0.0065$ & $7$ & $2.105$ \\
                     & 27 & --- & --- & --- & --- & --- \\
                     & 81 & --- & --- & --- & --- & --- \\
        \hline
        \texttt{SCALA2D} & 3   & $[0.0010,0.0150]$ & $4.1662\times10^{1}$  & $1.9875\pm0.0186$ & $7$ & $0.259$ \\
                         & 9   & $[0.0040,0.0060]$ & $1.4432\times10^{4}$  & $4.1646\pm0.1021$ & $3$ & $0.134$ \\
                         & 15  & $[0.0090,0.0175]$ & $2.8159\times10^{6}$  & $6.2865\pm0.3195$ & $4$ & $0.744$ \\
                         & 21  & $[0.0150,0.0250]$ & $9.7687\times10^{8}$  & $8.2183\pm0.2105$ & $5$ & $1.965$ \\
                         & 31  & $[0.0225,0.0375]$ & $2.0099\times10^{13}$ & $11.8283\pm0.1604$ & $7$ & $0.515$ \\
                         & 51  & --- & --- & --- & --- & --- \\
                         & 81  & --- & --- & --- & --- & --- \\
                         & 111 & --- & --- & --- & --- & --- \\
        \hline\hline
\end{tabular}
\end{table*}

\rev{The collapse of the large-$d$ curves in Fig.~\ref{fig:scala1d_screening} shows that, at the investigated $p=q=0.02$, increasing the system size does not extend the distance over which a tagged pair experiences appreciable attraction. The signal-mediated attraction therefore has a finite characteristic range over the investigated distances. This screening provides a possible dynamical mechanism for the downward-moving crossover points in Fig.~\ref{fig:scala1d_pheno_crossings}: increasing $d$ beyond the interaction range does not generate correspondingly longer-range attraction, consistent with the absence of a stable nonzero crossing over the investigated distance range.}

\rev{This interpretation is closely related to the confinement perspective of Ref.~\cite{lake2025localactiveerrorcorrection}. There, fault-tolerant local active error correction is obtained by engineering an effective confining interaction whose protection persists on increasing length scales. The construction uses geometrically local message passing but allows the local classical resources to grow polylogarithmically with system size. By contrast, the present \texttt{SCALA} signaling rule uses a fixed local state space. Our results illustrate the complementary failure mode: an attractive local interaction may not yield increasing protection with system size if that attraction is screened at a finite length scale.}

\rev{We show the screening diagnostic explicitly only for \texttt{SCALA1D}, where the separation of a tagged pair provides a direct scalar measure of the effective attraction. Since \texttt{SCALA2D} uses the same fixed local signaling resources in a finite-density background of competing defects, a related screening mechanism may also operate there. The absence of a stable common crossing in its phenomenological data is consistent with this interpretation, but establishing screening directly in two dimensions would require an analogous diagnostic.}

\section{Sub-threshold code-capacity fits}
\label{app:subthreshold_fits}

\rev{Here we provide the fit parameters and fit windows used for the sub-threshold code-capacity scaling shown in Figs.~\ref{fig:h2d_cc} and \ref{fig:scala2d_cc}. Where the accessible low-$p$ data approach the asymptotic regime, we fit the logical error rate to
\begin{equation}
    p_L=A(d)p^{\lambda(d)},
\end{equation}
using weighted least squares with the Monte Carlo standard errors as statistical uncertainties. Both $A(d)$ and $\lambda(d)$ are treated as free fit parameters. The resulting parameter uncertainties are obtained from the covariance matrix of the weighted fit. To avoid interpreting pre-asymptotic finite-$p$ slopes as asymptotic exponents, we report numerical fits only where the low-$p$ behavior yields a stable exponent under variations of the fit window. Curves that remain pre-asymptotic over the statistically resolved range are shown in the corresponding figures but omitted from the exponent fits. For \texttt{SCALA2D}, the fitted exponents are compared to the discrete prediction $w_0(d)$ of Eq.~\ref{eq:scala_discrete_w0}. The accepted \texttt{SCALA2D} fit windows additionally require an acceptable reduced $\chi^2$. Table~\ref{tab:code_capacity_tail_fits} reports the fit intervals and resulting parameters, with $N_{\mathrm{fit}}$ denoting the number of data points included in each fit.}

\rev{For Harrington's two-dimensional decoder, the accessible Monte Carlo data for $d=27$ and $81$ do not reach a sufficiently clear asymptotic regime to extract reliable numerical exponents. For $d=27$, the hierarchical construction predicts $w_{\min}=2^{\log_3(27)}=8$, while the statistically resolved data have not yet reached the corresponding $p_L\propto p^8$ regime. As a direct check of the hierarchical failure mechanism, we generate all $1458$ translated, reflected, and oriented configurations of the corresponding recursive weight-$8$ family, of which $648$ produce a logical failure under the full two-dimensional decoder. We additionally test all $11664$ distinct weight-$7$ configurations obtained by deleting one error from this family; all are corrected successfully, with no logical or unresolved outcomes. This is not an exhaustive search over all weight-$7$ configurations, but provides additional numerical support for the predicted hierarchical minimum-weight mechanism.}

\rev{For Harrington1D and \texttt{SCALA1D}, no corresponding numerical power-law fit is required in the code-capacity setting. Harrington1D is described exactly by concatenated majority voting, $p_L=p_{\mathrm{maj}}^m(p)$, as shown in Fig.~\ref{fig:har1d_cc}. Likewise, \texttt{SCALA1D} reproduces global majority voting in the code-capacity setting, so that its logical error rate is given analytically by Eq.~\ref{eq:pL_ML}, as demonstrated in Fig.~\ref{fig:scala1d_code_capa}.}

\section{Code-capacity threshold analyses}
\label{app:fss}

\rev{In this section, we provide additional details on the analyses used to determine the code-capacity thresholds reported in the main text. For Harrington2D and \texttt{SCALA1D}, we use the finite-size scaling analysis described below. For \texttt{SCALA2D}, the pairwise crossings exhibit appreciable finite-size drift and do not support a comparably clean single-parameter scaling collapse over the accessible system sizes. We therefore obtain a finite-size threshold estimate from the stability of the large-distance pairwise crossings, as described in App.~\ref{app:scala2d_crossings}.

For the finite-size scaling analyses, we use the single-parameter scaling ansatz
\begin{equation}
    p_L(p,d)
    =
    F\!\left[
        (p-p_c)d^{1/\nu}
    \right],
    \label{eq:fss_ansatz}
\end{equation}
where $p_c$ denotes the critical physical error rate, $\nu$ is the finite-size scaling exponent, and $F$ is a smooth scaling function. Close to the critical point, we approximate $F$ by a low-order polynomial and determine $p_c$, $\nu$, and the polynomial coefficients from a weighted nonlinear least-squares fit to the Monte Carlo data. Statistical uncertainties of $p_L$ are obtained from the corresponding binomial sampling statistics, $\sigma_{p_L}=\sqrt{p_L(1-p_L)/N}$. Unless stated otherwise, the quoted uncertainties of the fitted parameters are one-standard-deviation uncertainties obtained from the covariance matrix of the weighted nonlinear least-squares fit. We quantify the goodness of fit by the reduced chi-squared
\begin{equation}
    \chi^2_{\mathrm{red}}
    =
    \frac{1}{N_{\mathrm{data}}-N_{\mathrm{par}}}
    \sum_i
    \frac{\left[p_{L,i}-p_L^{\mathrm{fit}}(p_i,d_i)\right]^2}
         {\sigma_i^2},
\end{equation}
where $N_{\mathrm{data}}$ is the number of fitted data points, $N_{\mathrm{par}}$ the number of fitted parameters, and $\sigma_i$ the Monte Carlo standard error of $p_{L,i}$. Values of $\chi^2_{\mathrm{red}}$ of order unity indicate consistency of the residuals with the stated statistical uncertainties.}

\subsection{Harrington2D}
\label{app:fss_harrington2d}

\begin{figure}[hbtp]
    \centering
    \includegraphics[scale=0.7]{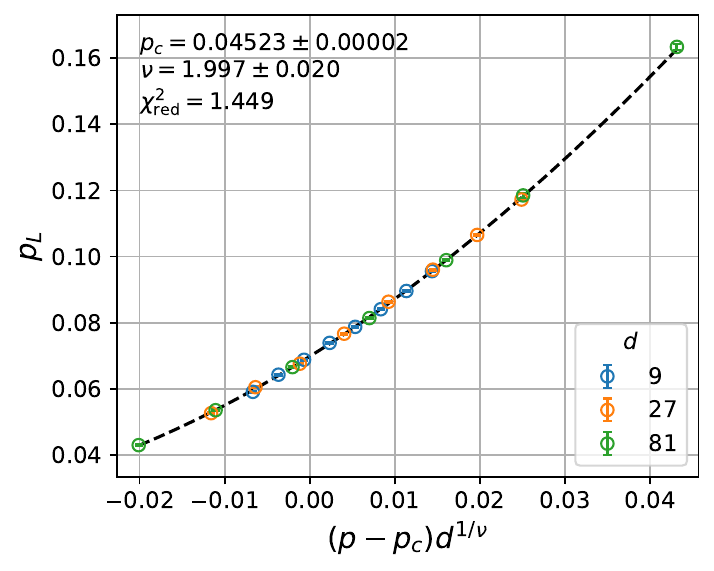}
    \caption{\rev{Finite-size scaling collapse of the Harrington2D code-capacity data for $d=9,27,81$, using the scaling variable $(p-p_c)d^{1/\nu}$. The dashed black line shows the fitted cubic approximation to the scaling function $F$. Error bars show the Monte Carlo standard errors. The distance $d=3$ is excluded because of substantial finite-size corrections.}}
    \label{fig:fss_harrington2d}
\end{figure}

\rev{For Harrington's two-dimensional decoder, we perform the finite-size scaling fit using code distances $d=9,27,81$ and physical error rates in the interval $p\in[0.043,0.050]$. The scaling function $F$ in Eq.~\eqref{eq:fss_ansatz} is approximated by a cubic polynomial, and all data points are weighted by their Monte Carlo standard errors. We obtain
\begin{equation}
    p_c = 0.04523\pm0.00002,
    \qquad
    \nu = 1.997\pm0.020,
\end{equation}
with reduced chi-squared $\chi^2_{\mathrm{red}}=1.449$. The corresponding collapse is shown in Fig.~\ref{fig:fss_harrington2d}.}

\rev{The smallest distance $d=3$ is excluded from the final fit because it exhibits substantial finite-size corrections and is not described by the same scaling collapse. In otherwise comparable fit windows, including $d=3$ increases the reduced chi-squared to values above $20$, whereas fits using $d=9,27,81$ give $\chi^2_{\mathrm{red}}$ of order unity. The fitted values of $p_c$ and $\nu$ are also stable under moderate variations of the fitting interval around the critical point.}

\subsection{\texttt{SCALA1D}}
\label{app:fss_scala1d}

\begin{figure}[hbtp]
    \centering
    \includegraphics[scale=0.7]{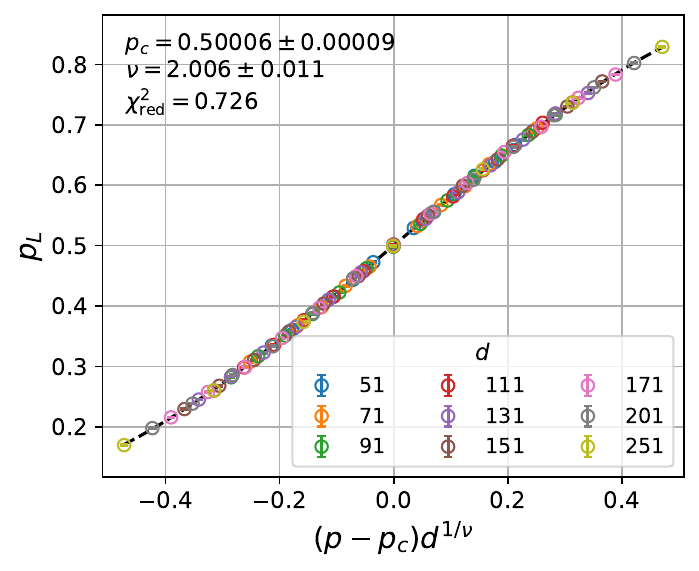}
    \caption{\rev{Finite-size scaling collapse of the \texttt{SCALA1D} code-capacity data using the scaling variable $(p-p_c)d^{1/\nu}$. The dashed black line shows the fitted cubic approximation to the scaling function $F$, and error bars show the Monte Carlo standard errors.}}
    \label{fig:fss_scala1d}
\end{figure}

\rev{For \texttt{SCALA1D}, we perform the finite-size scaling analysis using the ansatz in Eq.~\eqref{eq:fss_ansatz}. The final fit uses code distances $d\geq 51$ and physical error rates in the interval $p\in[0.47,0.53]$. The scaling function $F$ is approximated by a cubic polynomial and all data points are weighted by their Monte Carlo standard errors. We obtain
\begin{equation}
    p_c = 0.50006\pm0.00009,
    \qquad
    \nu = 2.006\pm0.010,
\end{equation}
with reduced chi-squared $\chi^2_{\mathrm{red}}=0.726$. The corresponding data collapse is shown in Fig.~\ref{fig:fss_scala1d}. The result is consistent with the exact majority-voting threshold $p_c=1/2$ and the majority-voting finite-size scaling exponent $\nu=2$.}

\rev{The fitted values are stable under moderate variations of both the minimum included code distance and the fitting interval around the transition.}

\subsection{\texttt{SCALA2D} pairwise-crossing analysis}
\label{app:scala2d_crossings}

\rev{For \texttt{SCALA2D}, the pairwise crossings of the code-capacity logical-error-rate curves show appreciable finite-size fluctuations for small and intermediate code distances. We therefore do not fit these crossings to a specific finite-size drift ansatz. Instead, we determine pairwise crossing positions and analyze their stability as the minimum included system size is increased. For each pair of code distances $(d_1,d_2)$, we characterize the crossing by the effective distance
\begin{equation}
    d_{\mathrm{eff}}=\sqrt{d_1d_2}.
\end{equation}
For each adjacent pair of simulated distances, we use the common data points in the crossing window $p\in[0.071,0.078]$ and apply an increasing isotonic regression to each curve, weighted by the inverse squared Monte Carlo standard errors. The crossing $p_\times$ is then obtained by linear interpolation between adjacent points at which the difference of the two fitted curves changes sign. Statistical uncertainties are estimated from $3000$ parametric bootstrap realizations, in which each $p_L$ is resampled from a normal distribution with its measured mean and Monte Carlo standard error and the complete crossing analysis is repeated; the reported uncertainty is the standard deviation of the resulting crossing positions.}

\begin{figure}[hbtp]
    \centering
    \includegraphics[scale=0.7]{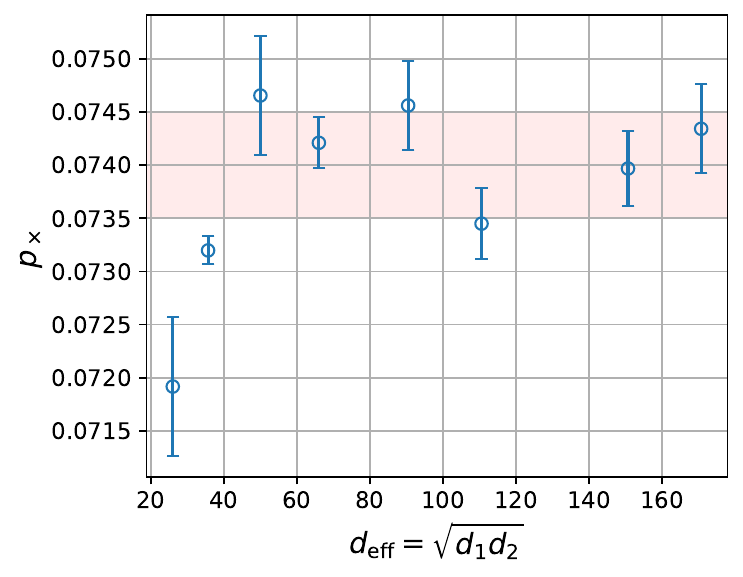}
    \caption{\rev{Representative pairwise crossing estimates $p_\times$ for the \texttt{SCALA2D} code-capacity logical-error-rate curves as a function of $d_{\mathrm{eff}}=\sqrt{d_1d_2}$. Error bars denote bootstrap uncertainties. The red dashed line shows $p=0.0740$, and the shaded region indicates the $\pm0.0005$ range assigned from the residual variation of the largest-distance crossings. Only a representative subset of pairwise crossings is shown for visual clarity; the large-distance stability analysis uses all available crossings satisfying the stated distance cut.}}
    \label{fig:scala2d_crossings}
\end{figure}

\rev{Figure~\ref{fig:scala2d_crossings} shows a representative subset of the resulting pairwise crossings spanning the simulated distance range. The small-distance crossings exhibit substantial finite-size drift, whereas the largest-distance crossings cluster around $p_\times\simeq0.074$. Restricting to crossings with $d_{\mathrm{eff}}\geq80$ leaves four crossing estimates and gives the inverse-variance-weighted mean
\begin{equation}
    \bar p_\times = 0.07402 \pm 0.00019 ,
\end{equation}
where the quoted uncertainty is statistical. The sample standard deviation of these four crossings is $4.9\times10^{-4}$. Increasing the lower cut to $d_{\mathrm{eff}}\geq100$ and $d_{\mathrm{eff}}\geq120$ gives weighted means $0.07387$ and $0.07412$, respectively, showing that the variation under reasonable large-distance cuts is comparable to the residual scatter among the largest-distance crossings. We therefore use
\begin{equation}
    p_c \simeq 0.0740(5)
\end{equation}
as our large-distance finite-size estimate of the code-capacity threshold. The quoted uncertainty characterizes the residual variation of the largest-distance crossings under the stated distance cuts; it should not be interpreted as a controlled extrapolation uncertainty in the thermodynamic limit.}

\section{Another measurement error process in Harrington's decoder}

\rev{In this section, we analyze an additional process by which measurement errors alone can cause logical failure in Harrington's decoder. This process is not dominant in the low-noise regime, since it requires multiple measurement errors to accumulate, but it illustrates another potential failure mechanism of Harrington's accumulation-and-signaling scheme.}

\rev{Under measurement noise, repeated faulty measurements of a representative cell can accumulate in its defect counter and create a \textit{ghost} defect. If a level-$k$ center cell accumulates at least $f_CU^k$ such events during a working period of length $U^k$, the decoder can erroneously infer a level-$k$ defect and execute the corresponding default correction chain. This process is most consequential at high hierarchy levels, where the resulting correction chains are longest. Since the measurement errors are independent, the number of accumulated ghost defects is binomially distributed, giving}
\begin{equation}
    \mathbb{P}_k = \sum_{l=f_CU^k}^{U^k}{U^k\choose{l}}\rev{q^l(1-q)^{U^k-l}}.
\end{equation}
\rev{For $q<f_C$, this binomial tail can be upper-bounded using a Chernoff bound \cite{chernoff_measure_1952},}
\begin{equation}\label{eq:upper_bound}
    \mathbb{P}_k \leq \mathbb{P}_k^{\uparrow}
    = \exp\bigg(-U^k D\bigg(f_C\parallel q\bigg)\bigg),
\end{equation}
\rev{where}
\begin{equation}
    D(a \parallel q)
    = a\ln\frac{a}{q} + (1-a)\ln\frac{1-a}{1-q}
\end{equation}
\rev{is the binary Kullback--Leibler divergence \cite{kullback_information_1951}. A corresponding lower bound is}
\begin{equation}\label{eq:lower_bound}
    \rev{
    \mathbb{P}_k \geq \mathbb{P}_k^{\downarrow}
    =
    \frac{1}{\sqrt{2U^k}}
    \exp\left[-U^k D\left(f_C\parallel q\right)\right].
    }
\end{equation}
\rev{Comparing successive hierarchy levels gives}
\begin{equation}
    R_{\text{max}} = \mathbb{P}_{k}^{\uparrow} / \mathbb{P}_{k-1}^{\downarrow}=\sqrt{2 U^{k-1}}\exp(-\Delta U_k D(f_C\parallel \rev{q})),
\end{equation}
and 
\begin{equation}
    R_{\text{min}} = \mathbb{P}_{k}^{\downarrow} / \mathbb{P}_{k-1}^{\uparrow}=\frac{1}{\sqrt{2 U^k}}\exp(-\Delta U_k D(f_C\parallel \rev{q})),
\end{equation}
\rev{where $\Delta U_k=U^k-U^{k-1}$ and $R_{\mathrm{min}}\leq\mathbb{P}_k/\mathbb{P}_{k-1}\leq R_{\mathrm{max}}$. For $q<f_C$, the exponential factor increasingly suppresses this channel at higher hierarchy levels. At $q=f_C$, this exponential suppression vanishes and only the algebraic prefactors remain.}

\begin{figure}[hbtp]
    \centering
    \includegraphics[scale=0.65]{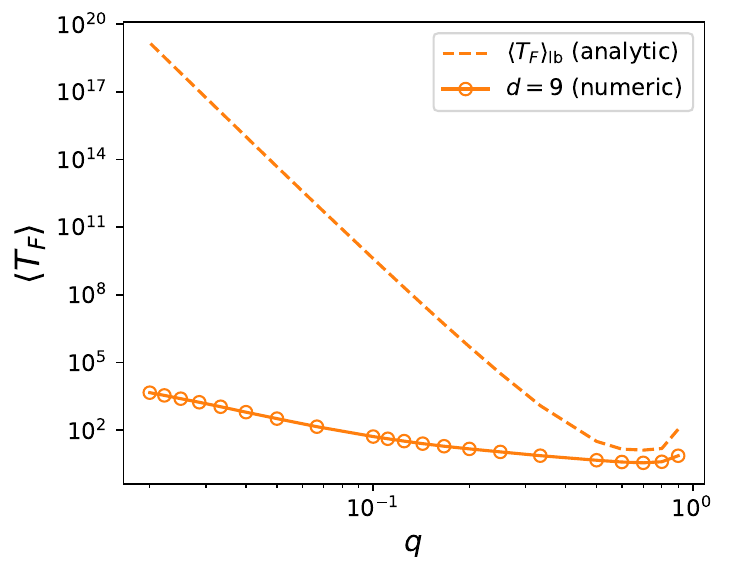}
    \caption{\rev{Average logical lifetime $\langle T_F\rangle$ of the one-dimensional Harrington decoder under pure measurement noise for $d=9$ (solid), compared with the characteristic lifetime $\langle T_F\rangle_{\mathrm{ghost}}$ associated with the level-$1$ ghost-defect accumulation channel (dashed).}}
    \label{fig:ll_hl_perf_msmt}
\end{figure}

\rev{To compare this channel with the observed logical lifetime, we use the upper bound $\mathbb{P}_1^\uparrow$ from Eq.~\ref{eq:upper_bound}. Since at least two erroneous level-$1$ chains are required to produce a logical error, the corresponding failure probability is $p_{\mathrm{maj}}(\mathbb{P}_1^\uparrow)$, where $p_{\mathrm{maj}}(p)=3p^2(1-p)+p^3$. With the level-$1$ update-cycle duration $\tau=U+Q$, the characteristic lifetime of this channel is}
\begin{equation}
    \langle T_F \rangle_{\mathrm{ghost}}
    := \frac{\tau}{p_{\mathrm{maj}}(\mathbb{P}_1^\uparrow)}.
\end{equation}
\rev{As shown in Fig.~\ref{fig:ll_hl_perf_msmt}, $\langle T_F\rangle_{\mathrm{ghost}}$ lies well above the observed logical lifetime. This accumulation channel is therefore subdominant; the shorter-timescale measurement-error mechanism is discussed in Sec.~\ref{har1dmeas}.}

\end{document}